\documentclass{article}
\usepackage{latexsym}

\DeclareMathAlphabet{\mbit}{OT1}{cmr}{bx}{it}
\DeclareMathAlphabet{\mssl}{OT1}{cmss}{m}{sl}
\DeclareMathAlphabet{\mssb}{OT1}{cmss}{bx}{n}
\newcommand{\ord}[1]{\langle {#1}, \leq_{\mathbf{{#1}}} \rangle}

\newcommand{\clsn}[1]{\langle \mathsf{inst}(\mathbf{{#1}}), \mathsf{typ}(\mathbf{{#1}}), \models_{\mathbf{{#1}}} \rangle}

\newcommand{\cl}[1]{\langle \mathsf{elem}(\mathbf{{#1}}), \leq_{\mathbf{{#1}}}, \bigwedge_{\mathbf{{#1}}}, \bigvee_{\mathbf{{#1}}} \rangle}
\newcommand{\clg}[1]{\langle \mathsf{lat}(\mathbf{{#1}}), \mathsf{inst}(\mathbf{{#1}}), \mathsf{typ}(\mathbf{{#1}}), \iota_{\mathbf{{#1}}}, \tau_{\mathbf{{#1}}} \rangle}
\newcommand{\clgmor}[3]{\langle \mathsf{left}(\mathbf{{#1}}), \mathsf{right}(\mathbf{{#1}}), \mathsf{inst}(\mathbf{{#1}}), \mathsf{typ}(\mathbf{{#1}}) \rangle : \mathbf{{#2}} \rightleftharpoons \mathbf{{#3}}}

\newtheorem{theorem}{Theorem}
\newtheorem{lemma}{Lemma}
\newtheorem{fact}{Fact}

\begin{document}

\title{Truth Factors}
\author{Robert E. Kent}
\maketitle

\begin{abstract}
Truth refers to the satisfaction relation used 
to define the semantics of model-theoretic languages.
The satisfaction relation for first order languages (truth classification),
and the preservation of truth by first order interpretations (truth infomorphism),
is a motivating example in the theory of Information Flow (IF) \cite{barwise:seligman:97}.
The abstract theory of satisfaction is the basis for the theory of institutions 
\cite{goguen:burstall:92}.
Factoring refers to categorical factorization systems.
The concept lattice, which is the central structure studied by the theory of Formal Concept Analysis (FCA) \cite{ganter:wille:99},
is constructed by a factorization. 
The study of classification structures (IF) 
and the study of conceptual structures (FCA)
aim (at least is part) to provide a principled foundation for 
the logical theory of knowledge representation and organization.
In an effort to unify these two areas,
the paper "Distributed Conceptual Structures" \cite{kent:02}
abstracted the basic theorem of FCA 
in order to established three levels of categorical equivalence 
between classification structures and conceptual structures.
In this paper we refine this approach 
by resolving the equivalence as the factorization of three isomorphic versions:
relation, function and Galois connection.
We develop the latter more algebraic version of the equivalence as the polar factorization of Galois connections.
We advocate this abstract adjunctive representation of classification and conceptual structures.
\end{abstract}

\tableofcontents

\section{Categories with Factorization Systems}

Let $\mathsf{C}$ be an arbitrary category.
Let $\mathsf{C}^{\mathbf{2}}$ denote the arrow category\footnote{Recall that $\mathbf{2}$ is the two-object category, pictured as $\bullet \rightarrow \bullet$, with one non-trivial morphism. The arrow category $\mathsf{C}^{\mathbf{2}}$ is (isomorphic to) the functor category $[\mathbf{2}, \mathsf{C}]$.} of $\mathsf{C}$.
An object of $\mathsf{C}^{\mathbf{2}}$ is a triple $(A, f, B)$,
where $f : A \rightarrow B$ is a $\mathsf{C}$-morphism.
A morphism of $\mathsf{C}^{\mathbf{2}}$,
$(a, b) : (A_1, f_1, B_1) \rightarrow (A_2, f_2, B_2)$,
is a pair of $\mathsf{C}$-morphisms $a : A_1 \rightarrow A_2$ and $b : B_1 \rightarrow B_2$ that form a commuting square
\begin{center}
\setlength{\unitlength}{0.8pt}
\begin{picture}(50,60)(0,-5)
\put(-50,25){\makebox(100,50){$A_1$}}
\put(0,25){\makebox(100,50){$B_1$}}
\put(-50,-25){\makebox(100,50){$A_2$}}
\put(0,-25){\makebox(100,50){$B_2$}}
\put(0,32){\makebox(50,50){\footnotesize{$f_1$}}}
\put(0,-32){\makebox(50,50){\footnotesize{$f_2$}}}
\put(-30,0){\makebox(50,50){\footnotesize{$a$}}}
\put(31,0){\makebox(50,50){\footnotesize{$b$}}}
\thicklines
\put(10,0){\vector(1,0){30}}
\put(10,50){\vector(1,0){30}}
\put(0,40){\vector(0,-1){30}}
\put(50,40){\vector(0,-1){30}}
\end{picture}
\end{center}
There are source and target projection functors
$\partial_0^{\mathsf{C}}, \partial_1^{\mathsf{C}} : \mathsf{C}^{\mathbf{2}} \rightarrow \mathsf{C}$
and an arrow natural transformation
$\alpha_{\mathsf{C}} : \partial_0^{\mathsf{C}} \Rightarrow \partial_0^{\mathsf{C}} : \mathsf{C}^{\mathbf{2}} \rightarrow \mathsf{C}$
with component
$\alpha_{\mathsf{C}}(A, f, B) = f : A \rightarrow B$
(background of Figure~\ref{factorization-equivalence}).
A \emph{factorization system} in $\mathsf{C}$ is a pair 
$\langle \mathsf{E}, \mathsf{M} \rangle$ of classes of $\mathsf{C}$-morphisms satisfying the following conditions. 
\begin{description}
\item [Subcategories:] All $\mathsf{C}$-isomorphisms are in $\mathsf{E} \cap \mathsf{M}$. Both $\mathsf{E}$ and $\mathsf{M}$ are closed under $\mathsf{C}$-composition.
\item [Existence:] Every $\mathsf{C}$-morphism $f : A \rightarrow B$ has an $\langle \mathsf{E}, \mathsf{M} \rangle$-factorization $(A, e, C, m, B)$, a quadruple where $e : A \rightarrow C$, $m : C \rightarrow B$, $f = e \cdot m$, $e \in \mathsf{E}$ and $m \in \mathsf{M}$.
\item [Uniqueness:] Any two $\langle \mathsf{E}, \mathsf{M} \rangle$-factorizations are isomorphic; that is, if $(A, e, C, m, B)$ and $(A, e^\prime, C^\prime, m^\prime, B)$ are two $\langle \mathsf{E}, \mathsf{M} \rangle$-factorizations of $f : A \rightarrow B$, then there is a $\mathsf{C}$-isomorphism $h : C \cong C^\prime$ with $e \cdot h = e^\prime$ and $h \cdot m^\prime = m$.
\end{description}
The uniqueness condition is equivalent to the following condition.
\begin{description}
\item [Diagonalization:] For every commutative square $e \cdot s = r \cdot m$ of $\mathsf{C}$-morphisms, with $e \in \mathsf{E}$ and $m \in \mathsf{M}$, 
there is a unique $\mathsf{C}$-morphism $d$ with $e \cdot d = r$ and $d \cdot m = s$.
\end{description}
Note that $\mathsf{E}$ form $\mathsf{M}$ are $\mathsf{C}$-subcategories with the same objects as $\mathsf{C}$.
Let $\mathsf{E}^{\mathbf{2}}$ denote the full subcategory of $\mathsf{C}^{\mathbf{2}}$
whose objects are the morphisms in $\mathsf{E}$.
Make the same definitions for $\mathsf{M}^{\mathbf{2}}$.
Just as for $\mathsf{C}^{\mathbf{2}}$,
the category $\mathsf{E}^{\mathbf{2}}$ has source and target projection functors
$\partial_0^\mathsf{E}, \partial_1^\mathsf{E} : \mathsf{E}^{\mathbf{2}} \rightarrow \mathsf{C}$
and arrow natural transformation
$\alpha_{\mathsf{E}} : \partial_0^\mathsf{E} \Rightarrow \partial_0^\mathsf{E} : \mathsf{E}^{\mathbf{2}} \rightarrow \mathsf{C}$
(foreground of Figure~\ref{factorization-equivalence}).
The same is true for $\mathsf{M}^{\mathbf{2}}$.

\begin{figure}
\begin{center}
\setlength{\unitlength}{0.8pt}
\begin{tabular}{c}
\begin{picture}(200,100)(-83,0)
\put(0,0){\begin{picture}(100,100)(0,0)
\put(5,75){\makebox(100,50){$\mathsf{E} {\odot_{\mathsf{C}}} \mathsf{M}$}}
\put(-50,25){\makebox(100,50){$\mathsf{E}^{\mathbf{2}}$}}
\put(50,25){\makebox(100,50){$\mathsf{M}^{\mathbf{2}}$}}
\put(-100,-25){\makebox(100,50){$\mathsf{C}$}}
\put(0,-25){\makebox(100,50){$\mathsf{C}$}}
\put(100,-25){\makebox(100,50){$\mathsf{C}$}}
\put(-32,56){\makebox(100,50){\footnotesize{$\pi_\mathsf{E}$}}}
\put(33,56){\makebox(100,50){\footnotesize{$\pi_\mathsf{M}$}}}
\put(-80,8){\makebox(100,50){\footnotesize{$\partial_0^\mathsf{E}$}}}
\put(-23,12){\makebox(100,50){\footnotesize{$\partial_1^\mathsf{E}$}}}
\put(23,12){\makebox(100,50){\footnotesize{$\partial_0^\mathsf{M}$}}}
\put(82,8){\makebox(100,50){\footnotesize{$\partial_1^\mathsf{M}$}}}
\put(-50,-35){\makebox(100,50){\footnotesize{$\mathrm{id}$}}}
\put(50,-35){\makebox(100,50){\footnotesize{$\mathrm{id}$}}}
\put(-48,-4){\makebox(100,50){\footnotesize{$\alpha_{\mathsf{E}}$}}}
\put(-48,-14){\makebox(100,50){\Large{$\Rightarrow$}}}
\put(52,-4){\makebox(100,50){\footnotesize{$\alpha_{\mathsf{M}}$}}}
\put(52,-14){\makebox(100,50){\Large{$\Rightarrow$}}}
\put(50,25){\begin{picture}(30,15)(0,-15)
\put(0,0){\line(-1,-1){15}}
\put(0,0){\line(1,-1){15}}
\end{picture}}
\thicklines
\put(40,90){\vector(-1,-1){30}}
\put(60,90){\vector(1,-1){30}}
\put(-10,40){\vector(-1,-1){30}}
\put(10,40){\vector(1,-1){30}}
\put(-30,0){\vector(1,0){60}}
\put(90,40){\vector(-1,-1){30}}
\put(110,40){\vector(1,-1){30}}
\put(70,0){\vector(1,0){60}}
\end{picture}}
\thicklines
\put(-21,108){\vector(1,0){40}}
\put(-49,90){\makebox(100,50){\footnotesize{$\div_{\mathsf{C}}$}}}
\put(-52,75.5){\makebox(100,50){\footnotesize{$\equiv$}}}
\put(-49,62){\makebox(100,50){\footnotesize{$\circ_{\mathsf{C}}$}}}
\put(19,94){\vector(-1,0){40}}
\put(-35,90){\line(2,-1){44}}
\put(21,62){\line(2,-1){43}}
\put(79,33){\line(2,-1){15}}
\put(109,18){\vector(2,-1){28}}
\put(-50,85){\vector(0,-1){70}}
\put(-98,75){\makebox(100,50){$\mathsf{C}^{\mathbf{2}}$}}
\put(-109,24){\makebox(100,50){\footnotesize{$\partial_0^{\mathsf{C}}$}}}
\put(-7,36){\makebox(100,50){\footnotesize{$\partial_1^{\mathsf{C}}$}}}
\put(-75,45){\makebox(100,50){\footnotesize{$\alpha_{\mathsf{C}}$}}}
\put(-75,35){\makebox(100,50){\Large{$\Rightarrow$}}}

\end{picture}

\\ \\ \\

$\begin{array}{rcl}
\div_{\mathsf{C}} \circ\, \circ_{\mathsf{C}} & = & \mathsf{id}_{\mathsf{C}} 
\\
\circ_{\mathsf{C}} \,\circ \div_{\mathsf{C}} & \cong & \mathsf{id}_{\mathsf{E} {\odot_{\mathsf{C}}} \mathsf{M}} 
\\
\div_{\mathsf{C}} 
\left(
\pi_\mathsf{E} \, \alpha_{\mathsf{E}}
\bullet
\pi_\mathsf{M} \, \alpha_{\mathsf{M}}
\right)
& = &
\alpha_{\mathsf{C}}

\end{array}$

\end{tabular}
\end{center}
\caption{Categorical Factorization Equivalence}
\label{factorization-equivalence}
\end{figure}

Let $\mathsf{E} {\odot_{\mathsf{C}}} \mathsf{M}$ denote the category
of $\langle \mathsf{E}, \mathsf{M} \rangle$-factorizations
(top foreground of Figure~\ref{factorization-equivalence}),
whose objects are $\langle \mathsf{E}, \mathsf{M} \rangle$-factorizations $(A, e, C, m, B)$ (composable pairs $e : A \rightarrow C$ and $m : C \rightarrow B$ for $e \in \mathsf{E}$ and $m \in \mathsf{M}$),
and whose morphisms $(a, c, b) : (A_1, e_1, C_1, m_1, B_1) \rightarrow (A_2, e_2, C_2, m_2, B_2)$
\begin{center}
\setlength{\unitlength}{0.8pt}
\begin{picture}(100,60)(0,-5)
\put(-50,25){\makebox(100,50){$A_1$}}
\put(0,25){\makebox(100,50){$C_1$}}
\put(50,25){\makebox(100,50){$B_1$}}
\put(-50,-25){\makebox(100,50){$A_2$}}
\put(0,-25){\makebox(100,50){$C_2$}}
\put(50,-25){\makebox(100,50){$B_2$}}
\put(0,32){\makebox(50,50){\footnotesize{$e_1$}}}
\put(50,32){\makebox(50,50){\footnotesize{$m_1$}}}
\put(0,-32){\makebox(50,50){\footnotesize{$e_2$}}}
\put(50,-32){\makebox(50,50){\footnotesize{$m_2$}}}
\put(-30,0){\makebox(50,50){\footnotesize{$a$}}}
\put(20,0){\makebox(50,50){\footnotesize{$c$}}}
\put(70,0){\makebox(50,50){\footnotesize{$b$}}}
\put(10,50){\vector(1,0){30}}
\put(60,50){\vector(1,0){30}}
\put(10,0){\vector(1,0){30}}
\put(60,0){\vector(1,0){30}}
\put(0,40){\vector(0,-1){30}}
\put(50,40){\vector(0,-1){30}}
\put(100,40){\vector(0,-1){30}}
\end{picture}
\end{center}
are $\mathsf{C}$-morphism triples,
where $(a, c) : (A_1, e_1, C_1) \rightarrow (A_2, e_2, C_2)$ 
is an $\mathsf{E}^{\mathbf{2}}$-morphism
and $(c, b) : (C_1, m_1, B_1) \rightarrow (C_2, m_2, B_2)$ 
is an $\mathsf{M}^{\mathbf{2}}$-morphism.
$\mathsf{E} {\odot_{\mathsf{C}}} \mathsf{M}
= \mathsf{E}^{\mathbf{2}} \times_{|\mathsf{C}|} \mathsf{M}^{\mathbf{2}}$
is the pullback (in the category of categories)
of the $1^{\mathrm{st}}$-projection of $\mathsf{E}^{\mathbf{2}}$
and the $0^{\mathrm{th}}$-projection of $\mathsf{M}^{\mathbf{2}}$.
There is a composition functor
$\circ_{\mathsf{C}} : \mathsf{E} {\odot_{\mathsf{C}}} \mathsf{M} \rightarrow \mathsf{C}^{\mathbf{2}}$,
which is defined on objects as
$\circ_{\mathsf{C}}(A, e, C, m, B) = (A, e \circ_{\mathsf{C}} m, B)$
and on morphisms as
$\circ_{\mathsf{C}}(a, c, b) = (a, b)$.

An $\langle \mathsf{E}, \mathsf{M} \rangle$-factorization system with choice has a specified factorization for each $\mathsf{C}$-morphism;
that is, there is a choice function from the class of $\mathsf{C}$-morphisms to the class of $\langle \mathsf{E}, \mathsf{M} \rangle$-factorizations.
When choice is specified,
there is a factorization functor
$\div_{\mathsf{C}} : \mathsf{C}^{\mathbf{2}} \rightarrow \mathsf{E} {\odot_{\mathsf{C}}} \mathsf{M}$,
which is defined on objects as the chosen $\langle \mathsf{E}, \mathsf{M} \rangle$-factorization $\div_{\mathsf{C}}(e, m) = (e, C, m)$
and on morphisms as
$\div_{\mathsf{C}}(a, b) = (a, c, b)$
where $c$ is defined by diagonalization
($\div_{\mathsf{C}}$ is functorial by uniqueness of factorization).
Clearly,
factorization followed by composition is the identity
$\div_{\mathsf{C}} \circ\, \circ_{\mathsf{C}} 
= \mathsf{id}_{\mathsf{C}}$.
By uniqueness of factorication (up to iosomorphism)
composition followed by factorization is an isomorphism
$\circ_{\mathsf{C}} \,\circ \div_{\mathsf{C}} 
\cong \mathsf{id}_{\mathsf{E} {\odot_{\mathsf{C}}} \mathsf{M}}$.

\section{The Category of Adjunctions}\label{sec:category:adjunctions}

\subsection{Monotonic Functions}\label{subsec:monotonic:functions}

A \emph{monotonic function} (\emph{isotonic function})
$f : \mathbf{A} \rightarrow \mathbf{B}$
from source preorder $\mathbf{A} = \langle A, \leq_{\mathsf{A}} \rangle$ to target preorder $\mathbf{B} = \langle B, \leq_{\mathsf{B}} \rangle$
is a function $f : A \rightarrow B$ that preserves (preserves and respects) order:
if $a_1 \leq_{\mathsf{A}} a_2$ then $f(a_1) \leq_{\mathsf{A}} f(a_2)$ 
($a_1 \leq_{\mathsf{A}} a_2$ iff $f(a_1) \leq_{\mathsf{A}} f(a_2)$ 
for any two source elements $a_1, a_2 \in A$.
For any monotonic function
$f : \mathbf{A} \rightarrow \mathbf{B}$,
the \emph{kernel} preorder of $f$
is the preorder
$\mathsf{ker}(f) = \langle A, \leq_{f} \rangle$
whose elements are the source elements of $f$
and whose order relation is define by
$a_1 \leq_{f} a_2$ when $f(a_1) \leq_{\mathsf{B}} f(a_2)$.
Three facts are clear from the definition of the kernel order relation:
the source order relation is a subrelation of the kernel order relation
$\leq_{\mathsf{A}} \,\subseteq\, \leq_{f}$, 
and hence the identity function $\mathrm{id}_{A}$ is an injective monotonic function
$\mathrm{id}_{A} : \mathbf{A} \rightarrow \mathsf{ker}(f)$; 
the function $f$ is an isotonic function
$f : \mathsf{ker}(f) \rightarrow \mathbf{B}$; and
the original monotonic function factors as the composition of these two.
It is also clear that a monotonic function is isotonic iff the kernel preorder is the source $\mathbf{A} = \mathsf{ker}(f)$.

\begin{figure}
\begin{center}

\begin{tabular}{c@{\hspace{20pt}}c}

\begin{tabular}[b]{c}
\\
\begin{picture}(90,90)(0,0)
\put(-15,33.75){\makebox(30,22.5){$\mathsf{Set}$}}
\put(30,78.75){\makebox(30,22.5){$\mathsf{Set}$}}
\put(34,-11.25){\makebox(30,22.5){$\mathsf{Set}^{\mathrm{op}}$}}
\put(80,33.75){\makebox(30,22.5){$\mathsf{Set}^{\mathrm{op}}$}}
\put(30,33.75){\makebox(30,22.5){$\mathsf{Adj}$}}
\put(5,65){\makebox(20,12.5){\footnotesize{$\exists$}}}
\put(-4,7.5){\makebox(20,12.5){\scriptsize{$( {(-)}^{\mathrm{-1}} )^{\mathrm{op}}$}}}
\put(44,58.25){\makebox(20,12.5){\footnotesize{$\mathsf{left}$}}}
\put(25,19.25){\makebox(20,12.5){\footnotesize{$\mathsf{right}^{\mathrm{op}}$}}}
\put(13.5,45){\makebox(20,12.5){\small{$\mathsf{dir}$}}}
\put(58.5,45){\makebox(20,12.5){\small{$\mathsf{inv}$}}}
\put(72,68){\makebox(20,12.5){\scriptsize{${(-)}^{\mathrm{-1}}$}}}
\put(68,12){\makebox(20,12.5){\footnotesize{$\exists^{\mathrm{op}}$}}}
\put(10,55){\vector(1,1){25}}
\put(80,55){\vector(-1,1){25}}
\put(10,35){\vector(1,-1){25}}
\put(80,35){\vector(-1,-1){25}}
\put(10,45){\vector(1,0){25}}
\put(80,45){\vector(-1,0){25}}
\put(45,56.25){\vector(0,1){22.5}}
\put(45,33.75){\vector(0,-1){22.5}}
\end{picture}
\\ \\
\small{$\begin{array}[t]{r@{\hspace{5pt}}c@{\hspace{5pt}}l}
\mathsf{inv} \circ \mathsf{right}^{\mathrm{op}} & = & \exists^{\mathrm{op}} \\ 
\mathsf{inv} \circ \mathsf{left}                & = & {(-)}^{\mathrm{-1}} \\
\mathsf{dir} \circ \mathsf{right}^{\mathrm{op}} & = & ( {(-)}^{\mathrm{-1}} )^{\mathrm{op}} \\ 
\mathsf{dir} \circ \mathsf{left}                & = & \exists
\end{array}$}
\\ \\
\underline{components and power}
\end{tabular}

&

\begin{tabular}[b]{c}
\\
\begin{picture}(90,90)(-12,0)
\put(7.5,78.75){\makebox(30,22.5){$\mathsf{Set}$}}
\put(-15,33.75){\makebox(30,22.5){$\mathsf{Adj}^{\mathrm{op}}$}}
\put(35,33.75){\makebox(30,22.5){$\mathsf{Adj}$}}
\put(12.5,-11.25){\makebox(30,22.5){$\mathsf{Set}^{\mathrm{op}}$}}
\put(-8,65){\makebox(20,12.5){\small{${\mathsf{inv}}^{\mbox{\tiny{op}}}$}}}
\put(17.5,80){\vector(-1,-2){11.25}}
\put(27.5,80){\vector(1,-2){11.25}}
\put(12.5,51.0){\makebox(20,12.5){\small{$\propto$}}}
\put(32,65){\makebox(20,12.5){\small{$\mathsf{dir}$}}}
\put(10,52){\vector(1,0){25}}
\put(7,38){\makebox(30,12.5){{$\cong$}}}
\put(35,38){\vector(-1,0){25}}
\put(16,26){\makebox(20,12.5){\small{${\propto}^{\mbox{\tiny{op}}}$}}}
\put(-8,11){\makebox(20,12.5){\small{${\mathsf{dir}}^{\mbox{\tiny{op}}}$}}}
\put(17.5,10){\vector(-1,2){11.25}}
\put(27.5,10){\vector(1,2){11.25}}
\put(31,11){\makebox(20,12.5){\small{$\mathsf{inv}$}}}
\end{picture} 
\\ \\
\small{$\begin{array}[t]{r@{\hspace{5pt}}c@{\hspace{5pt}}l}
{\mathsf{inv}}^{\mbox{\tiny{op}}} \circ \propto & = & \mathsf{dir} \\
{\mathsf{dir}}^{\mbox{\tiny{op}}} \circ \propto & = & \mathsf{inv}
\end{array}$}
\\ \\
\underline{power and involution}
\end{tabular}

\\ & \\

\begin{tabular}[b]{c}
\\
\begin{picture}(65,90)(-32,0)
\put(-35,73.5){\vector(2,1){25}}
\put(-10,50){\vector(-2,1){25}}
\put(-60,56){\makebox(30,22.5){$\mathsf{Set}^{\mathrm{op}}$}}
\put(-24,67){\makebox(20,12.5){\footnotesize{$\uparrow$}}}
\put(-25,58){\makebox(20,12.5){\Large{$\Leftarrow$}}}
\put(-40,82){\makebox(30,12.5){\scriptsize{${(-)}^{\mathrm{-1}}$}}}
\put(-40,43){\makebox(30,12.5){\footnotesize{${\mathsf{right}}^{\mathrm{op}}$}}}
\put(10,40){\vector(2,-1){25}}
\put(35,17.5){\vector(-2,-1){25}}
\put(30,11.25){\makebox(30,22.5){$\mathsf{Set}$}}
\put(-15,78.75){\makebox(30,22.5){$\mathsf{Set}$}}
\put(-15,33.75){\makebox(30,22.5){$\mathsf{Adj}$}}
\put(-11,-11.25){\makebox(30,22.5){$\mathsf{Set}^{\mathrm{op}}$}}
\put(0,58.25){\makebox(20,12.5){\footnotesize{$\mathsf{left}$}}}
\put(0,56.25){\vector(0,1){22.5}}
\put(-26,19.25){\makebox(30,12.5){\footnotesize{$\mathsf{right}^{\mathrm{op}}$}}}
\put(0,33.75){\vector(0,-1){22.5}}
\put(14,34){\makebox(30,12.5){\footnotesize{$\mathsf{left}$}}}
\put(26,-2){\makebox(30,12.5){\scriptsize{$( {(-)}^{\mathrm{-1}} )^{\mathrm{op}}$}}}
\put(5,22){\makebox(20,12.5){\footnotesize{$\downarrow^{\mathrm{op}}$}}}
\put(1,13){\makebox(30,12.5){\Large{$\Leftarrow$}}}
\end{picture} 
\\ \\
\small{$\begin{array}[t]{r@{\hspace{5pt}}c@{\hspace{5pt}}l}
\mathsf{right}^{\mathrm{op}} \circ {(-)}^{\mathrm{-1}} & \stackrel{\uparrow}{\Leftarrow} & \mathsf{left} \\
\mathsf{left}^{\mathrm{op}} \circ {(-)}^{\mathrm{-1}} & \stackrel{\downarrow}{\Leftarrow} & \mathsf{right} \\
\uparrow & = & {\propto}^{\mbox{\tiny{op}}} \circ \downarrow \\
\downarrow & = & {\propto} \circ \uparrow
\end{array}$}
\\ \\
\underline{above ($\uparrow$) and below ($\downarrow$)}
\end{tabular}

&

\end{tabular}

\end{center}
\caption{The Component Architecture for Adjunctions}
\label{adjunction-component-architecture}
\end{figure}

\subsection{Galois Connections}\label{subsec:galois:connections}

A \emph{Galois connection} (adjoint pair of monotonic functions) 
$\mathbf{g}
= \langle \mathsf{left}(\mathbf{g}), \mathsf{right}(\mathbf{g}) \rangle 
= \langle \check{\mathbf{g}}, \hat{\mathbf{g}} \rangle
: \mathbf{A} \rightleftharpoons \mathbf{B}$
from preorder $\mathbf{A} = \ord{A}$ to preorder $\mathbf{B} = \ord{B}$
is a contravariant pair of functions, 
a monotonic function 
$\mathsf{left}(\mathbf{g}) = \check{\mathbf{g}} : A \rightarrow B$ in the forward direction 
and a monotonic function 
$\mathsf{right}(\mathbf{g}) = \hat{\mathbf{g}} : A \leftarrow B$ in the backward direction, 
satisfying the fundamental adjointness property 
$\check{\mathbf{g}}(a) \leq_{\mathbf{B}} b$ 
\underline{iff} 
$a \leq_{\mathbf{A}} \hat{\mathbf{g}}(b)$
for every source element $a \in A$ and every target element $b \in B$,
or equivalently the pair of inequalities
$\check{\mathbf{g}}(\hat{\mathbf{g}}(b)) \leq_{\mathbf{B}} b$ for all $b \in B$
and
$a \leq_{\mathbf{A}} \hat{\mathbf{g}}(\check{\mathbf{g}}(a))$ for all $a \in A$.
Preorders and Galois connections form the category $\mathsf{Adj}$. 
Posets and Galois connections form the full subcategory $\mathsf{Adj}_{=} \subset \mathsf{Adj}$. 
Projecting to the left and right gives rise to two component functors,
$\mathsf{left} : \mathsf{Adj} \rightarrow \mathsf{Set}$ 
and $\mathsf{right} : \mathsf{Adj}^{\mathrm{op}} \rightarrow \mathsf{Set}$. 

\subsubsection{Closure and Interior.}

Let $\mathbf{g} : \mathbf{A} \rightleftharpoons \mathbf{B}$ be any Galois connection.
The closure of $\mathbf{g}$ is the monotonic function
$(\mbox{-})^{\bullet_{\mathbf{g}}}
= \check{\mathbf{g}} \cdot \hat{\mathbf{g}}
= \mathsf{left}(\mathbf{g}) \cdot \mathsf{right}(\mathbf{g})
: \mathbf{A} \rightarrow \mathbf{A}$.
Closure is increasing $\mathrm{id}_{\mathbf{A}} \leq (\mbox{-})^{\bullet_{\mathbf{g}}}$ and idempotent $(\mbox{-})^{\bullet_{\mathbf{g}}} \cdot (\mbox{-})^{\bullet_{\mathbf{g}}} \equiv (\mbox{-})^{\bullet_{\mathbf{g}}}$.
Idempotency is implied by the fact that
$\check{\mathbf{g}} \cdot \hat{\mathbf{g}} \cdot \check{\mathbf{g}} \equiv \check{\mathbf{g}}$.
A source element $a \in A$ is a closed element of $\mathbf{g}$ when it is equivalent to its closure $a \equiv_{\mathbf{A}} a^{\bullet_{\mathbf{g}}}$.
A source element $a \in A$ is closed iff it is equivalent $a \equiv_{\mathbf{A}} \hat{\mathbf{g}}(b)$ to the right adjoint image of some target element $b \in B$.   
Let $\mathsf{clo}(\mathbf{g}) \subseteq A$ denote the set of closed elements of $\mathbf{g}$.
The interior of $\mathbf{g}$ is the monotonic function
$(\mbox{-})^{\circ_{\mathbf{g}}}
= \hat{\mathbf{g}} \cdot \check{\mathbf{g}}
= \mathsf{right}(\mathbf{g}) \cdot \mathsf{left}(\mathbf{g})
: \mathbf{B} \rightarrow \mathbf{B}$.
Interior is decreasing $\mathrm{id}_{\mathbf{A}} \geq (\mbox{-})^{\circ_{\mathbf{g}}}$ and idempotent $(\mbox{-})^{\circ_{\mathbf{g}}} \cdot (\mbox{-})^{\circ_{\mathbf{g}}} \equiv (\mbox{-})^{\circ_{\mathbf{g}}}$.
Idempotency is implied by the fact that
$\hat{\mathbf{g}} \cdot \check{\mathbf{g}} \cdot \hat{\mathbf{g}} \equiv \hat{\mathbf{g}}$.
A target element $b \in B$ is an open element of $\mathbf{g}$ when it is equivalent to its interior $b \equiv_{\mathbf{B}} b^{\circ_{\mathbf{g}}}$.
A target element $b \in B$ is open iff it is equivalent $b \equiv_{\mathbf{B}} 
\check{\mathbf{g}}(a)$ to the left adjoint image of some source element $a \in A$.
Let $\mathsf{open}(\mathbf{g}) \subseteq B$ denote the set of open elements of $\mathbf{g}$.

\subsubsection{Reflections and Coreflections.}

A reflection is a Galois connection 
$\mathbf{g} : \mathbf{A} \rightleftharpoons \mathbf{B}$
that satisfies the condition
$\mathrm{id}_{B} 
\equiv (\mbox{-})^{\circ_{\mathbf{g}}}$. 
Since
$\hat{\mathbf{g}}(b_1) \leq_{\mathbf{A}} \hat{\mathbf{g}}(b_2)$
implies
$b_1 \equiv_{\mathbf{B}} \check{\mathbf{g}}(\hat{\mathbf{g}}(b_1)) \leq_{\mathbf{B}} \check{\mathbf{g}}(\hat{\mathbf{g}}(b_2)) \equiv_{\mathbf{B}} b_2$,
the right adjoint $\hat{\mathbf{g}}$ of a reflection is an isotonic function.
A coreflection is a Galois connection 
$\mathbf{g} : \mathbf{A} \rightleftharpoons \mathbf{B}$
that satisfies the condition
$\mathrm{id}_{A} 
\equiv (\mbox{-})^{\bullet_{\mathbf{g}}}$. 
Since
$\check{\mathbf{g}}(a_1) \leq_{\mathbf{B}} \check{\mathbf{g}}(a_2)$
implies
$a_1 \equiv_{\mathbf{A}} \hat{\mathbf{g}}(\check{\mathbf{g}}(a_1)) \leq_{\mathbf{A}} \hat{\mathbf{g}}(\check{\mathbf{g}}(a_2)) \equiv_{\mathbf{A}} a_2$,
the left adjoint $\check{\mathbf{g}}$ of a coreflection is an isotonic function.

\begin{theorem}\label{induce:lattice}
The following properties hold.
\begin{itemize}
\item Let
$\mathbf{g}
= \langle \check{\mathbf{g}}, \hat{\mathbf{g}} \rangle
: \mathbf{A} \rightleftharpoons \mathbf{B}$
be a reflection.
If the source $\mathbf{A}$ is a poset,
then the target $\mathbf{B}$ is also a poset.
If the source $\mathbf{A}$ is a complete lattice,
then the target $\mathbf{B}$ is a complete lattice
with the definitions
$\bigvee_{\mathbf{B}}B 
= \check{\mathbf{g}}\left( \bigvee_{\mathbf{A}} \hat{\mathbf{g}}[B] \right)$
and
$\bigwedge_{\mathbf{B}}B 
= \check{\mathbf{g}}\left( \bigwedge_{\mathbf{A}} \hat{\mathbf{g}}[B] \right)$
for any target subset $B$.
Also,
the following identities hold:
$\hat{\mathbf{g}} \left( \bigvee_{\mathbf{B}}B \right)
= {\left( \bigvee_{\mathbf{A}} \hat{\mathbf{g}}[B] \right)}^{\bullet}$
and
$\hat{\mathbf{g}}\left( \bigwedge_{\mathbf{B}}B \right)
= \bigwedge_{\mathbf{A}} \hat{\mathbf{g}}[B]$.
\item Let
$\mathbf{g}
= \langle \check{\mathbf{g}}, \hat{\mathbf{g}} \rangle
: \mathbf{A} \rightleftharpoons \mathbf{B}$
be a coreflection.
If the target $\mathbf{B}$ is a poset,
then the source $\mathbf{A}$ is also a poset.
If the target $\mathbf{B}$ is a complete lattice,
then the source $\mathbf{A}$ is a complete lattice
with the definitions
$\bigwedge_{\mathbf{A}}A 
= \hat{\mathbf{g}}\left( \bigwedge_{\mathbf{B}} \check{\mathbf{g}}[A] \right)$
and
$\bigvee_{\mathbf{A}}A 
= \hat{\mathbf{g}}\left( \bigvee_{\mathbf{B}} \check{\mathbf{g}}[A] \right)$
for any source subset $A$.
Also,
the following identities hold:
$\check{\mathbf{g}} ( \bigwedge_{\mathbf{A}}A )
= {\left( \bigwedge_{\mathbf{B}} \check{\mathbf{g}}[A] \right)}^{\circ}$
and
$\check{\mathbf{g}}(\bigvee_{\mathbf{A}}A)
= \bigvee_{\mathbf{B}} \check{\mathbf{g}}[A]$.
\end{itemize}
\end{theorem}
{\bfseries Proof.} 
We prove the property for coreflections.
The property for reflections is dual.
The following argument shows that the definition for meet is well-defined.
$\bigwedge_{\mathbf{B}} \check{\mathbf{g}}[A] \leq_{\mathbf{B}} \hat{\mathbf{g}}(a)$
for all $a \in A$.
This implies that
$\hat{\mathbf{g}}\left( \bigwedge_{\mathbf{B}} \check{\mathbf{g}}[A] \right) \leq_{\mathbf{A}} a$
for all $a \in A$.
Hence,
$\hat{\mathbf{g}}\left( \bigwedge_{\mathbf{B}} \check{\mathbf{g}}[A] \right) \leq_{\mathbf{A}} \bigwedge_{\mathbf{A}}A$.
Assume $x \leq_{\mathbf{A}} a$ for all $a \in A$. 
Since the left adjoint is monotonic,
$\check{\mathbf{g}}(x) \leq_{\mathbf{B}} \check{\mathbf{g}}(a)$ for all $a \in A$.
Hence,
$\check{\mathbf{g}}(x) \leq_{\mathbf{B}} \bigwedge_{\mathbf{B}} \check{\mathbf{g}}[A]$.
Since the right adjoint is monotonic and right inverse,
$x = \hat{\mathbf{g}}(\check{\mathbf{g}}(x)) \leq_{\mathbf{A}} \hat{\mathbf{g}}\left( \bigwedge_{\mathbf{B}} \check{\mathbf{g}}[A] \right)$.
Dually,
the following argument shows that the definition for join is well-defined.
$\check{\mathbf{g}}(a) \leq_{\mathbf{A}} \bigvee_{\mathbf{B}} \check{\mathbf{g}}[A]$
for all $a \in A$.
This implies that
$a \leq_{\mathbf{A}} \hat{\mathbf{g}}\left( \bigvee_{\mathbf{B}} \check{\mathbf{g}}[A] \right)$
for all $a \in A$.
Hence,
$\bigvee_{\mathbf{A}}A \leq_{\mathbf{A}} \hat{\mathbf{g}}\left( \bigvee_{\mathbf{B}} \check{\mathbf{g}}[A] \right)$.
Assume $a \leq_{\mathbf{A}} x$ for all $a \in A$. 
Since the left adjoint is monotonic,
$\check{\mathbf{g}}(a) \leq_{\mathbf{B}} \check{\mathbf{g}}(x)$ for all $a \in A$.
Hence,
$\bigvee_{\mathbf{B}} \check{\mathbf{g}}[A] \leq_{\mathbf{B}} \check{\mathbf{g}}(x)$.
Since the right adjoint is monotonic and right inverse,
$\hat{\mathbf{g}}\left( \bigvee_{\mathbf{B}} \check{\mathbf{g}}[A] \right)
\leq_{\mathbf{A}} \hat{\mathbf{g}}(\check{\mathbf{g}}(x)) = x$.
The second identity is trivial,
since $\check{\mathbf{g}}$ is a left adjoint.
By monotonicity of the left adjoint,
$\check{\mathbf{g}} \left( \bigwedge_{\mathbf{A}}A \right)
\leq_{\mathbf{B}} \check{\mathbf{g}}(a)$ for each element $a \in A$.
Hence,
$\check{\mathbf{g}} \left( \bigwedge_{\mathbf{A}}A \right)
\leq_{\mathbf{B}} \bigwedge_{\mathbf{B}} \check{\mathbf{g}}[A]$.
By idempotency and monotonicity of the $\mathbf{g}$-interior,
$\check{\mathbf{g}} ( \bigwedge_{\mathbf{A}}A)
= {\left( \check{\mathbf{g}} ( \bigwedge_{\mathbf{A}}A) \right)}^{\circ}
\leq_{\mathbf{B}} {\left( \bigwedge_{\mathbf{B}} \check{\mathbf{g}}[A] \right)}^{\circ}$.
For any element $a \in A$,
$\bigwedge_{\mathbf{B}} \check{\mathbf{g}}[A] \leq _{\mathbf{B}} \check{\mathbf{g}}(a)$.
Since the right adjoint is right inverse and monotonic,
$\hat{\mathbf{g}} \left( \bigwedge_{\mathbf{B}} \check{\mathbf{g}}[A] \right)
\leq _{\mathbf{A}} \hat{\mathbf{g}}(\check{\mathbf{g}}(a)) = a$.
Hence,
$\hat{\mathbf{g}} \left( \bigwedge_{\mathbf{B}} \check{\mathbf{g}}[A] \right)
\leq_{\mathbf{B}} 
\bigwedge_{\mathsf{A}}A$.
By monotonicity of the left adjoint,
${\left( \bigwedge_{\mathbf{B}} \check{\mathbf{g}}[A] \right)}^{\circ}
= \check{\mathbf{g}} ( \hat{\mathbf{g}} \left( \bigwedge_{\mathbf{B}} \check{\mathbf{g}}[A] \right) )
\leq_{\mathbf{B}}
\check{\mathbf{g}} \left( \bigwedge_{\mathsf{A}}A \right)$. \rule{3pt}{8pt}

\subsubsection{Examples.}

There is an existential direct image functor
$\exists : \mathsf{Set} \rightarrow \mathsf{Set}$,
where the existential direct image $\exists{A}$ of any set $A$ is the power set ${\wp}\,{A}$, 
and the existential direct image $\exists{h} : {\wp}\,A_1 \rightarrow {\wp}\,A_2$ of any function $h : A_1 \rightarrow A_2$ maps a source subset $X_1 \subseteq A_1$ to the target subset $\{ a_2 \in A_2 \mid \exists a_1 \in X_1, h(a_1) = a_2 \} \subseteq A_2$.
There is an inverse image functor
${(-)}^{-1} : \mathsf{Set}^{\mathrm{op}} \rightarrow \mathsf{Set}$,
where the inverse image ${A}^{-1}$ of any set $A$ is the power set ${\wp}\,{A}$, 
and the inverse image $h^{-1} : {\wp}\,A_1 \leftarrow {\wp}\,A_2$ of any function $h : A_1 \rightarrow A_2$ maps a target subset $X_2 \subseteq A_2$ to the source subset $\{ a_1 \in A_1 \mid h(a_1) \in X_2 \} \subseteq A_1$. 
There is a direct image functor
$\mathsf{dir} : \mathsf{Set} \rightarrow \mathsf{Adj}$,
where 
$\mathsf{dir} \circ \mathsf{left} = \exists$
and
$\mathsf{dir} \circ \mathsf{right}^{\mathrm{op}} = {(-)}^{-1}$;
that is,
where the direct image $\mathsf{dir}(A)$ of any set $A$ is the power preorder (poset) 
${\wp}\,{A} = \langle {\wp}\,{A}, \subseteq \rangle$, 
and the direct image $\mathsf{dir}(A) = \langle \exists{h}, \rangle : {\wp}\,A_1 \rightleftharpoons {\wp}\,A_2$ of any function $h : A_1 \rightarrow A_2$ is the Galois connection with existential direct image as left adjoint
and inverse image as right adjoint.

\subsubsection{Polar Factorization.}

Let $\mathbf{g}
= \langle \mathsf{left}(\mathbf{g}), \mathsf{right}(\mathbf{g}) \rangle 
= \langle \check{\mathbf{g}}, \hat{\mathbf{g}} \rangle
: \mathbf{A} \rightleftharpoons \mathbf{B}$
be a Galois connection.
A \emph{bipole} (bipolar pair) $(a, b)$ is a pair consisting of
a closed element $a \in \mathsf{clo}(\mathbf{g})$ and 
an open element $b \in \mathsf{open}(\mathbf{g})$,
where $a \equiv_{\mathbf{A}} \hat{\mathbf{g}}(b)$ 
(equivalently, $b \equiv_{\mathbf{B}} \check{\mathbf{g}}(a))$.
Let $\mathsf{elem}(\mathbf{g})$ denote the set of bipoles of $\mathbf{g}$.
Define the bipolar order
$(a_1, b_1) \leq_{\mathbf{g}} (a_2, b_2)$
when $a_1 \leq_{\mathbf{A}} a_2$
(equivalently, when $b_1 \leq_{\mathbf{B}} b_2$).
The preorder 
$\mathsf{axis}(\mathbf{g}) = \diamondsuit(\mathbf{g}) = \langle \mathsf{elem}(\mathbf{g}), \leq_{\mathbf{g}} \rangle$ is called the axis of $\mathbf{g}$.
There is an obvious pair of projection (monotonic) functions:
the source projection 
$\mathsf{proj}_0(\mathbf{g}) = \pi_0^{\mathbf{g}} 
: \mathsf{elem}(\mathbf{g}) \rightarrow A$,
and
the target projection 
$\mathsf{proj}_1(\mathbf{g}) = \pi_1^{\mathbf{g}} 
: \mathsf{elem}(\mathbf{g}) \rightarrow B$.
There is also a pair of ``embedding'' (monotonic) functions in the opposite direction:
the source embedding
$\mathsf{embed}_0(\mathbf{g}) = \xi_0^{\mathbf{g}} 
: A \rightarrow \mathsf{elem}(\mathbf{g})$
defined by
$\xi_0^{\mathbf{g}}(a) 
\doteq (a^{\bullet_{\mathbf{g}}}, \check{\mathbf{g}}(a))
= (\hat{\mathbf{g}}(\check{\mathbf{g}}(a), \check{\mathbf{g}}(a))$
for all $a \in A$,
and the target embedding
$\mathsf{embed}_1(\mathbf{g}) = \xi_1^{\mathbf{g}} 
: B \rightarrow \mathsf{elem}(\mathbf{g})$
defined by
$\xi_1^{\mathbf{g}}(b) 
\doteq (\hat{\mathbf{g}}(b), b^{\circ_{\mathbf{g}}})
= (\hat{\mathbf{g}}(b), \check{\mathbf{g}}(\hat{\mathbf{g}}(b)))$
for all $b \in B$.
It is straightforward to check that
the source embedding and projection form a reflection
$\mathsf{refl}(\mathbf{g})
= \langle \xi_0^{\mathbf{g}}, \pi_0^{\mathbf{g}} \rangle
: \mathbf{A} \rightleftharpoons \diamondsuit(\mathbf{g})$,
the target projection and embedding form a coreflection
$\mathsf{corefl}(\mathbf{g})
= \langle \pi_1^{\mathbf{g}}, \xi_1^{\mathbf{g}} \rangle
: \diamondsuit(\mathbf{g}) \rightleftharpoons \mathbf{A}$,
and the original Galois connection factors as
$\mathbf{g} = \mathsf{refl}(\mathbf{g})\circ \mathsf{corefl}(\mathbf{g})$.
This is called the polar (or reflection-coreflection) factorization of $\mathbf{g}$.
If both source $\mathbf{A}$ and target $\mathbf{B}$ are posets, 
then $\mathsf{axis}(\mathbf{g})$ is also a poset.
If both source $\mathbf{A}$ and target $\mathbf{B}$ are complete lattices, 
then $\mathsf{axis}(\mathbf{g})$ is also a complete lattices
with meet and join define by $\bigwedge \mathcal{C} = $ and $\bigvee \mathcal{C} = $.
\begin{lemma}
Assume that we are given a commutative square
\begin{center}
\setlength{\unitlength}{0.85pt}
\begin{picture}(50,50)
\put(-50,25){\makebox(100,50){$\mathbf{A}$}}
\put(0,25){\makebox(100,50){$\mathbf{B}$}}
\put(-50,-25){\makebox(100,50){$\mathbf{C}$}}
\put(0,-25){\makebox(100,50){$\mathbf{D}$}}
\put(0,30){\makebox(50,50){\footnotesize{$\mathbf{e}$}}}
\put(0,-31){\makebox(50,50){\footnotesize{$\mathbf{m}$}}}
\put(-30,0){\makebox(50,50){\footnotesize{$\mathbf{r}$}}}
\put(31,0){\makebox(50,50){\footnotesize{$\mathbf{s}$}}}
\put(-5,5){\makebox(50,50){\footnotesize{$\mathbf{h}$}}}
\thicklines
\put(10,0){\vector(1,0){30}}
\put(10,50){\vector(1,0){30}}
\put(0,40){\vector(0,-1){30}}
\put(50,40){\vector(0,-1){30}}
\put(40,40){\vector(-1,-1){30}}
\end{picture}
\end{center}
of Galois connections between posets,
with reflection $\mathbf{e}$ and coreflection $\mathbf{m}$.
Then there is a unique Galois connection 
$\mathbf{h} : \mathbf{B} \rightleftharpoons \mathbf{C}$
with $\mathbf{e} \circ \mathbf{h} = \mathbf{r}$ and $\mathbf{h} \circ \mathbf{m} = \mathbf{s}$.
\end{lemma}
{\bfseries Proof.} 
The fundamental adjointness property,
the special conditions for (co)reflections
and the above commutative diagram,
resolve into the following identities and inequalities:
$\hat{\mathbf{e}} \cdot \check{\mathbf{e}} = \mathrm{id}_{B}$,
$\mathrm{id}_{A} \leq \check{\mathbf{e}} \cdot \hat{\mathbf{e}}$,
$\hat{\mathbf{s}} \cdot \check{\mathbf{s}} \leq \mathrm{id}_{D}$,
$\mathrm{id}_{B} \leq \check{\mathbf{s}} \cdot \hat{\mathbf{s}}$,
$\hat{\mathbf{r}} \cdot \check{\mathbf{r}} \leq \mathrm{id}_{C}$,
$\mathrm{id}_{A} \leq \check{\mathbf{r}} \cdot \hat{\mathbf{r}}$,
$\hat{\mathbf{m}} \cdot \check{\mathbf{m}} \leq \mathrm{id}_{D}$,
$\mathrm{id}_{C} = \check{\mathbf{m}} \cdot \hat{\mathbf{m}}$,
$\check{\mathbf{e}} \cdot \check{\mathbf{m}} =
\check{\mathbf{r}} \cdot \check{\mathbf{m}}$,
and
$\hat{\mathbf{m}} \cdot \hat{\mathbf{r}} =
\hat{\mathbf{s}} \cdot \hat{\mathbf{e}}$.
By suitable pre- and post-composition we can prove the identities:
$\check{\mathbf{e}} \cdot \check{\mathbf{s}} \cdot \hat{\mathbf{m}} 
= \check{\mathbf{r}}$ and
$\hat{\mathbf{m}} \cdot \hat{\mathbf{r}} \cdot \check{\mathbf{e}} 
= \hat{\mathbf{s}}$,
(and then)
$\check{\mathbf{s}} \cdot \hat{\mathbf{m}} 
= \hat{\mathbf{e}} \cdot \check{\mathbf{r}}$ and
$\hat{\mathbf{r}} \cdot \check{\mathbf{e}} 
= \check{\mathbf{m}} \cdot \hat{\mathbf{s}}$.
[{\bfseries Existence}]
Define the monotonic functions
$\check{\mathbf{h}}
\doteq 
\check{\mathbf{s}} \cdot \hat{\mathbf{m}} 
= \hat{\mathbf{e}} \cdot \check{\mathbf{r}}$
and
$\check{\mathbf{h}}
\doteq 
\hat{\mathbf{r}} \cdot \check{\mathbf{e}} 
= \check{\mathbf{m}} \cdot \hat{\mathbf{s}}$.
It is straightforward to check the equalities
$\hat{\mathbf{h}} \cdot \check{\mathbf{h}}
= \check{\mathbf{m}} \cdot \hat{\mathbf{s}} \cdot \hat{\mathbf{e}} \cdot \check{\mathbf{r}}
= \check{\mathbf{m}} \cdot \hat{\mathbf{m}} \cdot \hat{\mathbf{r}} \cdot \check{\mathbf{r}}
= \hat{\mathbf{r}} \cdot \check{\mathbf{r}}
\leq = \mathrm{id}_{C}$,
$\check{\mathbf{h}} \cdot \hat{\mathbf{h}}
= \check{\mathbf{s}} \cdot \hat{\mathbf{m}} \cdot \hat{\mathbf{r}} \cdot \check{\mathbf{e}}
= \check{\mathbf{s}} \cdot \hat{\mathbf{s}} \cdot \hat{\mathbf{e}} \cdot \check{\mathbf{e}}
= \check{\mathbf{s}} \cdot \hat{\mathbf{s}}
\geq = \mathrm{id}_{B}$,
$\check{\mathbf{h}} \cdot \check{\mathbf{m}}
= \hat{\mathbf{e}} \cdot \check{\mathbf{r}} \cdot \check{\mathbf{m}}
= \hat{\mathbf{e}} \cdot \check{\mathbf{e}} \cdot \check{\mathbf{s}}
= \check{\mathbf{s}}$,
$\hat{\mathbf{m}} \cdot \hat{\mathbf{h}}
= \hat{\mathbf{m}} \cdot \hat{\mathbf{r}} \cdot \check{\mathbf{e}}
= \hat{\mathbf{s}} \cdot \hat{\mathbf{e}} \cdot \check{\mathbf{e}}
= \hat{\mathbf{s}}$,
$\check{\mathbf{e}} \cdot \check{\mathbf{h}}
= \check{\mathbf{e}} \cdot \check{\mathbf{s}} \cdot \hat{\mathbf{m}}
= \check{\mathbf{r}} \cdot \check{\mathbf{m}} \cdot \hat{\mathbf{m}}
= \check{\mathbf{r}}$
and
$\hat{\mathbf{h}} \cdot \hat{\mathbf{e}}
= \check{\mathbf{m}} \cdot \hat{\mathbf{s}} \cdot \hat{\mathbf{e}}
= \check{\mathbf{m}} \cdot \hat{\mathbf{m}} \cdot \hat{\mathbf{r}}
= \hat{\mathbf{r}}$.
These show that
$\mathbf{h} = \langle \check{\mathbf{h}}, \hat{\mathbf{h}} \rangle
: \mathbf{B} \rightleftharpoons \mathbf{C}$
is a Galois connection satisfying the require identities
$\mathbf{e} \circ \mathbf{h} = \mathbf{r}$
and $\mathbf{h} \circ \mathbf{m} = \mathbf{s}$.
[{\bfseries Uniqueness}]
Suppose $\mathbf{k} = \langle \check{\mathbf{k}}, \hat{\mathbf{k}} \rangle
: \mathbf{B} \rightleftharpoons \mathbf{C}$
is another Galois connection satisfying the require identities
$\mathbf{e} \circ \mathbf{k} = \mathbf{r}$
and $\mathbf{k} \circ \mathbf{m} = \mathbf{s}$.
These identities resolve to the identities
$\check{\mathbf{e}} \cdot \check{\mathbf{k}} = \check{\mathbf{r}}$,
$\hat{\mathbf{k}} \cdot \hat{\mathbf{e}} = \hat{\mathbf{r}}$,
$\check{\mathbf{k}} \cdot \check{\mathbf{m}} = \check{\mathbf{s}}$,
and
$\hat{\mathbf{m}} \cdot \hat{\mathbf{k}} = \hat{\mathbf{s}}$.
Hence,
$\check{\mathbf{k}}
= \check{\mathbf{k}} \cdot \check{\mathbf{m}} \cdot \hat{\mathbf{m}}
= \check{\mathbf{s}} \cdot \hat{\mathbf{m}}
= \check{\mathbf{k}}$,
$\hat{\mathbf{k}}
= \hat{\mathbf{k}} \cdot \hat{\mathbf{e}} \cdot \check{\mathbf{e}}
= \hat{\mathbf{r}} \cdot \check{\mathbf{e}}
= \hat{\mathbf{h}}$
and thus
$\mathbf{k} = \mathbf{h}$. \rule{3pt}{8pt}

\begin{theorem}
The classes of reflections and coreflections form a factorization system for $\mathsf{Adj}_{\,=}$ the category of posets and Galois connections.
The axis construction makes this a factorization system with choice.
\end{theorem}
{\bfseries Proof.} The previous discussion and lemma. \rule{3pt}{8pt}

\begin{figure}
\begin{center}
\begin{tabular}{c@{\hspace{25pt}}c}
\setlength{\unitlength}{1.3pt}
\begin{picture}(60,60)(0,-15)
\put(-30,15){\makebox(60,30){$\mathbf{A}$}}
\put(1,15){\makebox(60,30){$\diamondsuit(\mathbf{g})$}}
\put(30,15){\makebox(60,30){$\mathbf{B}$}}
\put(1,45){\makebox(60,30){$\mathsf{ker}(\check{\mathbf{g}})$}}
\put(1,-15){\makebox(60,30){$\mathsf{ker}(\hat{\mathbf{g}})$}}
\put(-18,33){\makebox(30,30)[r]{\footnotesize{$\mathsf{clo}_{\mathbf{g}}$}}}
\put(49,33){\makebox(30,30)[l]{\footnotesize{${\mathsf{lift}}_{\mathbf{g}}^{0}$}}}
\put(-18,-3){\makebox(30,30)[r]{\footnotesize{${\mathsf{lift}}_{\mathbf{g}}^{1}$}}}
\put(50,-3){\makebox(30,30)[l]{\footnotesize{$\mathsf{int}_{\mathbf{g}}$}}}
\put(9,20){\makebox(30,30)[l]{\footnotesize{$\mathsf{ref}_{\mathbf{g}}$}}}
\put(40,20){\makebox(30,30)[l]{\footnotesize{$\mathsf{ref}_{\mathbf{g}}^{\propto}$}}}
\put(32,30){\makebox(30,30)[l]{\footnotesize{${\underline{\mathsf{ref}}}_{\mathbf{g}}$}}}
\put(17.5,1){\makebox(30,30)[l]{\footnotesize{${\underline{\mathsf{ref}}}_{\mathbf{g}}^{\propto}$}}}
\thicklines
\put(7,30){\vector(1,0){12}}
\put(41,30){\vector(1,0){12}}
\put(30,52){\vector(0,-1){14}}
\put(30,22){\vector(0,-1){14}}
\put(4.5,37){\vector(1,1){16}}
\put(39.5,53){\vector(1,-1){16}}
\put(4.5,23){\vector(1,-1){16}}
\put(39.5,7){\vector(1,1){16}}
\end{picture}
&
\setlength{\unitlength}{1.2pt}
\begin{picture}(100,100)
\put(-50,25){\makebox(100,50){$\mathbf{A}$}}
\put(-8,49){\vector(1,0){0}}
\put(-10,39){\oval(20,20)[tl]}
\put(-10,39){\oval(20,20)[b]}
\put(0,25){\makebox(100,50){$\diamondsuit(\mathbf{g})$}}
\put(50,25){\makebox(100,50){$\mathbf{B}$}}
\put(108,49){\vector(-1,0){0}}
\put(110,39){\oval(20,20)[tr]}
\put(110,39){\oval(20,20)[b]}
\put(-19.5,27){\makebox(25,25){\footnotesize{${\mathbf{g}}^{\bullet}$}}}
\put(99,27){\makebox(25,25){\footnotesize{${\mathbf{g}}^{\circ}$}}}
\put(0,75){\makebox(100,50){$\mathsf{ker}(\check{\mathbf{g}})$}}
\put(0,-25){\makebox(100,50){$\mathsf{ker}(\hat{\mathbf{g}})$}}
\put(-5,50){\begin{picture}(50,50)
\put(3,21){\makebox(25,25){\footnotesize{$\mathrm{id}_{A}$}}}
\put(20,9){\makebox(25,25){\footnotesize{${\mathbf{g}}^{\bullet}$}}}
\put(19,14){\makebox(25,25){\tiny{$(\mathrm{iso})$}}}
\put(9,17){\vector(1,1){24}}
\put(39,35){\vector(-1,-1){24}}
\end{picture}}
\put(55,50){\begin{picture}(50,50)
\put(20,20){\makebox(25,25){\footnotesize{$\check{\mathbf{g}}$}}}
\put(19,15){\makebox(25,25){\tiny{$(\mathrm{iso})$}}}
\put(8,8){\makebox(25,25){\footnotesize{$\hat{\mathbf{g}}$}}}
\put(17,41){\vector(1,-1){24}}
\put(35,11){\vector(-1,1){24}}
\end{picture}}
\put(28,60){\makebox(25,25){\footnotesize{$\pi_0^{\mathbf{g}}$}}}
\put(48,60){\makebox(25,25){\footnotesize{$\xi_0^{\mathbf{g}}$}}}
\put(48,55){\makebox(25,25){\tiny{$(\mathrm{iso})$}}}
\put(28.5,16){\makebox(25,25){\footnotesize{$\xi_1^{\mathbf{g}}$}}}
\put(27.5,11){\makebox(25,25){\tiny{$(\mathrm{iso})$}}}
\put(48.5,16){\makebox(25,25){\footnotesize{$\pi_1^{\mathbf{g}}$}}}
\put(46,62){\vector(0,1){26}}
\put(54,88){\vector(0,-1){26}}
\put(46,12){\vector(0,1){26}}
\put(54,38){\vector(0,-1){26}}
\put(15,54){\vector(1,0){20}}
\put(35,46){\vector(-1,0){20}}
\put(65,54){\vector(1,0){20}}
\put(85,46){\vector(-1,0){20}}
\put(13,46.5){\makebox(25,25){\footnotesize{$\xi_0^{\mathbf{g}}$}}}
\put(12,36){\makebox(25,25){\tiny{$(\mathrm{iso})$}}}
\put(13,28.5){\makebox(25,25){\footnotesize{$\pi_0^{\mathbf{g}}$}}}
\put(63,46.5){\makebox(25,25){\footnotesize{$\pi_1^{\mathbf{g}}$}}}
\put(62,39.5){\makebox(25,25){\tiny{$(\mathrm{iso})$}}}
\put(63,28.5){\makebox(25,25){\footnotesize{$\xi_1^{\mathbf{g}}$}}}
\put(-5,0){\begin{picture}(50,50)
\put(20,19){\makebox(25,25){\footnotesize{$\check{\mathbf{g}}$}}}
\put(8,7){\makebox(25,25){\footnotesize{$\hat{\mathbf{g}}$}}}
\put(7,2){\makebox(25,25){\tiny{$(\mathrm{iso})$}}}
\put(17,41){\vector(1,-1){24}}
\put(35,11){\vector(-1,1){24}}
\end{picture}}
\put(55,0){\begin{picture}(50,50)
\put(6.5,19.5){\makebox(25,25){\footnotesize{${\mathbf{g}}^{\circ}$}}}
\put(5.5,14.5){\makebox(25,25){\tiny{$(\mathrm{iso})$}}}
\put(21,8){\makebox(25,25){\footnotesize{$\mathrm{id}_{B}$}}}
\put(9,17){\vector(1,1){24}}
\put(39,35){\vector(-1,-1){24}}
\end{picture}}
\end{picture}
\\ \\ 
Galois connection (iconic) & Galois connection (detailed)
\\ \\
$\begin{array}[b]{r@{\hspace{5pt}}c@{\hspace{5pt}}l}
\mathbf{g} & \doteq & \langle \check{\mathbf{g}} \dashv \hat{\mathbf{g}} \rangle
: \mathbf{A} \rightleftharpoons \mathbf{B} \\
\mathsf{ref}_{\mathbf{g}} & \doteq & \langle \xi_0^{\mathbf{g}} \dashv \pi_0^{\mathbf{g}} \rangle : \mathbf{A} \rightleftharpoons \diamondsuit(\mathbf{g}) \\
\mathsf{ref}_{\mathbf{g}}^{\propto} & \doteq & \langle \pi_1^{\mathbf{g}} \dashv \xi_1^{\mathbf{g}} \rangle : \diamondsuit(\mathbf{g}) \rightleftharpoons \mathbf{B} \\
\underline{\mathsf{ref}}_{\mathbf{g}} & \doteq & \langle \xi_0^{\mathbf{g}} \dashv \pi_0^{\mathbf{g}} \rangle : \mathsf{ker}(\check{\mathbf{g}}) \rightleftharpoons \diamondsuit(\mathbf{g}) \\
\underline{\mathsf{ref}}_{\mathbf{g}}^{\propto} & \doteq & \langle \pi_1^{\mathbf{g}} \dashv \xi_1^{\mathbf{g}} \rangle : \diamondsuit(\mathbf{g}) \rightleftharpoons \mathsf{ker}(\hat{\mathbf{g}}) \\
{\mathsf{lift}}^{0}_{\mathbf{g}} & \doteq & \langle \check{\mathbf{g}} \dashv \hat{\mathbf{g}} \rangle : \mathsf{ker}(\check{\mathbf{g}}) \rightleftharpoons \mathbf{B} \\
{\mathsf{lift}}^{1}_{\mathbf{g}} & \doteq & \langle \check{\mathbf{g}} \dashv \hat{\mathbf{g}} \rangle : \mathbf{A} \rightleftharpoons \mathsf{ker}(\hat{\mathbf{g}}) \\
{\mathsf{lift}}_{\mathbf{g}} & \doteq & \langle \check{\mathbf{g}} \dashv \hat{\mathbf{g}} \rangle : \mathsf{ker}(\check{\mathbf{g}}) \rightleftharpoons \mathsf{ker}(\hat{\mathbf{g}}) \\
\mathsf{clo}_{\mathbf{g}} & \doteq & \langle \mathrm{id}_{A} \dashv {\mathbf{g}}^{\bullet} \rangle : \mathbf{A} \rightleftharpoons \mathsf{ker}(\check{\mathbf{g}}) \\
\mathsf{int}_{\mathbf{g}} & \doteq & \langle {\mathbf{g}}^{\circ} \dashv \mathrm{id}_{B} \rangle : \mathsf{ker}(\hat{\mathbf{g}}) \rightleftharpoons \mathbf{B}
\end{array}$
&
$\begin{array}[b]{r@{\hspace{5pt}}c@{\hspace{5pt}}l}
\mathsf{ref}_{\mathbf{g}} \circ \mathsf{ref}_{\mathbf{g}}^{\propto} & = & \mathbf{g} \\
\mathsf{clo}_{\mathbf{g}} \circ \underline{\mathsf{ref}}_{\mathbf{g}} & = & \mathsf{ref}_{\mathbf{g}} \\
\underline{\mathsf{ref}}_{\mathbf{g}}^{\propto} \circ \mathsf{int}_{\mathbf{g}} & = & \mathsf{ref}_{\mathbf{g}}^{\propto} \\
\underline{\mathsf{ref}}_{\mathbf{g}} \circ \underline{\mathsf{ref}}_{\mathbf{g}}^{\propto} & = & {\mathsf{lift}}_{\mathbf{g}} \\
\underline{\mathsf{ref}}_{\mathbf{g}} \circ \mathsf{ref}_{\mathbf{g}}^{\propto} & = & {\mathsf{lift}}^{0}_{\mathbf{g}} \\
\mathsf{ref}_{\mathbf{g}} \circ \underline{\mathsf{ref}}_{\mathbf{g}}^{\propto} & = & {\mathsf{lift}}^{1}_{\mathbf{g}} \\
\mathsf{clo}_{\mathbf{g}} \circ \mathsf{lift}^0_{\mathbf{g}} & = & \mathbf{g} \\
\mathsf{clo}_{\mathbf{g}} \circ {\mathsf{lift}}_{\mathbf{g}} & = & \mathsf{lift}^1_{\mathbf{g}} \\
\mathsf{lift}^1_{\mathbf{g}} \circ \mathsf{int}_{\mathbf{g}} & = & \mathbf{g} \\
\mathsf{lift}_{\mathbf{g}} \circ \mathsf{int}_{\mathbf{g}} & = & \mathsf{lift}^0_{\mathbf{g}}
\end{array}$
\end{tabular}
\end{center}
\caption{Combined Factorization of Galois Connections}
\label{factorization-galois-connections}
\end{figure}

\subsection{Quartets of Galois Connections}\label{subsec:quartets:galois:connections}

\begin{figure}
\begin{center}
\begin{tabular}{c@{\hspace{50pt}}c}
\begin{picture}(80,40)(0,-10)
\put(-40,20){\makebox(80,40){$\mathbf{A}_1$}}
\put(1,20){\makebox(80,40){$\mathsf{ker}(\check{\mathbf{g}}_1)$}}
\put(40,20){\makebox(80,40){$\mathbf{B}_1$}}
\put(-40,-20){\makebox(80,40){$\mathbf{A}_2$}}
\put(1,-20){\makebox(80,40){$\mathsf{ker}(\check{\mathbf{g}}_2)$}}
\put(40,-20){\makebox(80,40){$\mathbf{B}_2$}}
\put(23,57){\makebox(40,20){\footnotesize{$\mathbf{g}_1$}}}
\put(1,38){\makebox(40,20){\footnotesize{${\mathsf{clo}}(\mathbf{g}_1)$}}}
\put(44,38){\makebox(40,20){\footnotesize{$\mathsf{lift}(\mathbf{g}_1)$}}}
\put(1,-18){\makebox(40,20){\footnotesize{${\mathsf{clo}}(\mathbf{g}_2)$}}}
\put(44,-18){\makebox(40,20){\footnotesize{$\mathsf{lift}(\mathbf{g}_2)$}}}
\put(23,-37){\makebox(40,20){\footnotesize{$\mathbf{g}_2$}}}
\put(-46,1){\makebox(80,40){\footnotesize{$\mathbf{a}$}}}
\put(-6,1){\makebox(80,40){\footnotesize{$\mathbf{c}$}}}
\put(34,1){\makebox(80,40){\footnotesize{$\mathbf{b}$}}}
\thicklines
\put(13,61){\line(-1,-1){10}}
\put(13,61){\line(1,0){54}}
\put(67,61){\vector(1,-1){10}}
\put(8,40){\vector(1,0){16}}
\put(56,40){\vector(1,0){16}}
\put(8,0){\vector(1,0){16}}
\put(56,0){\vector(1,0){16}}
\put(0,33){\vector(0,-1){26}}
\put(40,33){\vector(0,-1){26}}
\put(80,33){\vector(0,-1){26}}
\put(13,-21){\line(-1,1){10}}
\put(13,-21){\line(1,0){54}}
\put(67,-21){\vector(1,1){10}}
\end{picture}
&
\begin{picture}(120,100)
\put(-60,30){\makebox(120,60){$\mathbf{A}_1$}}
\put(1,30){\makebox(120,60){$\mathsf{ker}(\check{\mathbf{g}}_1)$}}
\put(60,30){\makebox(120,60){$\mathbf{B}_1$}}
\put(-60,-30){\makebox(120,60){$\mathbf{A}_2$}}
\put(1,-30){\makebox(120,60){$\mathsf{ker}(\check{\mathbf{g}}_2)$}}
\put(60,-30){\makebox(120,60){$\mathbf{B}_2$}}
\put(-50,0){\makebox(120,60){\footnotesize{$\check{\mathbf{a}}$}}}
\put(-70,0){\makebox(120,60){\footnotesize{$\hat{\mathbf{a}}$}}}
\put(10,0){\makebox(120,60){\footnotesize{$\check{\mathbf{c}}$}}}
\put(-10,0){\makebox(120,60){\footnotesize{$\hat{\mathbf{c}}$}}}
\put(70,0){\makebox(120,60){\footnotesize{$\check{\mathbf{b}}$}}}
\put(50,0){\makebox(120,60){\footnotesize{$\hat{\mathbf{b}}$}}}
\put(0,56){\makebox(120,60){\footnotesize{$\check{\mathbf{g}}_1$}}}
\put(-32,42){\makebox(120,60){\footnotesize{$\mathrm{id}_{A_1}$}}}
\put(33,42){\makebox(120,60){\footnotesize{$\check{\mathbf{g}}_1$}}}
\put(-32,18){\makebox(120,60){\footnotesize{${\mathbf{g}}_1^{\bullet}$}}}
\put(37,18){\makebox(120,60){\footnotesize{$\hat{\mathbf{g}}_1$}}}
\put(-32,-18){\makebox(120,60){\footnotesize{$\mathrm{id}_{A_2}$}}}
\put(33,-18){\makebox(120,60){\footnotesize{$\check{\mathbf{g}}_2$}}}
\put(-32,-42){\makebox(120,60){\footnotesize{${\mathbf{g}}_2^{\bullet}$}}}
\put(37,-42){\makebox(120,60){\footnotesize{$\hat{\mathbf{g}}_2$}}}
\put(0,-56){\makebox(120,60){\footnotesize{$\hat{\mathbf{g}}_2$}}}
\put(15,80){\line(-1,-1){10}}
\put(15,80){\line(1,0){90}}
\put(105,80){\vector(1,-1){10}}
\put(12,65){\vector(1,0){30}}
\put(78,65){\vector(1,0){30}}
\put(42,55){\vector(-1,0){30}}
\put(108,55){\vector(-1,0){30}}
\put(12,5){\vector(1,0){30}}
\put(78,5){\vector(1,0){30}}
\put(42,-5){\vector(-1,0){30}}
\put(108,-5){\vector(-1,0){30}}
\put(15,-20){\vector(-1,1){10}}
\put(15,-20){\line(1,0){90}}
\put(105,-20){\line(1,1){10}}
\put(5,45){\vector(0,-1){30}}
\put(-5,15){\vector(0,1){30}}
\put(65,45){\vector(0,-1){30}}
\put(55,15){\vector(0,1){30}}
\put(125,45){\vector(0,-1){30}}
\put(115,15){\vector(0,1){30}}
\end{picture}
\\ \\ \\ \\
\multicolumn{2}{c}{Morphism of Reflections}
\\ \\
\begin{picture}(80,40)(0,-10)
\put(-40,20){\makebox(80,40){$\mathbf{A}_1$}}
\put(1,20){\makebox(80,40){$\mathsf{ker}(\hat{\mathbf{g}}_1)$}}
\put(40,20){\makebox(80,40){$\mathbf{B}_1$}}
\put(-40,-20){\makebox(80,40){$\mathbf{A}_2$}}
\put(1,-20){\makebox(80,40){$\mathsf{ker}(\hat{\mathbf{g}}_2)$}}
\put(40,-20){\makebox(80,40){$\mathbf{B}_2$}}
\put(23,57){\makebox(40,20){\footnotesize{$\mathbf{g}_1$}}}
\put(1,38){\makebox(40,20){\footnotesize{${\mathsf{lift}}(\mathbf{g}_1)$}}}
\put(44,38){\makebox(40,20){\footnotesize{$\mathsf{int}(\mathbf{g}_1)$}}}
\put(1,-18){\makebox(40,20){\footnotesize{${\mathsf{lift}}(\mathbf{g}_2)$}}}
\put(44,-18){\makebox(40,20){\footnotesize{$\mathsf{int}(\mathbf{g}_2)$}}}
\put(23,-37){\makebox(40,20){\footnotesize{$\mathbf{g}_2$}}}
\put(-46,1){\makebox(80,40){\footnotesize{$\mathbf{a}$}}}
\put(-6,1){\makebox(80,40){\footnotesize{$\mathbf{d}$}}}
\put(34,1){\makebox(80,40){\footnotesize{$\mathbf{b}$}}}
\thicklines
\put(13,61){\line(-1,-1){10}}
\put(13,61){\line(1,0){54}}
\put(67,61){\vector(1,-1){10}}
\put(8,40){\vector(1,0){16}}
\put(56,40){\vector(1,0){16}}
\put(8,0){\vector(1,0){16}}
\put(56,0){\vector(1,0){16}}
\put(0,33){\vector(0,-1){26}}
\put(40,33){\vector(0,-1){26}}
\put(80,33){\vector(0,-1){26}}
\put(13,-21){\line(-1,1){10}}
\put(13,-21){\line(1,0){54}}
\put(67,-21){\vector(1,1){10}}
\end{picture}
&
\begin{picture}(120,100)
\put(-60,30){\makebox(120,60){$\mathbf{A}_1$}}
\put(1,30){\makebox(120,60){$\mathsf{ker}(\hat{\mathbf{g}}_1)$}}
\put(60,30){\makebox(120,60){$\mathbf{B}_1$}}
\put(-60,-30){\makebox(120,60){$\mathbf{A}_2$}}
\put(1,-30){\makebox(120,60){$\mathsf{ker}(\hat{\mathbf{g}}_2)$}}
\put(60,-30){\makebox(120,60){$\mathbf{B}_2$}}
\put(-50,0){\makebox(120,60){\footnotesize{$\check{\mathbf{a}}$}}}
\put(-70,0){\makebox(120,60){\footnotesize{$\hat{\mathbf{a}}$}}}
\put(10,0){\makebox(120,60){\footnotesize{$\check{\mathbf{d}}$}}}
\put(-10,0){\makebox(120,60){\footnotesize{$\hat{\mathbf{d}}$}}}
\put(70,0){\makebox(120,60){\footnotesize{$\check{\mathbf{b}}$}}}
\put(50,0){\makebox(120,60){\footnotesize{$\hat{\mathbf{b}}$}}}
\put(0,56){\makebox(120,60){\footnotesize{$\check{\mathbf{g}}_1$}}}
\put(-32,42){\makebox(120,60){\footnotesize{$\check{\mathbf{g}}_1$}}}
\put(33,42){\makebox(120,60){\footnotesize{${\mathbf{g}}_1^{\circ}$}}}
\put(-32,18){\makebox(120,60){\footnotesize{$\hat{\mathbf{g}}_1$}}}
\put(37,18){\makebox(120,60){\footnotesize{$\mathrm{id}_{B_1}$}}}
\put(-32,-18){\makebox(120,60){\footnotesize{$\check{\mathbf{g}}_2$}}}
\put(33,-18){\makebox(120,60){\footnotesize{${\mathbf{g}}_2^{\circ}$}}}
\put(-32,-42){\makebox(120,60){\footnotesize{$\hat{\mathbf{g}}_2$}}}
\put(37,-42){\makebox(120,60){\footnotesize{$\mathrm{id}_{B_2}$}}}
\put(0,-56){\makebox(120,60){\footnotesize{$\hat{\mathbf{g}}_2$}}}
\put(15,80){\line(-1,-1){10}}
\put(15,80){\line(1,0){90}}
\put(105,80){\vector(1,-1){10}}
\put(12,65){\vector(1,0){30}}
\put(78,65){\vector(1,0){30}}
\put(42,55){\vector(-1,0){30}}
\put(108,55){\vector(-1,0){30}}
\put(12,5){\vector(1,0){30}}
\put(78,5){\vector(1,0){30}}
\put(42,-5){\vector(-1,0){30}}
\put(108,-5){\vector(-1,0){30}}
\put(15,-20){\vector(-1,1){10}}
\put(15,-20){\line(1,0){90}}
\put(105,-20){\line(1,1){10}}
\put(5,45){\vector(0,-1){30}}
\put(-5,15){\vector(0,1){30}}
\put(65,45){\vector(0,-1){30}}
\put(55,15){\vector(0,1){30}}
\put(125,45){\vector(0,-1){30}}
\put(115,15){\vector(0,1){30}}
\end{picture}
\\ \\ \\ \\
\multicolumn{2}{c}{Morphism of Coreflections}
\end{tabular}
\end{center}
\caption{Factorization of Quartets (with Poset 0-cells)}
\label{quartet-factorization}
\end{figure}

\begin{flushleft}
\begin{tabular}{@{}p{280pt}c}
A \emph{quartet of Galois connections}
$\langle \mathbf{a}, \mathbf{b} \rangle : \mathbf{g}_1 \Rightarrow \mathbf{g}_2$
from vertical source Galois connection
$\mathbf{g}_1
= \langle \check{\mathbf{g}}_1, \hat{\mathbf{g}}_1 \rangle
: \mathbf{A}_1 \rightleftharpoons \mathbf{B}_1$
to vertical target Galois connection
$\mathbf{g}_2
= \langle \check{\mathbf{g}}_2, \hat{\mathbf{g}}_2 \rangle
: \mathbf{A}_2 \rightleftharpoons \mathbf{B}_2$,
is a pair of Galois connections,
the horizontal source
$\mathbf{a}
= \langle \check{\mathbf{a}}, \hat{\mathbf{a}} \rangle
: \mathbf{A}_1 \rightleftharpoons \mathbf{A}_2$
and the horizontal target
$\mathbf{b}
= \langle \check{\mathbf{b}}, \hat{\mathbf{b}} \rangle
: \mathbf{B}_1 \rightleftharpoons \mathbf{B}_2$,
which form a commuting square in the category $\mathsf{Adj}$:
$\mathbf{g}_1 \circ \mathbf{b} = \mathbf{a} \circ \mathbf{g}_2$.
In detail, 
this quartet condition means that
$\check{\mathbf{g}}_1 \cdot \check{\mathbf{b}} = \check{\mathbf{a}} \cdot \check{\mathbf{g}}_2$
and
$\hat{\mathbf{g}}_2 \cdot \hat{\mathbf{a}} = \hat{\mathbf{b}} \cdot \hat{\mathbf{g}}_1$.
&
\setlength{\unitlength}{0.8pt}
\begin{picture}(0,0)(20,70)
\thicklines
\put(10,0){\vector(1,0){30}}
\put(10,50){\vector(1,0){30}}
\put(0,40){\vector(0,-1){30}}
\put(50,40){\vector(0,-1){30}}
\put(-50,25){\makebox(100,50){$\mathbf{A}_1$}}
\put(0,25){\makebox(100,50){$\mathbf{B}_1$}}
\put(-50,-25){\makebox(100,50){$\mathbf{A}_2$}}
\put(0,-25){\makebox(100,50){$\mathbf{B}_2$}}
\put(0,35){\makebox(50,50){\footnotesize{$\mathbf{g}_1$}}}
\put(0,-35){\makebox(50,50){\footnotesize{$\mathbf{g}_2$}}}
\put(-35,0){\makebox(50,50){\footnotesize{$\mathbf{a}$}}}
\put(35,0){\makebox(50,50){\footnotesize{$\mathbf{b}$}}}
\end{picture}
\\
\multicolumn{2}{@{}p{340pt}}{Let $\Box(\mathsf{Adj})$ denote the double category,
whose objects (0-cells) are preorders,
whose vertical and horizontal arrows (1-cells) are Galois connections,
and whose 2-cells are quartets of Galois connections.}
\end{tabular}
\end{flushleft}

\subsubsection{Morphisms of Reflections.}

This paragraph is applicable to the extentional aspect of conceptual structures.
A morphism of reflections is a morphism in the full vertical subcategory whose horizontal arrows are reflections.
When $\langle \mathbf{a}, \mathbf{b} \rangle : \mathbf{g}_1 \Rightarrow \mathbf{g}_2$
is a morphism of reflections between posets,
the quartet condition implies the special conditions
$\check{\mathbf{b}} = \hat{\mathbf{g}}_1 \cdot \check{\mathbf{a}} \cdot \check{\mathbf{g}}_2$
and
$\hat{\mathbf{b}} = \hat{\mathbf{g}}_2 \cdot \hat{\mathbf{a}} \cdot \check{\mathbf{g}}_1$
(note that only the source needs to be a reflection);
that is,
that the Galois connection $\mathbf{b}$ is defined by the Galois connection $\mathbf{a}$.
In this case,
just as a reflection factors horizontally
$\mathbf{g}
= {\mathsf{clo}}(\mathbf{g}) \circ \mathsf{lift}(\mathbf{g})
: \mathbf{A} \rightleftharpoons \mathsf{ker}(\check{\mathbf{g}}) \rightleftharpoons \mathbf{B}$
in terms of the kernel of its left adjoint,
so also a morphism of reflections factors horizontally
$\langle \mathbf{a}, \mathbf{b} \rangle
= \langle \mathbf{a}, \mathbf{c} \rangle \circ \langle \mathbf{c}, \mathbf{b} \rangle$
in terms of a Galois connection between the kernel of the left adjoint of source and target 
(see the top of Figure~\ref{quartet-factorization});
that is,
given any morphism of reflections 
$\langle \mathbf{a}, \mathbf{b} \rangle : \mathbf{g}_1 \Rightarrow \mathbf{g}_2$,
there is a Galois connection
$\mathbf{c}
= \langle \check{\mathbf{c}}, \hat{\mathbf{c}} \rangle
: \mathsf{ker}(\check{\mathbf{g}}_1) \rightleftharpoons \mathsf{ker}(\check{\mathbf{g}}_2)$
such that
$\langle \mathbf{a}, \mathbf{c} \rangle : {\mathsf{clo}}(\mathbf{g}_1) \Rightarrow {\mathsf{clo}}(\mathbf{g}_2)$
and
$\langle \mathbf{c}, \mathbf{b} \rangle : \mathsf{lift}(\mathbf{g}_1) \Rightarrow \mathsf{lift}(\mathbf{g}_2)$
are quartets of Galois connections,
and
$\langle \mathbf{a}, \mathbf{b} \rangle
= \langle \mathbf{a}, \mathbf{c} \rangle \circ \langle \mathbf{c}, \mathbf{b} \rangle$
(horizontal composition of quartets).
The quartet conditions for $\langle \mathbf{a}, \mathbf{c} \rangle$ require that
$\mathrm{id}_{A_1} \cdot \check{\mathbf{c}} = \check{\mathbf{a}} \cdot \mathrm{id}_{A_2}$
and 
${\mathbf{g}}_2^{\bullet} \cdot \hat{\mathbf{a}} = \hat{\mathbf{c}} \cdot {\mathbf{g}}_1^{\bullet}$.
The first gives the definition 
$\check{\mathbf{c}} \doteq \check{\mathbf{a}}$.
Define $\hat{\mathbf{c}} \doteq {\mathbf{g}}_2^{\bullet} \cdot \hat{\mathbf{a}}$.
The second holds, since
$\underline{{\mathbf{g}}_2^{\bullet} \cdot \hat{\mathbf{a}}}$
= $\check{\mathbf{g}}_2 \cdot \hat{\mathbf{g}}_2 \cdot \hat{\mathbf{a}}$
= (by the second quartet condition for $\langle \mathbf{a}, \mathbf{b} \rangle$)
$\check{\mathbf{g}}_2 \cdot \hat{\mathbf{b}} \cdot \hat{\mathbf{g}}_1$
= (by a special property of reflection morphisms)
$\check{\mathbf{g}}_2 \cdot \hat{\mathbf{g}}_2 \cdot \hat{\mathbf{a}} \cdot \check{\mathbf{g}}_1 \cdot \hat{\mathbf{g}}_1
= {\mathbf{g}}_2^{\bullet} \cdot \hat{\mathbf{a}} \cdot {\mathbf{g}}_1^{\bullet}
= \underline{\hat{\mathbf{c}} \cdot {\mathbf{g}}_1^{\bullet}}$.
The quartet conditions for $\langle \mathbf{b}, \mathbf{c} \rangle$ require that
$\check{\mathbf{g}}_1 \cdot \check{\mathbf{b}} = \check{\mathbf{c}} \cdot \check{\mathbf{g}}_2$
and
$\hat{\mathbf{g}}_2 \cdot \hat{\mathbf{c}} = \hat{\mathbf{b}} \cdot \hat{\mathbf{g}}_1$.
The first holds,
since
$\check{\mathbf{g}}_1 \cdot \check{\mathbf{b}}$ 
= (by the first quartet condition for $\langle \mathbf{a}, \mathbf{b} \rangle$)
$\check{\mathbf{a}} \cdot \check{\mathbf{g}}_2
= \check{\mathbf{c}} \cdot \check{\mathbf{g}}_2$.
The second holds,
since
$\hat{\mathbf{g}}_2 \cdot \hat{\mathbf{c}} 
= \hat{\mathbf{g}}_2 \cdot {\mathbf{g}}_2^{\bullet} \cdot \hat{\mathbf{a}}$
= (by idempotency)
$\hat{\mathbf{g}}_2 \cdot \hat{\mathbf{a}}$
= (by the second quartet condition for $\langle \mathbf{a}, \mathbf{b} \rangle$)
$\hat{\mathbf{b}} \cdot \hat{\mathbf{g}}_1$.
For the Galois connection
$\mathbf{c}
= \langle \check{\mathbf{c}}, \hat{\mathbf{c}} \rangle
: \mathsf{ker}(\check{\mathbf{g}}_1) \rightleftharpoons \mathsf{ker}(\check{\mathbf{g}}_2)$,
we prove that the fundamental condition holds by chasing it around the left hand commutative square in the top of Figure~\ref{quartet-factorization}:
$\check{\mathbf{c}}(a_1) \leq_{\check{\mathbf{g}}_2} a_2$ 
iff (by the fundamental condition for $\mathsf{clo}(\mathbf{g}_2)$)
$\check{\mathbf{c}}(a_1) \leq_{\mathbf{B}_1} {\mathbf{g}}_2^{\bullet}(a_2)$
iff
$\check{\mathbf{a}}(a_1) \leq_{\mathbf{B}_1} {\mathbf{g}}_2^{\bullet}(a_2)$
iff (by the fundamental condition for $\mathbf{a}$) 
$a_1 \leq_{\mathbf{A}_1} \hat{\mathbf{a}}({\mathbf{g}}_2^{\bullet}(a_2))$
iff (by the underlined equality above) 
$a_1 \leq_{\mathbf{A}_1} {\mathbf{g}}_1^{\bullet}(\hat{\mathbf{c}}(a_2))$ 
iff (by the fundamental condition for $\mathsf{clo}(\mathbf{g}_1)$) 
$a_1 \leq_{\check{\mathbf{g}}_1} \hat{\mathbf{c}}(a_2)$
for every source element $a_1 \in A_1$ and every target element $a_2 \in A_2$.

Refletions and their morphism form a category $\mathsf{Refl}$
with a horizontal source component functor
$\partial_0^{\mathrm{h}} : \mathsf{Clg}_\tau^{\mathrm{op}} \rightarrow \mathsf{Adj}$,
a horizontal target component functor
$\partial_1^{\mathrm{h}} : \mathsf{Clg}_\tau \rightarrow \mathsf{Set}$,
and a natural transformation
$\mathsf{refl} 
: \partial_0^{\mathrm{h}} \Rightarrow \partial_1^{\mathrm{h}}
: \mathsf{Refl} \rightarrow \mathsf{Adj}$.
The category $\mathsf{Refl}$ is a subcategory of 
${\mathsf{Adj}}^{\mathbf{2}}
= {\Box(\mathsf{Adj})}_{\mathrm{vert}}$ the category of Galois connections (as objects),
the vertical category of 
the double cateogy $\Box(\mathsf{Adj})$ of quartets of Galous connections,
with inclusion functor
$\mathsf{incl} : \mathsf{Refl} \rightarrow {\mathsf{Adj}}^{\mathbf{2}}$.
The components are related as
$\partial_0^{\mathrm{h}} = \mathsf{incl} \circ \partial_0^{\mathrm{h}}$,
$\partial_1^{\mathrm{h}} = \mathsf{incl} \circ \partial_1^{\mathrm{h}}$
and $\mathsf{refl} = \mathsf{incl} \circ \mathsf{galcon}$.

\subsubsection{Morphisms of Coreflections.}\label{par:morphisms:coreflections}

This paragraph is applicable to the intensional aspect of conceptual structures.
A morphism of coreflections is a morphism in the full vertical subcategory whose horizontal arrows are coreflections.
When $\langle \mathbf{a}, \mathbf{b} \rangle : \mathbf{g}_1 \Rightarrow \mathbf{g}_2$
is a morphism of coreflections between posets,
the quartet condition implies the special conditions
$\check{\mathbf{a}} = \check{\mathbf{g}}_1 \cdot \check{\mathbf{b}} \cdot \hat{\mathbf{g}}_2$
and
$\hat{\mathbf{a}} = \check{\mathbf{g}}_2 \cdot \hat{\mathbf{b}} \cdot \hat{\mathbf{g}}_1$
(note that only the target needs to be a coreflection);
that is,
that the Galois connection $\mathbf{a}$ is defined by the Galois connection $\mathbf{b}$.
In this case,
just as a coreflection factors horizontally
$\mathbf{g}
= {\mathsf{lift}}(\mathbf{g}) \circ \mathsf{int}(\mathbf{g})
: \mathbf{A} \rightleftharpoons \mathsf{ker}(\hat{\mathbf{g}}) \rightleftharpoons \mathbf{B}$
in terms of the kernel of its right adjoint,
so also a morphism of coreflections factors horizontally
$\langle \mathbf{a}, \mathbf{b} \rangle
= \langle \mathbf{a}, \mathbf{d} \rangle \circ \langle \mathbf{d}, \mathbf{b} \rangle$
in terms of a Galois connection between the kernel of the right adjoint of source and target 
(see the bottom of Figure~\ref{quartet-factorization});
that is,
given any morphism of coreflections 
$\langle \mathbf{a}, \mathbf{b} \rangle : \mathbf{g}_1 \Rightarrow \mathbf{g}_2$,
there is a Galois connection
$\mathbf{d}
= \langle \check{\mathbf{d}}, \hat{\mathbf{d}} \rangle
: \mathsf{ker}(\hat{\mathbf{g}}_1) \rightleftharpoons \mathsf{ker}(\hat{\mathbf{g}}_2)$
such that
$\langle \mathbf{a}, \mathbf{d} \rangle : {\mathsf{lift}}(\mathbf{g}_1) \Rightarrow {\mathsf{lift}}(\mathbf{g}_2)$
and
$\langle \mathbf{d}, \mathbf{b} \rangle : \mathsf{int}(\mathbf{g}_1) \Rightarrow \mathsf{int}(\mathbf{g}_2)$
are quartets of Galois connections,
and
$\langle \mathbf{a}, \mathbf{b} \rangle
= \langle \mathbf{a}, \mathbf{d} \rangle \circ \langle \mathbf{d}, \mathbf{b} \rangle$
(horizontal composition of quartets).
The quartet conditions for $\langle \mathbf{d}, \mathbf{b} \rangle$ require that
${\mathbf{g}}_1^{\circ} \cdot \check{\mathbf{b}} = \check{\mathbf{d}} \cdot {\mathbf{g}}_2^{\circ}$
and $\mathrm{id}_{B_2} \cdot \hat{\mathbf{d}} = \hat{\mathbf{b}} \cdot \mathrm{id}_{B_1}$.
The second gives the definition 
$\hat{\mathbf{d}} \doteq \hat{\mathbf{b}}$.
Define $\check{\mathbf{d}} \doteq {\mathbf{g}}_1^{\circ} \cdot \check{\mathbf{b}}$.
The first holds, since
$\underline{{\mathbf{g}}_1^{\circ} \cdot \check{\mathbf{b}}}$
= $\hat{\mathbf{g}}_1 \cdot \check{\mathbf{g}}_1 \cdot \check{\mathbf{b}}$
= (by the first quartet condition for $\langle \mathbf{a}, \mathbf{b} \rangle$)
$\hat{\mathbf{g}}_1 \cdot \check{\mathbf{a}} \cdot \check{\mathbf{g}}_2$
= (by a special property of coreflection morphisms)
$\hat{\mathbf{g}}_1 \cdot \check{\mathbf{g}}_1 \cdot \check{\mathbf{b}} \cdot \hat{\mathbf{g}}_2 \cdot \check{\mathbf{g}}_2
= {\mathbf{g}}_1^{\circ} \cdot \check{\mathbf{b}} \cdot {\mathbf{g}}_2^{\circ}
= \underline{\check{\mathbf{d}} \cdot {\mathbf{g}}_2^{\circ}}$.
The quartet conditions for $\langle \mathbf{a}, \mathbf{d} \rangle$ require that
$\check{\mathbf{g}}_1 \cdot \check{\mathbf{d}} = \check{\mathbf{a}} \cdot \check{\mathbf{g}}_2$
and
$\hat{\mathbf{g}}_2 \cdot \hat{\mathbf{a}} = \hat{\mathbf{d}} \cdot \hat{\mathbf{g}}_1$.
The first holds,
since
$\check{\mathbf{g}}_1 \cdot \check{\mathbf{d}} 
= \check{\mathbf{g}}_1 \cdot {\mathbf{g}}_1^{\circ} \cdot \check{\mathbf{b}}$
= (by idempotency)
$\check{\mathbf{g}}_1 \cdot \check{\mathbf{b}}$
= (by the first quartet condition for $\langle \mathbf{a}, \mathbf{b} \rangle$)
$\check{\mathbf{a}} \cdot \check{\mathbf{g}}_2$.
The second holds,
since
$\hat{\mathbf{g}}_2 \cdot \hat{\mathbf{a}}$ 
= (by the second quartet condition for $\langle \mathbf{a}, \mathbf{b} \rangle$)
$\hat{\mathbf{b}} \cdot \hat{\mathbf{g}}_1
= \hat{\mathbf{d}} \cdot \hat{\mathbf{g}}_1$.
For the Galois connection
$\mathbf{d}
= \langle \check{\mathbf{d}}, \hat{\mathbf{d}} \rangle
: \mathsf{ker}(\hat{\mathbf{g}}_1) \rightleftharpoons \mathsf{ker}(\hat{\mathbf{g}}_2)$,
we prove that the fundamental condition holds by chasing it around the right hand commutative square in the bottom of Figure~\ref{quartet-factorization}:
$\check{\mathbf{d}}(b_1) \leq_{\hat{\mathbf{g}}_2} b_2$ 
iff (by the fundamental condition for $\mathsf{int}(\mathbf{g}_2)$)
${\mathbf{g}}_2^{\circ}(\check{\mathbf{d}}(b_1)) \leq_{\mathbf{B}_2} b_2$ 
iff (by the underlined equality above) 
$\check{\mathbf{b}}({\mathbf{g}}_1^{\circ}(b_1)) \leq_{\mathbf{B}_2} b_2$
iff (by the fundamental condition for $\mathbf{b}$) 
${\mathbf{g}}_1^{\circ}(b_1) \leq_{\mathbf{B}_1} \hat{\mathbf{b}}(b_2)$
iff
${\mathbf{g}}_1^{\circ}(b_1) \leq_{\mathbf{B}_1} \hat{\mathbf{d}}(b_2)$
iff (by the fundamental condition for $\mathsf{int}(\mathbf{g}_1)$) $b_1 \leq_{\hat{\mathbf{g}}_1} \hat{\mathbf{d}}(b_2)$
for every source element $b_1 \in B_1$ and every target element $b_2 \in B_2$.

\section{The Category of Classifications}\label{sec:category:classifications}

\subsection{Classifications}\label{subsec:classification:category:objects}

Classification structure $\mathbf{A}$ has three isomorphic versions:
a relation version, a function version and a Galois connection version.

The relation version of classification structure \cite{barwise:seligman:97}
consists of 
a set $\mathsf{inst}(\mathbf{A})$ of things to be classified, called the \emph{instances} of $\mathbf{A}$,
a set $\mathsf{typ}(\mathbf{A})$ of things used to classify the instances, called the \emph{types} of $\mathbf{A}$,
and a binary \emph{classification} relation 
$\models_{\mathbf{A}} \subseteq \mathsf{inst}(\mathbf{A}) {\times} \mathsf{typ}(\mathbf{A})$.
The notation $x \models_{\mathbf{A}} y$ is read 
``instance $x$ has type $y$ in classification $\mathbf{A}$''. 
Classifications are known as formal contexts in the theory of formal concept analysis as describe in Ganter and Wille \cite{ganter:wille:99},
where instances are called formal objects,
types are called formal attributes,
and the classification relation is called an incidence relation.

\subsubsection{Extent and Intent.}

Let $\mathbf{A}$ be any classification.
The function version of classification structure
consists of a pair of dual functions.

For any type $y \in \mathsf{typ}(\mathbf{A})$, 
the extent (or instance set) of $y$ is the set 
$\mathsf{ext}_{\mathbf{A}}(y) 
= \{ x \in \mathsf{inst}(\mathbf{A}) \mid x \models_{\mathbf{A}} y \}$.
This defines the \emph{extent} function
$\mathsf{ext}_{\mathbf{A}} : \mathsf{typ}(\mathbf{A}) \rightarrow {\wp}\,\mathsf{inst}(\mathbf{A})$.
Types $y, y^{\prime} \in \mathsf{typ}(\mathbf{A})$ are coextensive in $\mathbf{A}$ 
when $\mathsf{ext}_{\mathbf{A}}(y) = \mathsf{ext}_{\mathbf{A}}(y^{\prime})$.
A classification $\mathbf{A}$ is extensional when there are no distinct coextensive types;
that is,
when the extent function $\mathsf{ext}_{\mathbf{A}}$ is injective.
Instance and power classifications are extensional.
For any infomorphism
$\mathbf{f} 
= \langle \mathsf{inst}(\mathbf{f}), \mathsf{typ}(\mathbf{f}) \rangle : \mathbf{A}_1 \rightleftharpoons \mathbf{A}_2$,
the following naturality condition holds:
$\mathsf{ext}_{\mathbf{A}_1} \cdot_{\mathsf{Set}} {\mathsf{inst}(\mathbf{f})}^{{-}1} 
= \mathsf{typ}(\mathbf{f}) \cdot_{\mathsf{Set}} \mathsf{ext}_{\mathbf{A}_2}$.
Hence,
there is an extent natural transformation
$\mathsf{ext} : \mathsf{typ} \Rightarrow \mathsf{inst}^{\mathrm{op}} \circ {(-)}^{-1} : \mathsf{Clsn} \rightarrow \mathsf{Set}$
(bottom left Figure~\ref{classification-component-architecture}).

Dually,
for any instance $x \in \mathsf{inst}(\mathbf{A})$, 
the intent (or type set) of $x$ is the set 
$\mathsf{int}_{\mathbf{A}}(x) 
= \{ y \in \mathsf{typ}(\mathbf{A}) \mid x \models_{\mathbf{A}} y \}$.
This defines the \emph{intent} function
$\mathsf{int}_{\mathbf{A}} : \mathsf{inst}(\mathbf{A}) \rightarrow {\wp}\,\mathsf{typ}(\mathbf{A})$.
Instances $x, x^{\prime} \in \mathsf{inst}(\mathbf{A})$ are indistinguishable in $\mathbf{A}$ 
when $\mathsf{int}_{\mathbf{A}}(x) = \mathsf{int}_{\mathbf{A}}(x^{\prime})$.
A classification $\mathbf{A}$ is separated when there are no distinct indistinguishable instances;
that is,
when the intent function $\mathsf{int}_{\mathbf{A}}$ is injective.
Instance and power classifications are separated.
For any infomorphism
$\mathbf{f} 
= \langle \mathsf{inst}(\mathbf{f}), \mathsf{typ}(\mathbf{f}) \rangle : \mathbf{A}_1 \rightleftharpoons \mathbf{A}_2$,
the following naturality condition holds:
$\mathsf{int}_{\mathbf{A}_2} \cdot_{\mathsf{Set}} {\mathsf{typ}(\mathbf{f})}^{{-}1} 
= \mathsf{inst}(\mathbf{f}) \cdot_{\mathsf{Set}} \mathsf{int}_{\mathbf{A}_1}$.
Hence,
there is an intent natural transformation
$\mathsf{int} : \mathsf{inst} \Rightarrow \mathsf{typ}^{\mathrm{op}} \circ {(-)}^{-1} : \mathsf{Clsn}^{\mathrm{op}} \rightarrow \mathsf{Set}$
(bottom left Figure~\ref{classification-component-architecture}).

\subsubsection{Derivation}

Let $\mathbf{A}$ be any classification.
The Galois connection version of classification structure
consists of a pair of derivation operators: 
$X^{\mathbf{A}} = \{ y \in \mathsf{typ}(\mathbf{A}) \mid x \models_{\mathbf{A}} y \;\mbox{for all}\; x \in X \}$ 
for any instance subset $X \subseteq \mathsf{inst}(\mathbf{A})$, 
and 
$Y^{\mathbf{A}} = \{ x \in \mathsf{inst}(\mathbf{A}) \mid x \models_{\mathbf{A}} y \;\mbox{for all}\; y \in Y \}$ 
for any type subset $Y \subseteq \mathsf{typ}(\mathbf{A})$. 
When $X : \mathsf{inst}(\mathbf{A}) \rightarrow 1$ and $Y : \mathsf{typ}(\mathbf{A}) \leftarrow 1$ are regarded as relations, 
derivation is seen to be residuation, 
$X^{\mathbf{A}} = X{\setminus}{\mathbf{A}}$ and $Y^{\mathbf{A}} = {\mathbf{A}}/Y$,
where the classification relation is represented as $\mathbf{A}$.  
Derivation forms a Galois connection
\[\langle ({-})^{\mathbf{A}}, ({-})^{\mathbf{A}} \rangle : {\wp}\,\mathsf{inst}(\mathbf{A}) \rightleftharpoons {{\wp}\,\mathsf{typ}(\mathbf{A})}^{\mathrm{op}}.\]
Hence,
there are closure operations
$({-})^{\bullet} = ({-})^{\mathbf{A}\mathbf{A}}$
for instances and types.
Derivation is expressed by the following conditions,
which indicate a natural duality between instance and types.
\begin{center}
\begin{tabular}{l}
$X_{1} \subseteq X_{2}$ implies $X_{2}^{\mathbf{A}} \subseteq X_{1}^{\mathbf{A}}$ for $X_{1}, X_{2} \subseteq \mathsf{inst}(\mathbf{A})$, \\
$Y_{1} \subseteq Y_{2}$ implies $Y_{2}^{\mathbf{A}} \subseteq Y_{1}^{\mathbf{A}}$ for $Y_{1}, Y_{2} \subseteq \mathsf{typ}(\mathbf{A})$, \\
$X \subseteq X^{\bullet}$ and $X^{\mathbf{A}}
= X^{\bullet\mathbf{A}}
= X^{\mathbf{A}\bullet}$ for $X \subseteq \mathsf{inst}(\mathbf{A})$, \\
$Y \subseteq Y^{\bullet}$ and 
$Y^{\mathbf{A}}
 = Y^{\bullet\mathbf{A}}
 = Y^{\mathbf{A}\bullet}$ 
for $Y \subseteq \mathsf{typ}(\mathbf{A})$, \\
$(\bigcup_{k \in K} X_k)^{\mathbf{A}} = \bigcap_{k \in K} X_k^{\mathbf{A}}$ for $X_k \subseteq \mathsf{inst}(\mathbf{A})$, \\ $(\bigcup_{k \in K} Y_k)^{\mathbf{A}} = \bigcap_{k \in K} Y_k^{\mathbf{A}}$ for $Y_k \subseteq \mathsf{typ}(\mathbf{A})$.
\end{tabular}
\end{center}

\paragraph{Examples.}

Systemic examples of classifications abound. 
Any preorder $\mathbf{A}$ is a classification $\mathsf{incl}(\mathbf{A})$,
whose instance and type sets are the underlying set
$\mathsf{inst}(\mathsf{incl}(\mathbf{A})) 
= \mathsf{typ}(\mathsf{incl}(\mathbf{A})) = A$
and whose incidence is the order
${\vdash_{\mathsf{incl}(\mathbf{A})}} = {\leq_{\mathbf{A}}}$.
Given any set $X$, 
the instance power classification 
$\check{\wp}\,{X} = \langle X, {\wp}{X}, \in \rangle$ associated with $X$, 
has elements of $X$ as instances, subsets of $X$ as types,
with membership serving as the classification relation.
Dually,
given any set $Y$,
the type power classification 
$\hat{\wp}\,{Y} = {\langle {\wp}{Y}, Y, \ni \rangle}$ 
associated with $Y$, 
has subsets of $Y$ as instances, elements of $Y$ as types, 
with membership transpose serving as the classification relation.
Given any classification $\mathbf{A}$,
the dual (or transpose) classification (or involution)
${\mathbf{A}}^{\propto} 
= \langle \mathsf{typ}(\mathbf{A}), \mathsf{inst}(\mathbf{A}), \models_{\mathbf{A}}^{\propto} \rangle$ 
has types of $\mathbf{A}$ as instances, instances of $\mathbf{A}$ as types, with classification being the transpose of the $\mathbf{A}$-classification.
The transpose operator is idempotent:
${\mathbf{A}}^{\propto\propto} = \mathbf{A}$.
The transpose of instance power is type power, and vice versa:
${(\check{\wp}\,{X})}^{\propto} = \hat{\wp}\,{X}$
and
${(\hat{\wp}\,{Y})}^{\propto} = \check{\wp}\,{Y}$.

\paragraph{Formal Concepts.}

A formal concept $c = (X, Y)$ is a pair of subsets, 
$X \subseteq \mathsf{inst}(\mathbf{A})$ and $Y \subseteq \mathsf{typ}(\mathbf{A})$,
that satisfies the closure properties $X = Y^{\mathbf{A}}$ and $Y = X^{\mathbf{A}}$. 
The subset $X = \mathsf{ext}_{\mathbf{A}}(c)$ is called the extent of $c$,
and the subset $Y = \mathsf{int}_{\mathbf{A}}(c)$ is called the intent of $c$.
Let $\mathsf{conc}(\mathbf{A})$ denote the set of all concepts of the classification $\mathbf{A}$.
There is a naturally defined concept order
$c_1 \leq c_2$ that holds when 
$\mathsf{ext}_{\mathbf{A}}(c_1) \subseteq \mathsf{ext}_{\mathbf{A}}(c_2)$,
or equivalently when 
$\mathsf{int}_{\mathbf{A}}(c_1) \supseteq \mathsf{int}_{\mathbf{A}}(c_2)$,
for any two concepts $c_1, c_2 \in \mathsf{conc}(\mathbf{A})$.
Then we say that $c_1$ is more specific than $c_2$ or dually that $c_2$ is more generic than $c_1$.
This partial order underlies a complete lattice 
$\mathsf{lat}(\mathbf{A}) 
= \langle \mathsf{conc}(\mathbf{A}), \leq_{\mathbf{A}}, \bigwedge_{\mathbf{A}}, \bigvee_{\mathbf{A}} \rangle$,
where the meet and join are defined by: 
$\bigwedge_{\mathbf{A}}C 
= \langle \bigcap_{c{\in}C} \, \mathsf{ext}_{\mathbf{A}}(c), \left( \bigcup_{c{\in}C} \, \mathsf{int}_{\mathbf{A}}(c) \right)^{\bullet} \rangle$ 
and 
$\bigvee_{\mathbf{A}}C = \langle \left( \bigcup_{c{\in}C} \, \mathsf{ext}_{\mathbf{A}}(c) \right)^{\bullet}, \bigcap_{c{\in}C} \, \mathsf{int}_{\mathbf{A}}(c) \rangle$ 
for any collection of concepts $C \subseteq \mathsf{conc}(\mathbf{A})$.

Define the instance embedding relation 
$\iota_{\mathbf{A}} \subseteq \mathsf{inst}(\mathbf{A}) \times \mathsf{conc}(\mathbf{A})$ as follows: 
for every instance $x \in \mathsf{inst}(\mathbf{A})$ and every formal concept $c \in \mathsf{conc}(\mathbf{A})$ 
the relationship $x\iota_{\mathbf{A}}c$ holds when $x \in \mathsf{ext}(c)$,
$x$ is in the extent of $c$. 
This relation is closed on the right with respect to lattice order:
$\iota_{\mathbf{A}} \circ \leq_{\mathbf{A}} = \iota_{\mathbf{A}}$. 
Instances are mapped into the lattice by the instance embedding function
$\iota_{\mathbf{A}} : \mathsf{inst}(\mathbf{A}) \rightarrow \mathsf{conc}(\mathbf{A})$,
where for each instance $x \in \mathsf{inst}(\mathbf{A})$, 
$\iota_{\mathbf{A}}(x) = \left( \{x\}^{\bullet}, \{x\}^{\mathbf{A}} \right)$ 
is defined to be the smallest concept with $x$ in its extent.
This (not necessarily injective) function is expressed in terms of the relation as the meet 
$\iota_{\mathbf{A}}(x) = \bigwedge_{\mathbf{A}} x\iota_{\mathbf{A}}$.
Conversely,
the relation can be expressed in terms of the function as:
$x\iota_{\mathbf{A}}c$ when $\iota_{\mathbf{A}}(x) \leq_{\mathbf{A}} c$.
Concepts in $\iota_{\mathbf{A}}[\mathsf{inst}(\mathbf{A})]$ are called instance concepts.
Any concept $c \in \mathsf{conc}(\mathbf{A})$ can be expressed as the join 
$c 
= \bigvee_{x{\in}\mathsf{ext}(c)} \iota_{\mathbf{A}}(x)
= \bigvee_{\mathbf{A}} \iota_{\mathbf{A}}[\mathsf{ext}_{\mathbf{A}}(c)]$ 
of a subset of instance concepts;
that is,
$\iota_{\mathbf{A}}[\mathsf{inst}(\mathbf{A})]$ is join-dense in $\mathsf{conc}(\mathbf{A})$. 
Subsets of instances are mapped into the lattice by the instance subset embedding function
$\mathsf{iota}_{\mathbf{A}} : {\wp}\,\mathsf{inst}(\mathbf{A}) \rightarrow \mathsf{conc}(\mathbf{A})$,
where
$\mathsf{iota}_{\mathbf{A}}(X)
= \langle X^{\bullet}, X^{\mathbf{A}} \rangle
= \bigwedge_{\mathbf{A}} \{ c \in \mathsf{conc}(\mathbf{A}) \mid X \subseteq \mathsf{ext}_{\mathbf{A}}(c) \}$
for each instance subset $X \subseteq \mathsf{inst}(\mathbf{A})$.

Dually, define the type embedding relation 
$\tau_{\mathbf{A}} \subseteq \mathsf{conc}(\mathbf{A}) \times \mathsf{typ}(\mathbf{A})$ as follows:
for every type $y \in \mathsf{typ}(\mathbf{A})$ and every formal concept $c \in \mathsf{conc}(\mathbf{A})$ the relationship $c\tau_{\mathbf{A}}y$ holds when $y \in \mathsf{int}(c)$,
$y$ is in the intent of $c$.
This relation is closed on the left respect to lattice order:
$\leq_{\mathbf{A}} \circ \tau_{\mathbf{A}} = \tau_{\mathbf{A}}$.
Types are mapped into the lattice by the type embedding function
$\tau_{\mathbf{A}} : \mathsf{typ}(\mathbf{A}) \rightarrow \mathsf{conc}(\mathbf{A})$,
where for each type $y \in \mathsf{typ}(\mathbf{A})$,
$\tau_{\mathbf{A}}(y) = \left( \{y\}^{\mathbf{A}}, \{y\}^{\bullet} \right)$
is defined to be the largest concept with $y$ in its intent.
This (not necessarily injective) function is expressed in terms of the relation as the join 
$\tau_{\mathbf{A}}(y) = \bigvee_{\mathbf{A}} \tau_{\mathbf{A}}y$.
Conversely,
the relation can be expressed in terms of the function as:
$c\tau_{\mathbf{A}}y$ when $c \leq_{\mathbf{A}} \tau_{\mathbf{A}}(x)$.
Concepts in $\tau_{\mathbf{A}}[\mathsf{typ}(\mathbf{A})]$ are called type concepts.
Any concept $c \in \mathsf{conc}(\mathbf{A})$ can be expressed as the meet
$c 
= \bigwedge_{y{\in}\mathsf{int}(c)} \tau_{\mathbf{A}}(y)
= \bigwedge_{\mathbf{A}} \tau_{\mathbf{A}}[\mathsf{int}_{\mathbf{A}}(c)]$ 
of a subset of type concepts;
that is,
$\tau_{\mathbf{A}}[\mathsf{typ}(\mathbf{A})]$ is meet-dense in $\mathsf{conc}(\mathbf{A})$.
Subsets of types are mapped into the lattice by the type subset embedding function
$\mathsf{T}_{\mathbf{A}} : {\wp}\,\mathsf{typ}(\mathbf{A}) \rightarrow \mathsf{conc}(\mathbf{A})$,
where
$\mathsf{T}_{\mathbf{A}}(Y)
= \langle Y^{\mathbf{A}}, Y^{\bullet} \rangle
= \bigvee_{\mathbf{A}} \{ c \in \mathsf{conc}(\mathbf{A}) \mid Y \subseteq \mathsf{int}_{\mathbf{A}}(c) \}$
for each type subset $Y \subseteq \mathsf{typ}(\mathbf{A})$.

The quintuple
$\mathsf{clg}(\mathbf{A}) 
= \langle \mathsf{lat}(\mathbf{A}), \mathsf{inst}(\mathbf{A}), \mathsf{typ}(\mathbf{A}), \iota_{\mathbf{A}}, \tau_{\mathbf{A}} \rangle$
is a concept lattice,
the concept lattice associated with the classification $\mathbf{A}$.

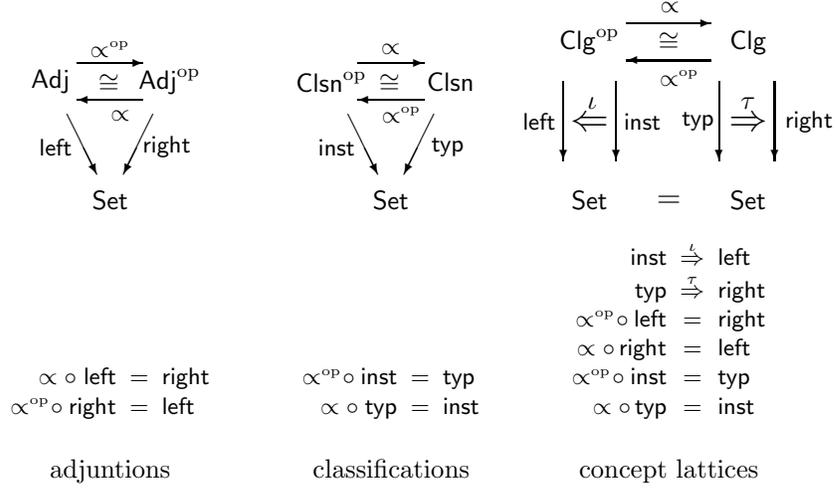
\begin{figure}
\begin{center}
\begin{tabular}{c@{\hspace{25pt}}c@{\hspace{25pt}}c}
\begin{picture}(45,60)(0,0)
\put(-15,33.75){\makebox(30,22.5){$\mathsf{Adj}$}}
\put(30,33.75){\makebox(30,22.5){$\mathsf{Adj}^{\mathrm{op}}$}}
\put(7.5,-11.25){\makebox(30,22.5){$\mathsf{Set}$}}
\put(12.5,51.0){\makebox(20,12.5){\small{${\propto}^{\mbox{\tiny{op}}}$}}}
\put(10,52){\vector(1,0){25}}
\put(7,38){\makebox(30,12.5){{$\cong$}}}
\put(35,38){\vector(-1,0){25}}
\put(16.5,26){\makebox(20,12.5){\small{$\propto$}}}
\put(-8,14){\makebox(20,12.5){\small{$\mathsf{left}$}}}
\put(6.25,32.5){\vector(1,-2){11.25}}
\put(38.75,32.5){\vector(-1,-2){11.25}}
\put(34,14){\makebox(20,12.5){\small{$\mathsf{right}$}}}
\end{picture}
&
\begin{picture}(45,60)(0,0)
\put(-15,33.75){\makebox(30,22.5){$\mathsf{Clsn}^{\mathrm{op}}$}}
\put(30,33.75){\makebox(30,22.5){$\mathsf{Clsn}$}}
\put(7.5,-11.25){\makebox(30,22.5){$\mathsf{Set}$}}
\put(12.5,51.0){\makebox(20,12.5){\small{$\propto$}}}
\put(10,52){\vector(1,0){25}}
\put(7,38){\makebox(30,12.5){{$\cong$}}}
\put(35,38){\vector(-1,0){25}}
\put(16.5,26){\makebox(20,12.5){\small{${\propto}^{\mbox{\tiny{op}}}$}}}
\put(-8,14){\makebox(20,12.5){\small{$\mathsf{inst}$}}}
\put(6.25,32.5){\vector(1,-2){11.25}}
\put(38.75,32.5){\vector(-1,-2){11.25}}
\put(34,14){\makebox(20,12.5){\small{$\mathsf{typ}$}}}
\end{picture}
&
\begin{picture}(60,60)(0,0)
\put(-15,48.75){\makebox(30,22.5){$\mathsf{Clg}^{\mathrm{op}}$}}
\put(45,48.75){\makebox(30,22.5){$\mathsf{Clg}$}}
\put(-15,-11.25){\makebox(30,22.5){$\mathsf{Set}$}}
\put(45,-11.25){\makebox(30,22.5){$\mathsf{Set}$}}
\put(20.5,67){\makebox(20,12.5){\small{$\propto$}}}
\put(14,67){\vector(1,0){32}}
\put(15,53.75){\makebox(30,12.5){{$\cong$}}}
\put(46,53){\vector(-1,0){32}}
\put(24,40){\makebox(20,12.5){\small{${\propto}^{\mbox{\tiny{op}}}$}}}
\put(-29,23.75){\makebox(20,12.5){\small{$\mathsf{left}$}}}
\put(-10,45){\vector(0,-1){30}}
\put(-14,30.75){\makebox(30,12.5){{$\iota$}}}
\put(-15,22.75){\makebox(30,12.5){{\Large{$\Leftarrow$}}}}
\put(10,45){\vector(0,-1){30}}
\put(10,23.75){\makebox(20,12.5){\small{$\mathsf{inst}$}}}
\put(32,23.75){\makebox(18,12.5){\small{$\mathsf{typ}$}}}
\put(49,45){\vector(0,-1){30}}
\put(45,30.75){\makebox(30,12.5){{$\tau$}}}
\put(45,22.75){\makebox(30,12.5){{\Large{$\Rightarrow$}}}}
\put(70,45){\vector(0,-1){30}}
\put(72,23.75){\makebox(22,12.5){\small{$\mathsf{right}$}}}
\put(15,-6.75){\makebox(30,12.5){{\large{$=$}}}}
\end{picture}
\\ \\
\small{$\begin{array}[b]{r@{\hspace{5pt}}c@{\hspace{5pt}}l}
\propto \circ\; \mathsf{left} & = & \mathsf{right} \\
{\propto}^{\mbox{\tiny{op}}}\! \circ \mathsf{right} & = & \mathsf{left}
\end{array}$}
&
\small{$\begin{array}[b]{r@{\hspace{5pt}}c@{\hspace{5pt}}l}
{\propto}^{\mbox{\tiny{op}}}\! \circ \mathsf{inst} & = & \mathsf{typ} \\
\propto \circ\; \mathsf{typ}                       & = & \mathsf{inst}
\end{array}$}
&
\small{$\begin{array}[b]{r@{\hspace{5pt}}c@{\hspace{5pt}}l}
\mathsf{inst} & \stackrel{\iota}{\Rightarrow} & \mathsf{left} \\
\mathsf{typ}  & \stackrel{\tau}{\Rightarrow} & \mathsf{right} \\
{\propto}^{\mbox{\tiny{op}}}\! \circ \mathsf{left} & = & \mathsf{right} \\
\propto \circ\, \mathsf{right}                     & = & \mathsf{left} \\
{\propto}^{\mbox{\tiny{op}}}\! \circ \mathsf{inst} & = & \mathsf{typ} \\
\propto \circ\, \mathsf{typ}                       & = & \mathsf{inst}
\end{array}$}
\\ \\
adjuntions & classifications & concept lattices 
\end{tabular}
\end{center}
\caption{Components and Involution}
\label{component-involution}
\end{figure}

\subsection{Infomorphisms}\label{sec:classification:category:morphisms}

\begin{figure}
\begin{center}

\begin{tabular}{c@{\hspace{20pt}}c}

\begin{tabular}[b]{c}
\\
\begin{picture}(65,90)(-32,0)
\put(-35,73.5){\vector(2,1){25}}
\put(-10,50){\vector(-2,1){25}}
\put(-60,56){\makebox(30,22.5){$\mathsf{Set}^{\mathrm{op}}$}}
\put(-24,67){\makebox(20,12.5){\footnotesize{$\mathsf{ext}$}}}
\put(-25,58){\makebox(20,12.5){\Large{$\Leftarrow$}}}
\put(-40,82){\makebox(30,12.5){\scriptsize{${(-)}^{\mathrm{-1}}$}}}
\put(-40,43){\makebox(30,12.5){\footnotesize{${\mathsf{inst}}^{\mathrm{op}}$}}}
\put(10,40){\vector(2,-1){25}}
\put(35,17.5){\vector(-2,-1){25}}
\put(30,11.25){\makebox(30,22.5){$\mathsf{Set}$}}
\put(-15,78.75){\makebox(30,22.5){$\mathsf{Set}$}}
\put(-15,33.75){\makebox(30,22.5){$\mathsf{Clsn}$}}
\put(-11,-11.25){\makebox(30,22.5){$\mathsf{Set}^{\mathrm{op}}$}}
\put(0,58.25){\makebox(20,12.5){\footnotesize{$\mathsf{typ}$}}}
\put(0,56.25){\vector(0,1){22.5}}
\put(-26,19.25){\makebox(30,12.5){\footnotesize{$\mathsf{inst}^{\mathrm{op}}$}}}
\put(0,33.75){\vector(0,-1){22.5}}
\put(14,34){\makebox(30,12.5){\footnotesize{$\mathsf{typ}$}}}
\put(26,-2){\makebox(30,12.5){\scriptsize{$( {(-)}^{\mathrm{-1}} )^{\mathrm{op}}$}}}
\put(5,22){\makebox(20,12.5){\footnotesize{$\mathsf{int}^{\mathrm{op}}$}}}
\put(1,13){\makebox(30,12.5){\Large{$\Leftarrow$}}}
\end{picture} 
\\ \\
\small{$\begin{array}[t]{r@{\hspace{5pt}}c@{\hspace{5pt}}l}
\mathsf{inst}^{\mathrm{op}} \circ {(-)}^{\mathrm{-1}} & \stackrel{\mathsf{ext}}{\Leftarrow} & \mathsf{typ} \\
\mathsf{typ}^{\mathrm{op}} \circ {(-)}^{\mathrm{-1}} & \stackrel{\mathsf{int}}{\Leftarrow} & \mathsf{inst} \\
\mathsf{ext} & = & {\propto}^{\mbox{\tiny{op}}} \circ \mathsf{int} \\
\mathsf{int} & = & {\propto} \circ \mathsf{ext}
\end{array}$}
\\ \\
\underline{extent and intent}
\end{tabular}

&

\end{tabular}

\end{center}
\caption{The Component Architecture for Classifications}
\label{classification-component-architecture}
\end{figure}

\begin{figure}
\begin{center}

\begin{tabular}{c@{\hspace{20pt}}c}

\begin{tabular}[b]{c}
\\
\begin{picture}(90,90)(0,0)
\put(-15,33.75){\makebox(30,22.5){$\mathsf{Set}$}}
\put(30,78.75){\makebox(30,22.5){$\mathsf{Set}$}}
\put(34,-11.25){\makebox(30,22.5){$\mathsf{Set}^{\mathrm{op}}$}}
\put(80,33.75){\makebox(30,22.5){$\mathsf{Set}^{\mathrm{op}}$}}
\put(30,33.75){\makebox(30,22.5){$\mathsf{Clsn}$}}
\put(5,70){\makebox(20,12.5){\footnotesize{$\mathrm{id}_{\mathsf{Set}}$}}}
\put(-4,7.5){\makebox(20,12.5){\scriptsize{$( {(-)}^{\mathrm{-1}} )^{\mathrm{op}}$}}}
\put(44,58.25){\makebox(20,12.5){\footnotesize{$\mathsf{typ}$}}}
\put(24,56.25){\makebox(20,12.5){\tiny{\emph{unit}}}}
\put(34,62.5){\oval(15,6)}
\put(25,19.25){\makebox(20,12.5){\footnotesize{$\mathsf{inst}^{\mathrm{op}}$}}}
\put(48,21.25){\makebox(20,12.5){\tiny{\emph{counit}}}}
\put(58,27.5){\oval(20,6)}
\put(13.5,45){\makebox(20,12.5){\small{$\hat{\wp}$}}}
\put(58.5,45){\makebox(20,12.5){\small{$\check{\wp}$}}}
\put(68,69){\makebox(20,12.5){\scriptsize{${(-)}^{\mathrm{-1}}$}}}
\put(68,8.5){\makebox(20,12.5){\footnotesize{$\mathrm{id}_{\mathsf{Set}^{\mathrm{op}}}$}}}
\put(10,55){\vector(1,1){25}}
\put(80,55){\vector(-1,1){25}}
\put(10,35){\vector(1,-1){25}}
\put(80,35){\vector(-1,-1){25}}
\put(10,45){\vector(1,0){25}}
\put(80,45){\vector(-1,0){25}}
\put(45,56.25){\vector(0,1){22.5}}
\put(45,33.75){\vector(0,-1){22.5}}
\end{picture}
\\ \\
\small{$\begin{array}[t]{r@{\hspace{5pt}}c@{\hspace{5pt}}l}
\check{\wp} \circ \mathsf{inst}^{\mathrm{op}} & = & \mathrm{id}_{\mathsf{Set}^{\mathrm{op}}} \\ 
\check{\wp} \circ \mathsf{typ}                & = & {(-)}^{\mathrm{-1}} \\
\hat{\wp} \circ \mathsf{inst}^{\mathrm{op}}   & = & ( {(-)}^{\mathrm{-1}} )^{\mathrm{op}} \\ 
\hat{\wp} \circ \mathsf{typ}                  & = & \mathrm{id}_{\mathsf{Set}}
\end{array}$}
\\ \\
\underline{power and components}
\end{tabular}

&

\begin{tabular}[b]{c}
\\
\begin{picture}(90,90)(-12,0)
\put(7.5,78.75){\makebox(30,22.5){$\mathsf{Set}$}}
\put(-15,33.75){\makebox(30,22.5){$\mathsf{Clsn}^{\mathrm{op}}$}}
\put(35,33.75){\makebox(30,22.5){$\mathsf{Clsn}$}}
\put(12.5,-11.25){\makebox(30,22.5){$\mathsf{Set}^{\mathrm{op}}$}}
\put(-5,65){\makebox(20,12.5){\small{${\check{\wp}}^{\mbox{\tiny{op}}}$}}}
\put(17.5,80){\vector(-1,-2){11.25}}
\put(27.5,80){\vector(1,-2){11.25}}
\put(12.5,51.0){\makebox(20,12.5){\small{$\propto$}}}
\put(29.5,65){\makebox(20,12.5){\small{$\hat{\wp}$}}}
\put(10,52){\vector(1,0){25}}
\put(7,38){\makebox(30,12.5){{$\cong$}}}
\put(35,38){\vector(-1,0){25}}
\put(16,26){\makebox(20,12.5){\small{${\propto}^{\mbox{\tiny{op}}}$}}}
\put(-5,11){\makebox(20,12.5){\small{${\hat{\wp}}^{\mbox{\tiny{op}}}$}}}
\put(17.5,10){\vector(-1,2){11.25}}
\put(27.5,10){\vector(1,2){11.25}}
\put(28,11){\makebox(20,12.5){\small{$\check{\wp}$}}}
\end{picture} 
\\ \\
\small{$\begin{array}[t]{r@{\hspace{5pt}}c@{\hspace{5pt}}l}
{\check{\wp}}^{\mbox{\tiny{op}}} \circ \propto & = & \hat{\wp} \\
{\hat{\wp}}^{\mbox{\tiny{op}}} \circ \propto & = & \check{\wp}
\end{array}$}
\\ \\
\underline{power and involution}
\end{tabular}

\\ \\

\begin{tabular}[b]{c}
\\
\begin{picture}(90,90)(0,0)
\put(-15,33.75){\makebox(30,22.5){$\mathsf{Set}$}}
\put(30,78.75){\makebox(30,22.5){$\mathsf{Set}$}}
\put(34,-11.25){\makebox(30,22.5){$\mathsf{Set}^{\mathrm{op}}$}}
\put(80,33.75){\makebox(30,22.5){$\mathsf{Set}^{\mathrm{op}}$}}
\put(30,33.75){\makebox(30,22.5){$\mathsf{Clg}$}}
\put(5,70){\makebox(20,12.5){\footnotesize{$\mathrm{id}_{\mathsf{Set}}$}}}
\put(-4,7.5){\makebox(20,12.5){\scriptsize{$( {(-)}^{\mathrm{-1}} )^{\mathrm{op}}$}}}
\put(44,58.25){\makebox(20,12.5){\footnotesize{$\mathsf{typ}$}}}
\put(24,56.25){\makebox(20,12.5){\tiny{\emph{unit}}}}
\put(34,62.5){\oval(15,6)}
\put(25,19.25){\makebox(20,12.5){\footnotesize{$\mathsf{inst}^{\mathrm{op}}$}}}
\put(48,21.25){\makebox(20,12.5){\tiny{\emph{counit}}}}
\put(58,27.5){\oval(20,6)}
\put(13.5,45){\makebox(20,12.5){\small{$\hat{\wp}$}}}
\put(58.5,45){\makebox(20,12.5){\small{$\check{\wp}$}}}
\put(68,69){\makebox(20,12.5){\scriptsize{${(-)}^{\mathrm{-1}}$}}}
\put(68,8.5){\makebox(20,12.5){\footnotesize{$\mathrm{id}_{\mathsf{Set}^{\mathrm{op}}}$}}}
\put(10,55){\vector(1,1){25}}
\put(80,55){\vector(-1,1){25}}
\put(10,35){\vector(1,-1){25}}
\put(80,35){\vector(-1,-1){25}}
\put(10,45){\vector(1,0){25}}
\put(80,45){\vector(-1,0){25}}
\put(45,56.25){\vector(0,1){22.5}}
\put(45,33.75){\vector(0,-1){22.5}}
\end{picture}
\\ \\
\small{$\begin{array}[t]{r@{\hspace{5pt}}c@{\hspace{5pt}}l}
\check{\wp} \circ \mathsf{inst}^{\mathrm{op}} & = & \mathrm{id}_{\mathsf{Set}^{\mathrm{op}}} \\ 
\check{\wp} \circ \mathsf{typ}                & = & {(-)}^{\mathrm{-1}} \\
\hat{\wp} \circ \mathsf{inst}^{\mathrm{op}}   & = & ( {(-)}^{\mathrm{-1}} )^{\mathrm{op}} \\ 
\hat{\wp} \circ \mathsf{typ}                  & = & \mathrm{id}_{\mathsf{Set}}
\end{array}$}
\\ \\
\underline{power and components}
\end{tabular}

&

\begin{tabular}[b]{c}
\\
\begin{picture}(90,90)(-12,0)
\put(7.5,78.75){\makebox(30,22.5){$\mathsf{Set}$}}
\put(-15,33.75){\makebox(30,22.5){$\mathsf{Clg}^{\mathrm{op}}$}}
\put(35,33.75){\makebox(30,22.5){$\mathsf{Clg}$}}
\put(12.5,-11.25){\makebox(30,22.5){$\mathsf{Set}^{\mathrm{op}}$}}
\put(-5,65){\makebox(20,12.5){\small{${\check{\wp}}^{\mbox{\tiny{op}}}$}}}
\put(17.5,80){\vector(-1,-2){11.25}}
\put(27.5,80){\vector(1,-2){11.25}}
\put(12.5,51.0){\makebox(20,12.5){\small{$\propto$}}}
\put(29.5,65){\makebox(20,12.5){\small{$\hat{\wp}$}}}
\put(10,52){\vector(1,0){25}}
\put(7,38){\makebox(30,12.5){\large{$\cong$}}}
\put(35,38){\vector(-1,0){25}}
\put(16,26){\makebox(20,12.5){\small{${\propto}^{\mbox{\tiny{op}}}$}}}
\put(-5,11){\makebox(20,12.5){\small{${\hat{\wp}}^{\mbox{\tiny{op}}}$}}}
\put(17.5,10){\vector(-1,2){11.25}}
\put(27.5,10){\vector(1,2){11.25}}
\put(28,11){\makebox(20,12.5){\small{$\check{\wp}$}}}
\end{picture} 
\\ \\
\small{$\begin{array}[t]{r@{\hspace{5pt}}c@{\hspace{5pt}}l}
{\check{\wp}}^{\mbox{\tiny{op}}} \circ \propto & = & \hat{\wp} \\
{\hat{\wp}}^{\mbox{\tiny{op}}} \circ \propto & = & \check{\wp}
\end{array}$}
\\ \\
\underline{power and involution}
\end{tabular}

\end{tabular}

\end{center}
\caption{Power: Classifications and Concept Lattices}
\label{power}
\end{figure}

\paragraph{Basics.}

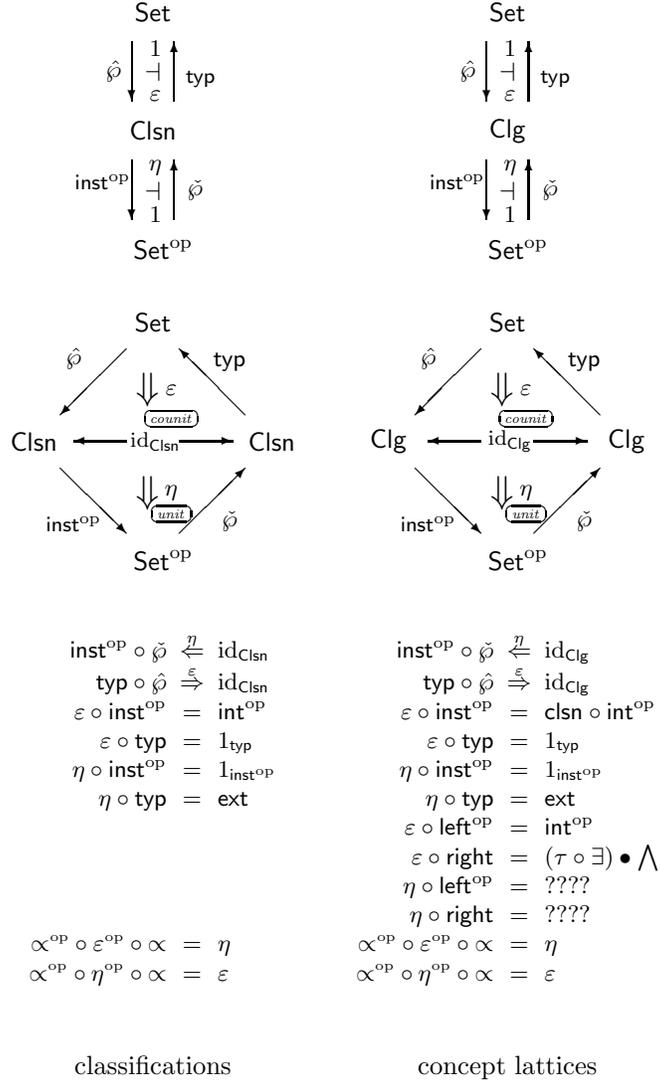
\begin{figure}
\begin{center}

\begin{tabular}{c@{\hspace{20pt}}c}

\begin{picture}(30,90)(-15,0)
\put(-15,78.75){\makebox(30,22.5){$\mathsf{Set}$}}
\put(-15,33.75){\makebox(30,22.5){$\mathsf{Clsn}$}}
\put(-11,-11.25){\makebox(30,22.5){$\mathsf{Set}^{\mathrm{op}}$}}
\put(-25,61.25){\makebox(20,12.5){\small{$\hat{\wp}$}}}
\put(-8,78.75){\vector(0,-1){22.5}}
\put(8,58.25){\makebox(20,12.5){\footnotesize{$\mathsf{typ}$}}}
\put(8,56.25){\vector(0,1){22.5}}
\put(-9,70){\makebox(20,12.5){\small{$1$}}}
\put(-15,61){\makebox(30,12.5){{$\dashv$}}}
\put(-9,52){\makebox(20,12.5){\small{$\varepsilon$}}}
\put(-34,19.25){\makebox(30,12.5){\footnotesize{$\mathsf{inst}^{\mathrm{op}}$}}}
\put(-8,33.75){\vector(0,-1){22.5}}
\put(6,16.25){\makebox(20,12.5){\small{$\check{\wp}$}}}
\put(8,11.25){\vector(0,1){22.5}}
\put(-9,25){\makebox(20,12.5){\small{$\eta$}}}
\put(-15,16){\makebox(30,12.5){{$\dashv$}}}
\put(-9,7){\makebox(20,12.5){\small{$1$}}}
\end{picture}

&

\begin{picture}(30,90)(-15,0)
\put(-15,78.75){\makebox(30,22.5){$\mathsf{Set}$}}
\put(-15,33.75){\makebox(30,22.5){$\mathsf{Clg}$}}
\put(-11,-11.25){\makebox(30,22.5){$\mathsf{Set}^{\mathrm{op}}$}}
\put(-25,61.25){\makebox(20,12.5){\small{$\hat{\wp}$}}}
\put(-8,78.75){\vector(0,-1){22.5}}
\put(8,58.25){\makebox(20,12.5){\footnotesize{$\mathsf{typ}$}}}
\put(8,56.25){\vector(0,1){22.5}}
\put(-9,70){\makebox(20,12.5){\small{$1$}}}
\put(-15,61){\makebox(30,12.5){{$\dashv$}}}
\put(-9,52){\makebox(20,12.5){\small{$\varepsilon$}}}
\put(-34,19.25){\makebox(30,12.5){\footnotesize{$\mathsf{inst}^{\mathrm{op}}$}}}
\put(-8,33.75){\vector(0,-1){22.5}}
\put(6,16.25){\makebox(20,12.5){\small{$\check{\wp}$}}}
\put(8,11.25){\vector(0,1){22.5}}
\put(-9,25){\makebox(20,12.5){\small{$\eta$}}}
\put(-15,16){\makebox(30,12.5){{$\dashv$}}}
\put(-9,7){\makebox(20,12.5){\small{$1$}}}
\end{picture}

\\ \\ \\

\begin{picture}(90,90)(0,0)
\put(-15,33.75){\makebox(30,22.5){$\mathsf{Clsn}$}}
\put(30,78.75){\makebox(30,22.5){$\mathsf{Set}$}}
\put(34,-11.25){\makebox(30,22.5){$\mathsf{Set}^{\mathrm{op}}$}}
\put(75,33.75){\makebox(30,22.5){$\mathsf{Clsn}$}}
\put(5,70){\makebox(20,12.5){\small{$\hat{\wp}$}}}
\put(5,7.5){\makebox(20,12.5){\footnotesize{$\mathsf{inst}^{\mathrm{op}}$}}}
\put(31,38.75){\makebox(30,12.5){\footnotesize{$\mathrm{id}_{\mathsf{Clsn}}$}}}
\put(64,69){\makebox(20,12.5){\small{$\mathsf{typ}$}}}
\put(64,8.5){\makebox(20,12.5){\small{$\check{\wp}$}}}
\put(28,58.25){\makebox(30,12.5){\Large{$\Downarrow$}}}
\put(42,58.25){\makebox(20,12.5){\small{$\varepsilon$}}}
\put(42,47.25){\makebox(20,12.5){\tiny{\emph{counit}}}}
\put(52,53.5){\oval(20,6)}
\put(28,19.25){\makebox(30,12.5){\Large{$\Downarrow$}}}
\put(42,19.25){\makebox(20,12.5){\small{$\eta$}}}
\put(42,11.25){\makebox(20,12.5){\tiny{\emph{unit}}}}
\put(52,17.5){\oval(15,6)}
\put(55,45){\vector(1,0){20}}
\put(35,45){\vector(-1,0){20}}
\put(35,80){\vector(-1,-1){25}}
\put(80,55){\vector(-1,1){25}}
\put(10,35){\vector(1,-1){25}}
\put(55,10){\vector(1,1){25}}
\end{picture}

&

\begin{picture}(90,90)(0,0)
\put(-15,33.75){\makebox(30,22.5){$\mathsf{Clg}$}}
\put(30,78.75){\makebox(30,22.5){$\mathsf{Set}$}}
\put(34,-11.25){\makebox(30,22.5){$\mathsf{Set}^{\mathrm{op}}$}}
\put(75,33.75){\makebox(30,22.5){$\mathsf{Clg}$}}
\put(5,70){\makebox(20,12.5){\small{$\hat{\wp}$}}}
\put(5,7.5){\makebox(20,12.5){\footnotesize{$\mathsf{inst}^{\mathrm{op}}$}}}
\put(31,38.75){\makebox(30,12.5){\footnotesize{$\mathrm{id}_{\mathsf{Clg}}$}}}
\put(64,69){\makebox(20,12.5){\small{$\mathsf{typ}$}}}
\put(64,8.5){\makebox(20,12.5){\small{$\check{\wp}$}}}
\put(28,58.25){\makebox(30,12.5){\Large{$\Downarrow$}}}
\put(42,58.25){\makebox(20,12.5){\small{$\varepsilon$}}}
\put(42,47.25){\makebox(20,12.5){\tiny{\emph{counit}}}}
\put(52,53.5){\oval(20,6)}
\put(28,19.25){\makebox(30,12.5){\Large{$\Downarrow$}}}
\put(42,19.25){\makebox(20,12.5){\small{$\eta$}}}
\put(42,11.25){\makebox(20,12.5){\tiny{\emph{unit}}}}
\put(52,17.5){\oval(15,6)}
\put(55,45){\vector(1,0){20}}
\put(35,45){\vector(-1,0){20}}
\put(35,80){\vector(-1,-1){25}}
\put(80,55){\vector(-1,1){25}}
\put(10,35){\vector(1,-1){25}}
\put(55,10){\vector(1,1){25}}
\end{picture}

\\ \\ \\

\small{$\begin{array}[t]{r@{\hspace{5pt}}c@{\hspace{5pt}}l}
\mathsf{inst}^{\mathrm{op}} \circ \check{\wp} & \stackrel{\eta}{\Leftarrow} & \mathrm{id}_{\mathsf{Clsn}} \\
\mathsf{typ} \circ \hat{\wp}                  & \stackrel{\varepsilon}{\Rightarrow} & \mathrm{id}_{\mathsf{Clsn}} \\
\varepsilon \circ \mathsf{inst}^{\mathrm{op}} & = & \mathsf{int}^{\mathrm{op}} \\
\varepsilon \circ \mathsf{typ}                & = & 1_{\mathsf{typ}} \\
\eta \circ \mathsf{inst}^{\mathrm{op}}        & = & 1_{\mathsf{inst}^{\mathrm{op}}} \\
\eta \circ \mathsf{typ}                       & = & \mathsf{ext} \\
&& \\
&& \\
&& \\
&& \\
{\propto}^{\mbox{\tiny{op}}} \circ {\varepsilon}^{\mbox{\tiny{op}}} \circ {\propto} & = & \eta \\
{\propto}^{\mbox{\tiny{op}}} \circ {\eta}^{\mbox{\tiny{op}}} \circ {\propto} & = & \varepsilon
\end{array}$}

&

\small{$\begin{array}[t]{r@{\hspace{5pt}}c@{\hspace{5pt}}l}
\mathsf{inst}^{\mathrm{op}} \circ \check{\wp} & \stackrel{\eta}{\Leftarrow} & \mathrm{id}_{\mathsf{Clg}} \\
\mathsf{typ} \circ \hat{\wp}                  & \stackrel{\varepsilon}{\Rightarrow} & \mathrm{id}_{\mathsf{Clg}} \\
\varepsilon \circ \mathsf{inst}^{\mathrm{op}} & = & \mathsf{clsn} \circ \mathsf{int}^{\mathrm{op}} \\
\varepsilon \circ \mathsf{typ}                & = & 1_{\mathsf{typ}} \\
\eta \circ \mathsf{inst}^{\mathrm{op}}        & = & 1_{\mathsf{inst}^{\mathrm{op}}} \\
\eta \circ \mathsf{typ}                       & = & \mathsf{ext} \\
\varepsilon \circ \mathsf{left}^{\mathrm{op}} & = & \mathsf{int}^{\mathrm{op}} \\
\varepsilon \circ \mathsf{right}              & = & (\tau \circ \exists) \bullet \bigwedge \\
\eta \circ \mathsf{left}^{\mathrm{op}} & = & \mbox{????} \\
\eta \circ \mathsf{right}              & = & \mbox{????} \\
{\propto}^{\mbox{\tiny{op}}} \circ {\varepsilon}^{\mbox{\tiny{op}}} \circ {\propto} & = & \eta \\
{\propto}^{\mbox{\tiny{op}}} \circ {\eta}^{\mbox{\tiny{op}}} \circ {\propto} & = & \varepsilon
\end{array}$}

\\ \\ \\

classifications & concept lattices

\end{tabular}
\end{center}
\caption{Component Adjunctions}
\label{component-adjunction}
\end{figure}

An infomorphism 
$\mathbf{f} 
= \langle \mathsf{inst}(\mathbf{f}), \mathsf{typ}(\mathbf{f}) \rangle : \mathbf{A}_1 \rightleftharpoons \mathbf{A}_2$
from source classification $\mathbf{A}_1$ to target classification $\mathbf{A}_2$ 
is defined by Barwise and Seligman \cite{barwise:seligman:97}
to consist of a contravariant pair of functions, 
a function 
$\mathsf{inst}(\mathbf{f}) : \mathsf{inst}(\mathbf{A}_1) \leftarrow \mathsf{inst}(\mathbf{A}_2)$ 
in the backward direction between instances
and a function 
$\mathsf{typ}(\mathbf{f}) : \mathsf{typ}(\mathbf{A}_1) \rightarrow \mathsf{typ}(\mathbf{A}_2)$ 
in the forward direction between types, 
satisfying the fundamental condition
$\mathsf{inst}(\mathbf{f})(x_2) \models_{\mathbf{A}_1} y_1$ iff $x_2 \models_{\mathbf{A}_2} \mathsf{typ}(\mathbf{f})(y_1)$
for each target instance $x_2 \in \mathsf{inst}(\mathbf{A}_2)$ and each source type $y_1 \in \mathsf{typ}(\mathbf{A}_1)$.
The fundamental condition for infomorphisms expresses the invariance of classification under change of notation. 
Given any two infomorphisms 
$\mathbf{f} : \mathbf{A} \rightleftharpoons \mathbf{B}$
and 
$\mathbf{g} : \mathbf{B} \rightleftharpoons \mathbf{C}$,
there is a composite infomorphism
$\mathbf{f} \circ \mathbf{g} : \mathbf{A} \rightleftharpoons \mathbf{C}$
defined by composing the instance and type functions:
$\mathsf{inst}(\mathbf{f} \circ \mathbf{g}) = \mathsf{inst}(\mathbf{g}) \cdot \mathsf{inst}(\mathbf{f})$
and
$\mathsf{typ}(\mathbf{f} \circ \mathbf{g}) = \mathsf{typ}(\mathbf{f}) \cdot \mathsf{typ}(\mathbf{g})$.
Given any classification $\mathbf{A}$,
there is an identity infomorphism
$\mathrm{id}_{\mathbf{A}} : \mathbf{A} \rightleftharpoons \mathbf{A}$
(with respect to composition)
defined in terms of the identity functions on types and instances:
$\mathsf{inst}(\mathrm{id}_{\mathbf{A}}) = \mathrm{id}_{\mathsf{inst}(\mathbf{A})}$
and
$\mathsf{typ}(\mathrm{id}_{\mathbf{A}}) = \mathrm{id}_{\mathsf{typ}(\mathbf{A})}$.
Using these notions of composition and identity,
classifications and infomorphisms form the category $\mathsf{Clsn}$. 
Classifications and infomorphisms resolve into components:
there is an instance functor
$\mathsf{inst} : \mathsf{Clsn}^{\mathrm{op}} \rightarrow \mathsf{Set}$ 
and
a type functor
$\mathsf{typ} : \mathsf{Clsn} \rightarrow \mathsf{Set}$.

\paragraph{Examples.}

Systemic examples of infomorphisms abound. 
Any Galois connection
$\mathbf{g} : \mathbf{A} \rightleftharpoons \mathbf{B}$
is an infomorphism 
$\mathsf{incl}(\mathbf{g}) 
: \mathsf{incl}(\mathbf{A}) \rightleftharpoons \mathsf{incl}(\mathbf{B})$,
whose instance and type functions are the left and right functions
$\mathsf{inst}(\mathsf{incl}(\mathbf{g})) = \mathsf{left}(\mathbf{g})$
and
$\mathsf{typ}(\mathsf{incl}(\mathbf{g})) = \mathsf{right}(\mathbf{g})$.
This fact is expressed as the functor 
$\mathsf{incl} : \mathsf{Adj}^{\mathrm{op}} \rightarrow \mathsf{Clsn}$ 
define by
$\mathsf{incl}^{\mathrm{op}} \circ \mathsf{inst} = \mathsf{left}$ 
and 
$\mathsf{incl} \circ \mathsf{typ} = \mathsf{right}$.
The fundamental property of infomorphisms is clearly related to the notion of adjointness.
For any function $f : X_1 \leftarrow X_2$,
it and its inverse image function $f^{-1} : {\wp}\,X_1 \rightarrow {\wp}\,X_2$ 
form an instance power infomorphism 
$\check{\wp}\,f = \langle f, f^{-1} \rangle : \check{\wp}\,X_1 \rightleftharpoons \check{\wp}\,X_2$ 
between the instance power classifications.
This defines an instance power functor
$\check{\wp} : \mathsf{Set}^{\mathrm{op}} \rightarrow \mathsf{Clsn}$,
where
$\check{\wp} \circ \mathsf{inst}^{\mathrm{op}} = \mathrm{id}_{\mathsf{Set}^{\mathrm{op}}}$
and
$\check{\wp} \circ \mathsf{typ} = {(-)}^{-1}$
(top left Figure~\ref{classification-component-architecture}).
For any function $g : Y_1 \rightarrow Y_2$,
it and its inverse image function $g^{-1} : {\wp}\,Y_1 \leftarrow {\wp}\,Y_2$ 
form a type power infomorphism $\hat{\wp}\,g = \langle g^{-1}, g \rangle : \hat{\wp}\,Y_1 \rightleftharpoons \hat{\wp}\,Y_2$ between the type power classifications.
This defines a type power functor
$\hat{\wp} : \mathsf{Set} \rightarrow \mathsf{Clsn}$,
where
$\hat{\wp} \circ \mathsf{inst}^{\mathrm{op}} = {( {(-)}^{-1} )}^\mathrm{op}$
and
$\hat{\wp} \circ \mathsf{typ} = \mathrm{id}_{\mathsf{Set}}$
(top left Figure~\ref{classification-component-architecture}).

Given any infomorphism
$\mathbf{f} 
= \langle \mathsf{inst}(\mathbf{f}), \mathsf{typ}(\mathbf{f}) \rangle : \mathbf{A}_1 \rightleftharpoons \mathbf{A}_2$,
the dual (or transpose) infomorphism (or involution)
${\mathbf{f}}^{\propto} 
= \langle \mathsf{typ}(\mathbf{f}), \mathsf{inst}(\mathbf{f}) \rangle : \mathbf{A}_2^{\propto} \rightleftharpoons \mathbf{A}_1^{\propto}$,
has the transpose of the target classification of $\mathbf{f}$ as source classification, 
has the transpose of the source classification of $\mathbf{f}$ as target classification, 
has the type function of $\mathbf{f}$ as instance function, and 
has the instance function of $\mathbf{f}$ as type function.
The transpose operator is idempotent:
${\mathbf{f}}^{\propto\propto} = \mathbf{f}$.
There is a transpose functor
$\propto : {\mathsf{Clsn}}^{\mathrm{op}} \rightarrow \mathsf{Clsn}$,
where 
${\propto}^\mathrm{op} \circ \mathsf{inst} = \mathsf{typ}$
and
${\propto} \circ \mathsf{typ} = \mathsf{inst}$
(top middle Figure~\ref{classification-component-architecture}).
Transpose is an isomorphism of categories, 
with
${\propto}^{-1} = {\propto}^{\mathrm{op}} : \mathsf{Clsn} \rightarrow {\mathsf{Clsn}}^{\mathrm{op}}$.
The transpose of instance power is type power, and vice versa:
${(\check{\wp}\,{f})}^{\propto} = \hat{\wp}\,{f}$ and
${(\hat{\wp}\,{g})}^{\propto} = \check{\wp}\,{g}$.
Instance and type power functors are related through involution:
${\hat{\wp}}^{\mathrm{op}} \circ \propto = \check{\wp}$ and
${\check{\wp}}^{\mathrm{op}} \circ \propto = \hat{\wp}$
(top right Figure~\ref{classification-component-architecture}).

\subsubsection{Unit and Counit.}

Any classification can be compared with the power of its instance.
Given any classification $\mathbf{A}$,
the eta infomorphism 
$\eta_{\mathbf{A}} 
= \langle \mathrm{id}_{\mathsf{inst}(\mathbf{A})}, \mathsf{ext}_{\mathbf{A}} \rangle 
: \mathbf{A} \rightleftharpoons \check{\wp}\,(\mathsf{inst}(\mathbf{A}))$ 
from $\mathbf{A}$ to the instance power classification of the instance set of $\mathbf{A}$,
is the identity function on instances and the extent function on types. 
For any infomorphism
$\mathbf{f} 
= \langle \mathsf{inst}(\mathbf{f}), \mathsf{typ}(\mathbf{f}) \rangle : \mathbf{A}_1 \rightleftharpoons \mathbf{A}_2$,
the following naturality conditions holds:
$\eta_{\mathbf{A}_1} \circ_{\mathsf{Clsn}} \check{\wp}\,{\mathsf{inst}(\mathbf{f})}
= \mathbf{f} \circ_{\mathsf{Clsn}} \eta_{\mathbf{A}_2}$.
Hence,
there is an eta natural transformation
$\eta : \mathrm{id}_{\mathsf{Clsn}} \Rightarrow \mathsf{inst}^{\mathrm{op}} \circ \check{\wp} : \mathsf{Clsn} \rightarrow \mathsf{Clsn}$
(bottom middle Figure~\ref{classification-component-architecture}),
with
$\eta \circ \mathsf{inst}^{\mathrm{op}} = 1_{\mathsf{inst}^{\mathrm{op}}}$
and
$\eta \circ \mathsf{typ} = \mathsf{ext}$.

Dually,
any classification can be compared with the power of its type.
Given any classification $\mathbf{A}$,
the epsilon infomorphism 
$\varepsilon_{\mathbf{A}} 
= \langle \mathsf{int}_{\mathbf{A}}, \mathrm{id}_{\mathsf{typ}(\mathbf{A})} \rangle 
: \hat{\wp}\,(\mathsf{typ}(\mathbf{A})) \rightleftharpoons \mathbf{A}$ 
to $\mathbf{A}$ from the type power classification of the type set of $\mathbf{A}$,
is the intent function on instances and the identity function on types.
For any infomorphism
$\mathbf{f} 
= \langle \mathsf{inst}(\mathbf{f}), \mathsf{typ}(\mathbf{f}) \rangle : \mathbf{A}_1 \rightleftharpoons \mathbf{A}_2$,
the following naturality conditions holds:
$\varepsilon_{\mathbf{A}_1} \circ_{\mathsf{Clsn}} \mathbf{f}
= \hat{\wp}\,{\mathsf{typ}(\mathbf{f})} \circ_{\mathsf{Clsn}} \varepsilon_{\mathbf{A}_2}$.
Hence,
there is an epsilon natural transformation
$\varepsilon : \mathsf{typ} \circ \hat{\wp} \Rightarrow \mathrm{id}_{\mathsf{Clsn}} : \mathsf{Clsn} \rightarrow \mathsf{Clsn}$
(bottom middle Figure~\ref{classification-component-architecture}),
with
$\varepsilon \circ \mathsf{inst}^{\mathrm{op}} = \mathsf{int}$
and
$\varepsilon \circ \mathsf{typ} = 1_{\mathsf{typ}}$.

\subsubsection{Concept Morphisms}

\begin{figure}
\begin{center}
\setlength{\unitlength}{0.85pt}
\begin{picture}(150,150)(0,0)
\put(-50,112.5){\makebox(100,75){$\mathsf{conc}(\mathbf{A}_1)$}}
\put(100,112.5){\makebox(100,75){$\mathsf{conc}(\mathbf{A}_2)$}}
\put(-50,37.5){\makebox(100,75){${\wp}\,\mathsf{inst}(\mathbf{A}_1) {\times} {\wp}\,\mathsf{typ}(\mathbf{A}_1)$}}
\put(100,37.5){\makebox(100,75){${\wp}\,\mathsf{inst}(\mathbf{A}_2) {\times} {\wp}\,\mathsf{typ}(\mathbf{A}_2)$}}
\put(-50,-37.5){\makebox(100,75){$\mathsf{conc}(\mathbf{A}_1)$}}
\put(100,-37.5){\makebox(100,75){$\mathsf{conc}(\mathbf{A}_2)$}}
\put(25,122.5){\makebox(100,75){\footnotesize{$\mathsf{right}(\mathbf{f})$}}}
\put(25,55){\makebox(100,75) {\footnotesize{${\mathsf{inst}(\mathbf{f})}^{-1}{\times}\left(\exists\mathsf{typ}(\mathbf{f})\circ\mathsf{clo}(\mathbf{A}_2)\right)$}}}
\put(25,20){\makebox(100,75) {\footnotesize{$\left(\exists\,\mathsf{inst}(\mathbf{f})\circ\mathsf{clo}(\mathbf{A}_1)\right){\times}\,{\mathsf{typ}(\mathbf{f})}^{-1}$}}}
\put(25,-47.5){\makebox(100,75){\footnotesize{$\mathsf{left}(\mathbf{f})$}}}
\put(35,150){\vector(1,0){80}}
\put(60,80){\vector(1,0){30}}
\put(90,70){\vector(-1,0){30}}
\put(115,0){\vector(-1,0){80}}
\put(0,132.5){\vector(0,-1){40}}
\put(3,132.5){\oval(6,6)[t]}
\put(150,132.5){\vector(0,-1){40}}
\put(153,132.5){\oval(6,6)[t]}
\put(0,17.5){\vector(0,1){40}}
\put(-3,17.5){\oval(6,6)[b]}
\put(150,17.5){\vector(0,1){40}}
\put(147,17.5){\oval(6,6)[b]}
\end{picture}
\end{center}
\caption{The Left-Right Galois Connection}
\label{left-right}
\end{figure}

\begin{sloppypar}
Let 
$\mathbf{f} 
= \langle \mathsf{inst}(\mathbf{f}), \mathsf{typ}(\mathbf{f}) \rangle : \mathbf{A}_1 \rightleftharpoons \mathbf{A}_2$
be any infomorphism between two classifications 
with instance function 
$\check{f} = \mathsf{inst}(\mathbf{f}) : \mathsf{inst}(\mathbf{A}_2) \rightarrow \mathsf{inst}(\mathbf{A}_1)$
and type function 
$\hat{f} = \mathsf{typ}(\mathbf{f}) : \mathsf{typ}(\mathbf{A}_1) \rightarrow \mathsf{typ}(\mathbf{A}_2)$.
Since for any concept $(X_1, Y_1) \in \mathsf{conc}(\mathbf{A}_1)$
the equality ${\check{f}}^{-1}[X_1] = {\hat{f}[Y_1]}^{\mathbf{A}_2}$ holds between direct and inverse images,
the mapping 
$(X_1, Y_1) 
\mapsto \left( {\check{f}}^{-1}[X_1], \left({\check{f}}^{-1}[X_1]\right)^{\mathbf{A}_2} \right)
= \left( {\check{f}}^{-1}[X_1], {\hat{f}[Y_1]}^{\bullet} \right)
= \left({\mathsf{inst}(\mathbf{f})}^{-1}{\times}\left(\exists\mathsf{typ}(\mathbf{f})\circ\mathsf{clo}(\mathbf{A}_2)\right)\right)(X_1, Y_1)
= \mathsf{right}(\mathbf{f})(X_1, Y_1)$
is a well-defined monotonic function
$\mathsf{right}(\mathbf{f}) : \mathsf{conc}(\mathbf{A}_1) \rightarrow \mathsf{conc}(\mathbf{A}_2)$ (Figure~\ref{left-right})
that preserves types in the sense that:
$\tau_{\mathbf{A}_1} \cdot \mathsf{right}(\mathbf{f}) = \mathsf{typ}(\mathbf{f}) \cdot \tau_{\mathbf{A}_2}$.
Dually,
since for any concept $(X_2, Y_2) \in \mathsf{conc}(\mathbf{A}_2)$
the equality ${\check{f}}[X_2]^{\mathbf{A}_1} = {\hat{f}}^{-1}[Y_2]$ holds,
the mapping
$(X_2, Y_2) 
\mapsto \left( {\hat{f}}^{-1}[Y_2]^{\mathbf{A}}, {\hat{f}}^{-1}[Y_2] \right)
= \left( {\check{f}[X_2]}^{\bullet}, {\hat{f}}^{-1}[Y_2] \right)
= \left(\left(\exists\,\mathsf{inst}(\mathbf{f})\circ\mathsf{clo}(\mathbf{A}_1)\right){\times}\,{\mathsf{typ}(\mathbf{f})}^{-1}\right)(X_2, Y_2)
= \mathsf{left}(\mathbf{f})(X_2, Y_2)$
is a well-defined monotonic function 
$\mathsf{left}(\mathbf{f}) : \mathsf{conc}(\mathbf{A}_2) \rightarrow \mathsf{conc}(\mathbf{A}_1)$ (Figure~\ref{left-right})
that preserves instance concepts in the sense that:
$\iota_{\mathbf{A}_2} \cdot \mathsf{left}(\mathbf{f}) = \mathsf{inst}(\mathbf{f}) \cdot \iota_{\mathbf{A}_1}$.

The left-right pair 
$\mathsf{adj}(\mathbf{f}) 
= \langle \mathsf{left}(\mathbf{f}), \mathsf{right}(\mathbf{f}) \rangle : \mathsf{conc}(\mathbf{A}_2) \rightleftharpoons \mathsf{conc}(\mathbf{A}_1)$
forms a Galois connection (pair of adjoint monotonic functions) between complete lattices.
The quadruple 
$\mathsf{clg}(\mathbf{f}) 
= \langle \mathsf{left}(\mathbf{f}), \mathsf{right}(\mathbf{f}), \mathsf{inst}(\mathbf{f}), \mathsf{typ}(\mathbf{f}) \rangle$ 
is a concept lattice morphism
$\mathsf{clg}(\mathbf{f}) : \mathsf{clg}(\mathbf{A}_1) \rightleftharpoons \mathsf{clg}(\mathbf{A}_2)$, 
called the concept lattice morphism of the infomorphism 
$\mathbf{f} : \mathbf{A}_1 \rightleftharpoons \mathbf{A}_2$.
\end{sloppypar}

\subsubsection{The Concept Lattice Functor}\label{subsubsec:concept:lattice:functor}

The concept lattice functor 
$\mathsf{clg} : \mathsf{Clsn} \rightarrow \mathsf{Clg}$,
from the category of classifications and infomorphisms
to the category of concept lattices and concept morphisms,
is defined below via its object and morphism class functions.

\begin{figure}
\begin{center}
\setlength{\unitlength}{0.8pt}
\begin{tabular}{c}
\begin{picture}(200,100)(-83,0)
\qbezier[100](18,147)(28,37)(-92,32)
\put(-160,60){\makebox(100,50){ {\itshape\large classification} }}
\put(-160,50){\makebox(100,50){ {\itshape\large structures} }}
\put(90,60){\makebox(100,50){ {\itshape\large conceptual} }}
\put(90,50){\makebox(100,50){ {\itshape\large structures} }}
\put(0,0){\begin{picture}(100,100)(0,0)
\put(5,75){\makebox(100,50){$\mathsf{Refl} {\odot}\, \mathsf{coRefl}$}}
\put(-50,25){\makebox(100,50){$\mathsf{Refl}^{\mathbf{2}}$}}
\put(50,25){\makebox(100,50){$\mathsf{coRefl}^{\mathbf{2}}$}}
\put(-100,-25){\makebox(100,50){$\mathsf{Adj}_{=}$}}
\put(0,-25){\makebox(100,50){$\mathsf{Adj}_{=}$}}
\put(100,-25){\makebox(100,50){$\mathsf{Adj}_{=}$}}
\put(-32,56){\makebox(100,50){\footnotesize{$\pi_0$}}}
\put(33,56){\makebox(100,50){\footnotesize{$\pi_1$}}}
\put(-80,8){\makebox(100,50){\footnotesize{$\partial_0$}}}
\put(-23,12){\makebox(100,50){\footnotesize{$\partial_1$}}}
\put(23,12){\makebox(100,50){\footnotesize{$\partial_0$}}}
\put(82,8){\makebox(100,50){\footnotesize{$\partial_1$}}}
\put(-50,-35){\makebox(100,50){\footnotesize{$\mathrm{id}$}}}
\put(50,-35){\makebox(100,50){\footnotesize{$\mathrm{id}$}}}
\put(-48,-4){\makebox(100,50){\footnotesize{$\alpha_{\mathsf{Refl}}$}}}
\put(-48,-14){\makebox(100,50){\Large{$\Rightarrow$}}}
\put(52,-4){\makebox(100,50){\footnotesize{$\alpha_{\mathsf{coRefl}}$}}}
\put(52,-14){\makebox(100,50){\Large{$\Rightarrow$}}}
\put(50,25){\begin{picture}(30,15)(0,-15)
\put(0,0){\line(-1,-1){15}}
\put(0,0){\line(1,-1){15}}
\end{picture}}
\thicklines
\put(40,90){\vector(-1,-1){30}}
\put(60,90){\vector(1,-1){30}}
\put(-10,40){\vector(-1,-1){30}}
\put(10,40){\vector(1,-1){30}}
\put(-30,0){\vector(1,0){60}}
\put(90,40){\vector(-1,-1){30}}
\put(110,40){\vector(1,-1){30}}
\put(70,0){\vector(1,0){60}}
\end{picture}}
\thicklines
\put(-21,108){\vector(1,0){40}}
\put(-49,90){\makebox(100,50){\footnotesize{$\div$}}}
\put(-49,75.5){\makebox(100,50){\footnotesize{$\equiv$}}}
\put(-49,62){\makebox(100,50){\footnotesize{$\circ$}}}
\put(19,94){\vector(-1,0){40}}
\put(-35,90){\line(2,-1){44}}
\put(21,62){\line(2,-1){43}}
\put(79,33){\line(2,-1){15}}
\put(109,18){\vector(2,-1){28}}
\put(-50,85){\vector(0,-1){70}}
\put(-98,75){\makebox(100,50){$\mathsf{Adj}_{=}^{\mathbf{2}}$}}
\put(-109,24){\makebox(100,50){\footnotesize{$\partial_0$}}}
\put(-7,36){\makebox(100,50){\footnotesize{$\partial_1$}}}
\put(-75,45){\makebox(100,50){\footnotesize{$\alpha_{\mathsf{Adj}_{=}}$}}}
\put(-75,35){\makebox(100,50){\Large{$\Rightarrow$}}}

\end{picture}

\\ \\ \\

$\begin{array}{rcl}
\div_{\mathsf{Adj}_{=}} \circ\, \circ_{\mathsf{Adj}_{=}} & = & \mathsf{id}_{\mathsf{Adj}_{=}} 
\\
\circ_{\mathsf{Adj}_{=}} \,\circ \div_{\mathsf{Adj}_{=}} & \cong & \mathsf{id}_{\mathsf{Refl} {\odot_{\mathsf{Adj}_{=}}} \mathsf{coRefl}} 
\\
\div_{\mathsf{Adj}_{=}} 
\left(
\pi_\mathsf{Refl} \, \alpha_{\mathsf{Refl}}
\bullet
\pi_\mathsf{coRefl} \, \alpha_{\mathsf{coRefl}}
\right)
& = &
\alpha_{\mathsf{Adj}_{=}}

\end{array}$

\end{tabular}
\end{center}
\caption{Equivalence between Classification and Conceptual Structures}
\label{basic-equivalence}
\end{figure}
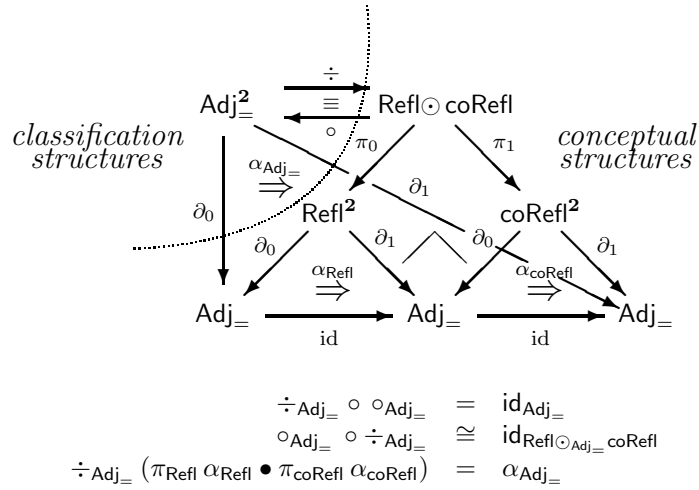

\section{The Category of Concept Lattices}\label{sec:category:concept:lattices}

\subsection{Concept Lattices}\label{subsec:concept:lattice:category:objects}

[Encyclopedia Brittanica] In logic, extension and intension are correlative words that indicate the reference of a concept: 
``intension'' or ``connotation'' indicates the internal content of a concept that constitutes its formal definition; and 
``extension'' or ``denotation'' indicates the range of applicability of a concept by naming the particular objects that it denotes.
Concept structure has both extensional and intensional aspects (points of view). 
Rudolf Wille's notion of a \emph{conceptual knowledge system} \cite{wille:92} is important in our development of the categorical analysis of conceptual structures.
To paraphrase Wille, a specification of conceptual knowledge is founded on three basic notions (\emph{instances} or formal objects, \emph{types} or formal attributes, and formal \emph{concepts}), which are linked by four basic relations:
\begin{center}
\begin{tabular}{r|c@{\hspace{5pt}}c}
            & \emph{type}        & \emph{conc}        \\ \hline
\rule[-1pt]{0pt}{12pt}\emph{inst} & $\mathcal{K}_{00}$ & $\mathcal{K}_{01}$ \\
\emph{conc} & $\mathcal{K}_{10}$ & $\mathcal{K}_{11}$
\end{tabular}
\end{center}
\begin{description}
\item[$\mathcal{K}_{00}$] 
The \emph{has} relation (``an instance has a type'') represents the case where things are classified by the various properties that they possess.\footnote{ Note: the boundaries between instances and concepts and between concepts and attributes is flexible;
in fact, we could think of all concepts as instances (on the rows) and types (on the columns), and then the whole matrix represents a has relation.} 
\item[$\mathcal{K}_{01}$] 
The \emph{instance-of} relation (``an instance belongs to a concept'') represents the case where things are classified by what they are thought to be.
\item[$\mathcal{K}_{10}$] 
The \emph{abstraction} relation (``a type abstracts from, and distinguishes, a concept''). 
The phrase ``property category'' corresponds to the concept generated by an attribute --- all concepts in the abstraction relation are subconcepts of this. 
\item[$\mathcal{K}_{11}$] 
The \emph{subconcept} relation (``a more specific concept is a subconcept of a more generic concept'').
In a conceptual knowledge system certain concepts are distinguished as being of relevance, and this choice corresponds to the kinds of classifications (categories) that tend to cleave the world
in meaningful and useful ways.
\end{description}

\subsubsection{Extension ($\mathcal{K}_{01}$ and $\mathcal{K}_{11}$).}

The extensional aspect of conceptual structure has three isomorphic versions:
a relation version, a function version and a reflection version.

The relation version of the extensional aspect of conceptual structure 
consists of a hierarchy of \emph{concepts} and a set of \emph{instances} linked through a \emph{instance-of} or \emph{membership} relation.
For example,
in a discussion about living things,
we might say that
``All fish are vertebrates'', ``Felix is a cat'', ``Nemo is a fish'' and ``That is a plant'' (pointing at something on the windowsill);
here, 
`Felix', `Nemo' and `That' (an indexical) represent instances (individuals or objects),
and `vertebrate', `cat', `fish' and `plant' represent concepts.
More precisely, 
the extensional aspect of a concept lattice $\mathbf{L}$ consists of 
a preorder 
$\mathsf{ord}(\mathbf{L}) 
= \langle \mathsf{elem}(\mathbf{L}), \leq_{\mathbf{L}} \rangle$
of elements called concepts in a generalization-specialization hierarchy (assumed to be reflexive and transitive) called the concept order,
a set of instances $\mathsf{inst}(\mathbf{L})$,
and a binary instance-of relation 
$\iota_{\mathbf{L}} \subseteq \mathsf{inst}(\mathbf{L}) {\times} \mathsf{elem}(\mathbf{L})$
that registers which instances belong to which concepts:
``$x \,\iota_{\mathbf{L}}\, c$'' 
means that instance $x \in \mathsf{inst}(\mathbf{L})$
belongs to concept $c \in \mathsf{elem}(\mathbf{L})$.
The fact that the concept order is a generalization-specialization hierarchy means that,
if instance $x$ belongs to concept $c_1$ and concept $c_1$ is more specialized that concept $c_2$,
then instance $x$ belongs to concept $c_1$;
that is,
the instance-of relation is closed on the right with respect to the concept order;
in symbols,
$x \,\iota_{\mathbf{L}}\, c_1$ and $c_1 \leq_{\mathbf{L}} c_2$ imply $x \,\iota_{\mathbf{L}}\, c_2$. 
The function version of the extensional aspect of conceptual structure 
consists of a hierarchy of concepts and a set of instances linked through an \emph{instance embedding} function 
$\iota_{\mathbf{L}} : \mathsf{inst}(\mathbf{L}) \rightarrow \mathsf{elem}(\mathbf{L})$.
The instance embedding function can be expressed in terms of the of-type relation as the meet $\iota_{\mathbf{L}}(x) = \bigwedge x\iota_{\mathbf{L}}$.
The of-type relation can be expressed in terms of the instance embedding function,
since $x \,\iota_{\mathbf{L}}\, c$ holds iff $\iota_{\mathbf{L}}(x) \leq_{\mathbf{L}} c$
for every instance $x \in \mathsf{inst}(\mathbf{L})$ and every element $c \in \mathsf{elem}(\mathbf{L})$.

The reflection version of the extensional aspect of conceptual structure consists of a pair of opposed functions (in opposite directions).
The \emph{extent} function
$\mathsf{ext}_{\mathbf{L}} : \mathsf{elem}(\mathbf{L}) \rightarrow {\wp}\,\mathsf{inst}(\mathbf{L})$
lists the instances of each concept $c \in \mathsf{elem}(\mathbf{L})$:
$\mathsf{ext}_{\mathbf{L}}(c) 
= \{ x \in \mathsf{inst}(\mathbf{L}) \mid x \,\iota_{\mathbf{L}}\, c \}$.
Clearly, given the extent function we can define the instance-of relation as
$x \,\iota_{\mathbf{L}}\, c$ when $x \in \mathsf{ext}_{\mathbf{L}}(c)$.
The instance concept generator (\emph{iota}) function
$\mathsf{iota}_{\mathbf{L}} 
: {\wp}\,\mathsf{inst}(\mathbf{L}) \rightarrow \mathsf{elem}(\mathbf{L})$
computes the most specific concept that contains all instances of a subset $X \subseteq \mathsf{inst} (\mathbf{L})$:
$\mathsf{iota}_{\mathbf{L}}(X) 
= \bigwedge\, \{ c \in \mathsf{elem}(\mathbf{L}) \mid \forall_{x \in X}\, (x \,\iota_{\mathbf{L}}\, c) \}
= \bigwedge\, \{ c \in \mathsf{elem}(\mathbf{L}) \mid X \subseteq \mathsf{ext}_{\mathbf{L}}(c) \}$.
Here we have assumed meets exist for all subsets of concepts.
The instance embedding function is the restriction of the iota function to single instances: $\iota_{\mathbf{L}}(x) = \mathsf{iota}_{\mathbf{L}}(\{x\})$
for any instance $x \in \mathsf{inst}(\mathbf{L})$.
The iota function can be expressed in terms of the instance embedding function by
$\mathsf{iota}_{\mathbf{L}}(X) 
= \bigvee_{\mathbf{L}} \{ \iota_{\mathbf{L}}(x) \mid x \in X \}
= \bigvee_{\mathbf{L}}\exists\iota_{\mathbf{L}}(X)$
for all instance subsets $X \subseteq \mathsf{inst}(\mathbf{L})$.
Clearly, given the iota function we can define the instance-of relation as
$x \,\iota_{\mathbf{L}}\, c$ when $\mathsf{iota}_{\mathbf{L}}(\{x\}) \leq_{\mathbf{L}} c$,
and we can define the extent function as
$\mathsf{ext}_{\mathbf{L}}(c) 
= \bigcup\, \{ X \subseteq \mathsf{inst}(\mathbf{L}) \mid \mathsf{iota}_{\mathbf{L}}(X) \leq_{\mathbf{L}} c \}$
for each concept $c \in \mathsf{elem}(\mathbf{L})$.
Hence, 
the instance-of relation, 
the extent function and 
the iota function 
are mutually equivalent.
\begin{fact}
The extensional aspect of conceptual structure is an \emph{extent} reflection 
(reflective Galois connection) from the instance power to the concept order
\[\mathsf{extent}_{\mathbf{L}} 
= \langle \mathsf{iota}_{\mathbf{L}}, \mathsf{ext}_{\mathbf{L}} \rangle
: {\wp}\,\mathsf{inst}(\mathbf{L}) \rightleftharpoons \mathsf{ord}(\mathbf{L}).\]
\end{fact}
This reflection defines the abstract extensional aspect of a concept lattice.
It is clear from the reflection part of Theorem~\ref{induce:lattice}
that we only need to assume the set of concepts forms a preorder;
antisymmetry, and existence of meets and joins follows from reflection properties and the fact that instance power is a complete lattice.

\subsubsection{Intension ($\mathcal{K}_{10}$ and $\mathcal{K}_{11}$).}

The intensional aspect of conceptual structure has three isomorphic versions:
a relation version, a function version and a reflection version.

The relation version of the intensional aspect of conceptual structure 
consists of a hierarchy of \emph{concepts} and a set of \emph{types} linked through a \emph{has} or \emph{of-type} relation.
For example,
in a discussion about living things,
we might say that
''All cats are mammals'', ``All fish swim'' and ``No plants are motile'';
here, 
`cat', `mammal', `fish' and `plant' represent concepts, 
and `swim' and `motile' represent types (properties or attributes).
More precisely, 
the intensional aspect of a concept lattice $\mathbf{L}$ consists of 
a preorder $\mathsf{ord}(\mathbf{L}) = \langle \mathsf{elem}(\mathbf{L}), \leq_{\mathbf{L}} \rangle$
of elements called concepts in a generalization-specialization hierarchy (assumed to be reflexive and transitive) called the concept order,
a set of types $\mathsf{typ}(\mathbf{L})$,
and a binary abstraction or of-type relation 
$\tau_{\mathbf{L}} \subseteq \mathsf{elem}(\mathbf{L}) {\times} \mathsf{typ}(\mathbf{L})$
that describes the concepts by recording the types of each:
``$c \,\tau_{\mathbf{L}}\, y$'' 
means that concept $c \in \mathsf{elem}(\mathbf{L})$ has type $y \in \mathsf{typ}(\mathbf{L})$.
The fact that the concept order is a generalization-specialization hierarchy means that,
if concept $c_1$ is more specialized that concept $c_2$ and concept $c_2$ has type $y$,
then concept $c_1$ also has type $y$;
that is,
the of-type relation is closed on the left with respect to the concept order;
in symbols,
$c_1 \leq_{\mathbf{L}} c_2$ and $c_2 \,\tau_{\mathbf{L}}\, y$ imply $c_1 \,\tau_{\mathbf{L}}\, y$. 
The function version of the intensional aspect of conceptual structure 
consists of a hierarchy of concepts and a set of types linked through 
a \emph{type embedding} function 
$\tau_{\mathbf{L}} : \mathsf{typ}(\mathbf{L}) \rightarrow \mathsf{elem}(\mathbf{L})$.
The type embedding function can be expressed in terms of the of-type relation as the join $\tau_{\mathbf{L}}(y) = \bigvee \tau_{\mathbf{L}}y$.
The of-type relation can be expressed in terms of the type embedding function,
since $c \,\tau_{\mathbf{L}}\, y$ holds iff $c \leq_{\mathbf{L}} \tau_{\mathbf{L}}(y)$
for every element $c \in \mathsf{elem}(\mathbf{L})$ and every type $y \in \mathsf{typ}(\mathbf{L})$.

We define two opposed functions (in opposite directions).
The \emph{intent} function
$\mathsf{int}_{\mathbf{L}} : \mathsf{elem}(\mathbf{L}) \rightarrow {\wp}\,\mathsf{typ}(\mathbf{L})$
collects the types possessed by each concept $c \in \mathsf{elem}(\mathbf{L})$:
$\mathsf{int}_{\mathbf{L}}(c) 
= \{ y \in \mathsf{typ}(\mathbf{L}) \mid c \,\tau_{\mathbf{L}}\, y \}$.
Clearly, given the intent function we can define the of-type relation as
$c \,\tau_{\mathbf{L}}\, y$ when $y \in \mathsf{int}_{\mathbf{L}}(c)$.
The type concept generator (\emph{tau}) function
$\mathsf{tau}_{\mathbf{L}} 
: {\wp}\,\mathsf{typ}(\mathbf{L}) \rightarrow \mathsf{elem}(\mathbf{L})$
computes the most generic concept that has all types of a subset $Y \subseteq \mathsf{typ} (\mathbf{L})$:
$\mathsf{tau}_{\mathbf{L}}(Y) 
= \bigvee\, \{ c \in \mathsf{elem}(\mathbf{L}) \mid \forall_{y \in Y}\, (c \,\tau_{\mathbf{L}}\, y) \}
= \bigvee\, \{ c \in \mathsf{elem}(\mathbf{L}) \mid \mathsf{int}_{\mathbf{L}}(c) \supseteq Y \}$.
Here we have assumed joins exist for all subsets of concepts.
The type embedding function is the restriction of the tau function to single types:
$\tau_{\mathbf{L}}(y) = \mathsf{tau}_{\mathbf{L}}(\{y\})$
for any type $y \in \mathsf{typ}(\mathbf{L})$.
The tau function can be expressed in terms of the type embedding function by
$\mathsf{tau}_{\mathbf{L}}(Y) 
= \bigwedge_{\mathbf{L}} \{ \tau_{\mathbf{L}}(y) \mid y \in Y \}
= \bigwedge_{\mathbf{L}}\exists\tau_{\mathbf{L}}(Y)$
for all type subsets $Y \subseteq \mathsf{typ}(\mathbf{L})$.
Clearly, given the tau function we can define the of-type relation as
$c \,\tau_{\mathbf{L}}\, y$ when $c \leq_{\mathbf{L}} \mathsf{tau}_{\mathbf{L}}(\{y\})$,
and we can define the intent function as
$\mathsf{int}_{\mathbf{L}}(c) 
= \bigcup\, \{ Y \subseteq \mathsf{typ}(\mathbf{L}) \mid c \leq_{\mathbf{L}} \mathsf{tau}_{\mathbf{L}}(Y) \}$
for each concept $c \in \mathsf{elem}(\mathbf{L})$.
Hence, 
the of-type relation, 
the intent function and 
the tau function 
are mutually equivalent.
\begin{fact}
The intensional aspect of conceptual structure is an \emph{intent} coreflection 
(coreflective Galois connection) from the concept order
to the type power opposite
\[\mathsf{intent}_{\mathbf{L}} 
= \langle \mathsf{int}_{\mathbf{L}}, \mathsf{tau}_{\mathbf{L}} \rangle
: \mathsf{ord}(\mathbf{L}) \rightleftharpoons {\wp}\,{\mathsf{typ}(\mathbf{L})}^{\mathrm{op}}.\]
\end{fact}
This coreflection gives the abstract intensional aspect of a concept lattice.
It is clear from the coreflection part of Theorem~\ref{induce:lattice} 
that we only need to assume the set of concepts forms a preorder;
antisymmetry, and existence of meets and joins follows from coreflection properties and the fact that type power is a complete lattice.

\begin{figure}
\begin{center}
\begin{tabular}{c}
\begin{tabular}{c@{\hspace{2pt}}|c@{\hspace{10pt}}|c|}
\multicolumn{1}{c}{\rule[-5mm]{0mm}{11mm}}
&\multicolumn{1}{c}{\begin{tabular}{c}\sffamily{Classification}\\\sffamily{Structures}\end{tabular}$[\mathbf{A}]$}
&\multicolumn{1}{c}{\begin{tabular}{c}\sffamily{Conceptual}\\\sffamily{Structures}\end{tabular}$[\mathbf{L}]$}
\\ \cline{2-3}
\rule[-3mm]{0mm}{8mm}
\makebox[0pt][r]{\begin{tabular}{c}\footnotesize\emph{Relations}\end{tabular}}
&
\footnotesize{$\mathsf{inst}(\mathbf{A})
\stackrel{\models_{\mathbf{A}}}{\rightharpoondown} 
\mathsf{typ}(\mathbf{A})$}
&
\footnotesize{$\mathsf{inst}(\mathbf{L})
\stackrel{\iota_{\mathbf{L}}}{\rightharpoondown} 
\mathsf{ord}(\mathbf{L})
\stackrel{\tau_{\mathbf{L}}}{\rightharpoondown} 
\mathsf{typ}(\mathbf{L})$}
\\ \cline{2-3}
\rule[-5mm]{0mm}{12mm}
\makebox[0pt][r]{\begin{tabular}{c}\footnotesize\emph{Functions}\end{tabular}}
&
\begin{tabular}{r@{\hspace{2pt}}c@{\hspace{2pt}}l}
\footnotesize{${\wp}\,\mathsf{inst}(\mathbf{A})$}
&$\stackrel{\;\mathsf{ext}_{\mathbf{A}}}{\leftarrow}$ 
&\footnotesize{$\mathsf{typ}(\mathbf{A})$}
\\
\footnotesize{$\mathsf{inst}(\mathbf{A})$}
&$\stackrel{\;\mathsf{int}_{\mathbf{A}}}{\rightarrow}$ 
&\footnotesize{${\wp}\,\mathsf{typ}(\mathbf{A})$}
\end{tabular}
&
\footnotesize{$\mathsf{inst}(\mathbf{L})
\stackrel{\iota_{\mathbf{L}}}{\rightarrow} 
\mathsf{ord}(\mathbf{L})
\stackrel{\tau_{\mathbf{L}}}{\leftarrow} 
\mathsf{typ}(\mathbf{L})$}
\\ \cline{2-3}
\rule[-5mm]{0mm}{12mm}
\makebox[0pt][r]{\begin{tabular}{c}\footnotesize\emph{Galois}\\\footnotesize\emph{Connections}\end{tabular}}
&
\footnotesize{${\wp}\,\mathsf{inst}(\mathbf{A}) 
\stackrel{\mathsf{deriv}_{\mathbf{A}}}{\rightleftharpoons} 
{\wp}\,\mathsf{typ}(\mathbf{A})^{\mathrm{op}}$}
&
\footnotesize{${\wp}\,\mathsf{inst}(\mathbf{L}) 
\stackrel{\mathsf{extent}_{\mathbf{L}}}{\rightleftharpoons} 
\mathsf{ord}(\mathbf{L})
\stackrel{\mathsf{intent}_{\mathbf{L}}}{\rightleftharpoons} 
{\wp}\,\mathsf{typ}(\mathbf{L})^{\mathrm{op}}$} \\ \cline{2-3}
\multicolumn{1}{c}{}&\multicolumn{1}{c}{}&\multicolumn{1}{c}{} \\
\multicolumn{1}{c}{}&
\multicolumn{1}{c}{\footnotesize{$\begin{array}[t]{r@{\hspace{5pt}}c@{\hspace{5pt}}l}
\models_{\mathbf{A}} 
& = & 
\iota_{\mathbf{L}} \circ \tau_{\mathbf{L}}
\end{array}$}}
&
\multicolumn{1}{c}{} \\
\multicolumn{1}{c}{}&\multicolumn{1}{c}{}&\multicolumn{1}{c}{} \\
\multicolumn{1}{c}{}&
\multicolumn{1}{c}{\footnotesize{$\begin{array}[t]{r@{\hspace{5pt}}c@{\hspace{5pt}}l}
\mathsf{ext}_{\mathbf{A}} 
& = & \tau_{\mathbf{L}} \cdot \mathsf{ext}_{\mathbf{L}}
\\
\mathsf{int}_{\mathbf{A}} 
& = & \iota_{\mathbf{L}} \cdot \mathsf{int}_{\mathbf{L}}
\end{array}$}}
&
\multicolumn{1}{c}{} \\
\multicolumn{1}{c}{}&\multicolumn{1}{c}{}&\multicolumn{1}{c}{} \\
\multicolumn{1}{c}{}&
\multicolumn{1}{c}{\footnotesize{$\begin{array}[t]{r@{\hspace{5pt}}c@{\hspace{5pt}}l}
\mathsf{deriv}_{\mathbf{A}}
& = & \langle {(\check{\mbox{-}})}^{\mathbf{A}}, {(\hat{\mbox{-}})}^{\mathbf{A}} \rangle
\\
{(\check{\mbox{-}})}^{\mathbf{A}} 
& : & {\wp}\,\mathsf{inst}(\mathbf{A}) \rightarrow {\wp}\,\mathsf{typ}(\mathbf{A})^{\mathrm{op}}
\\
{(\hat{\mbox{-}})}^{\mathbf{A}} 
& : & {\wp}\,\mathsf{typ}(\mathbf{A})^{\mathrm{op}} \rightarrow {\wp}\,\mathsf{inst}(\mathbf{A})
\\
\mathsf{deriv}_{\mathbf{A}}
& = & \mathsf{extent}_{\mathbf{L}} \circ \mathsf{intent}_{\mathbf{L}}
\\
{(\check{\mbox{-}})}^{\mathbf{A}}
& = & \mathsf{iota}_{\mathbf{L}} \cdot \mathsf{int}_{\mathbf{L}}
\\
{(\hat{\mbox{-}})}^{\mathbf{A}}
& = & \mathsf{tau}_{\mathbf{L}} \cdot \mathsf{ext}_{\mathbf{L}}
\end{array}$}}
&
\multicolumn{1}{c}{\footnotesize{$\begin{array}[t]{r@{\hspace{5pt}}c@{\hspace{5pt}}l}
\mathsf{extent}_{\mathbf{L}}
& = & 
\langle \mathsf{iota}_{\mathbf{L}}, \mathsf{ext}_{\mathbf{L}} \rangle
\\
\mathsf{iota}_{\mathbf{L}} 
& : & {\wp}\,\mathsf{inst}(\mathbf{L}) \rightarrow \mathsf{ord}(\mathbf{L})
\\
\mathsf{ext}_{\mathbf{L}} 
& : & \mathsf{ord}(\mathbf{L}) \rightarrow {\wp}\,\mathsf{inst}(\mathbf{L})
\\
\mathsf{intent}_{\mathbf{L}} 
& = & 
\langle \mathsf{int}_{\mathbf{L}}, \mathsf{tau}_{\mathbf{L}} \rangle
\\
\mathsf{int}_{\mathbf{L}} 
& : & \mathsf{ord}(\mathbf{L}) \rightarrow {\wp}\,\mathsf{typ}(\mathbf{L})^{\mathrm{op}}
\\
\mathsf{tau}_{\mathbf{L}} 
& : & {\wp}\,\mathsf{typ}(\mathbf{L})^{\mathrm{op}} \rightarrow \mathsf{ord}(\mathbf{L})
\end{array}$}}
\end{tabular}
\end{tabular}
\end{center}
\caption{Versions of Equivalent Structures}
\label{versions-equivalent-structures}
\end{figure}

\subsubsection{Abstraction.}

Hence,
in the complete correlated sense of conceptual structure,
a concept lattice $\mathbf{L}$ has two aspects:
an extensional reflection $\mathsf{extent}_{\mathbf{L}}$
and 
an intensional coreflection
$\mathsf{intent}_{\mathbf{L}}$
that are composable (as Galois connections).
We further assume that
the horizontal source preorder of the extent
and the horizontal target preorder of the intent
are free.
More formally,
a concept lattice 
$\mathbf{L} = 
\langle \mathsf{inst}(\mathbf{L}), \mathsf{typ}(\mathbf{L}), \mathsf{extent}_{\mathbf{L}}, \mathsf{intent}_{\mathbf{L}} \rangle$
consists of 
an instance set $\mathsf{inst}(\mathbf{L})$,
a type set $\mathsf{typ}(\mathbf{L})$,
a reflection $\mathsf{extent}_{\mathbf{L}}$ 
and a coreflection $\mathsf{intent}_{\mathbf{L}}$,
where
$\partial_0^{\mathrm{h}}(\mathsf{extent}_{\mathbf{L}})
= {\wp}\,\mathsf{inst}(\mathbf{L})$, 
$\partial_1^{\mathrm{h}}(\mathsf{extent}_{\mathbf{L}})
= \partial_0^{\mathrm{h}}(\mathsf{intent}_{\mathbf{L}})$
and
$\partial_1^{\mathrm{h}}(\mathsf{intent}_{\mathbf{L}})
= {{\wp}\,\mathsf{typ}(\mathbf{L})}^{\mathrm{op}}$.
If we make the definition
$\mathsf{ord}(\mathbf{L}) = \ord{L}
\doteq \partial_1^{\mathrm{h}}(\mathsf{extent}_{\mathbf{L}})
= \partial_0^{\mathrm{h}}(\mathsf{intent}_{\mathbf{L}})$,
we visualized a concept lattice $\mathbf{L}$ as
\[
{\wp}\,\mathsf{inst}(\mathbf{L}) 
\stackrel{\mathsf{extent}_{\mathbf{L}}}{\rightleftharpoons} 
\mathsf{ord}(\mathbf{L})
\stackrel{\mathsf{intent}_{\mathbf{L}}}{\rightleftharpoons}
{\wp}\,{\mathsf{typ}(\mathbf{L})}^{\mathrm{op}}
.\]

\begin{itemize}
\item
The extent reflection unpacks into the adjoint functions
\begin{center}
$\begin{array}{rcl}
\mathsf{ext}_{\mathbf{L}} 
= \mathsf{right}(\mathsf{extent}_{\mathbf{L}})
: \mathsf{elem}(\mathbf{L}) \rightarrow {\wp}\,\mathsf{inst}(\mathbf{L})
\\
\mathsf{iota}_{\mathbf{L}} 
= \mathsf{left}(\mathsf{extent}_{\mathbf{L}})
: {\wp}\,\mathsf{inst}(\mathbf{L}) \rightarrow \mathsf{elem}(\mathbf{L})
\end{array}$
\end{center}
called the extent function and the instance concept generator (iota) function,
respectively.
These satisfy the adjointness condition
$\mathsf{iota}_{\mathbf{L}}(X) \leq_{\mathbf{L}} c$
\underline{iff}
$X \subseteq \mathsf{ext}_{\mathbf{L}}(c)$
for any instance subset $X \subseteq \mathsf{inst} (\mathbf{L})$
and any concept $c \in \mathsf{elem}(\mathbf{L})$.
Using this equivalence for the special case of singleton instance sets,
we can define the binary instance-of relation 
\[\iota_{\mathbf{L}} \subseteq \mathsf{inst}(\mathbf{L}) {\times} \mathsf{elem}(\mathbf{L})\]
by
$x\iota_{\mathbf{L}}c$ 
when 
$\mathsf{iota}_{\mathbf{L}}(\{x\}) \leq_{\mathbf{L}} c$
iff
$x \in \mathsf{ext}_{\mathbf{L}}(c)$
for any instance $x \in \mathsf{inst}(\mathbf{L})$
and any concept $c \in \mathsf{elem}(\mathbf{L})$.
And, 
we can define the instance embedding function 
\[\iota_{\mathbf{L}} : \mathsf{inst}(\mathbf{L}) \rightarrow \mathsf{elem}(\mathbf{L})\]
by
$\iota_{\mathbf{L}}(x)
\doteq
\mathsf{iota}_{\mathbf{L}}(\{x\})$
for any instance $x \in \mathsf{inst}(\mathbf{L})$.

\item The intent coreflection unpacks with the function definitions
\begin{center}
$\begin{array}{rcl}
\mathsf{int}_{\mathbf{L}} 
= \mathsf{left}(\mathsf{intent}_{\mathbf{L}})
: \mathsf{elem}(\mathbf{L}) \rightarrow {\wp}\,\mathsf{typ}(\mathbf{L})
\\
\mathsf{tau}_{\mathbf{L}} 
= \mathsf{right}(\mathsf{intent}_{\mathbf{L}})
: {\wp}\,\mathsf{typ}(\mathbf{L}) \rightarrow \mathsf{elem}(\mathbf{L})
\end{array}$
\end{center}
called the intent function and the type concept generator (tau) function,
respectively.
These satisfy the conditions,
\begin{center}
$\mathsf{int}_{\mathbf{L}}(c) \supseteq Y$
\underline{iff}
$c \leq_{\mathbf{L}} \mathsf{tau}_{\mathbf{L}}(Y)$
\end{center}
for any concept $c \in \mathsf{elem}(\mathbf{L})$
and any type subset $Y \subseteq \mathsf{typ} (\mathbf{L})$.

\end{itemize}
Since $\mathsf{ext}_{\mathbf{L}}$ and $\mathsf{int}_{\mathbf{L}}$ are injective,
the elements (formal concepts) in $\mathsf{elem}(\mathbf{L})$ are determined by both their extent and their intent; 
that is, 
two distinct elements cannot have the same extent,
and
two distinct elements cannot have the same intent.

Applying Theorem~\ref{induce:lattice} to the intent coreflection,
the meets and joins in the complete lattice $\mathsf{lat}(\mathbf{L})$ are determined by the following expressions
\begin{center}
\begin{tabular}{cc}
$\begin{array}{rcl}
\mathsf{ext}_{\mathbf{L}}( \bigwedge_{\mathbf{L}}C ) 
& = &
\bigcap_{c{\in}C} \, \mathsf{ext}_{\mathbf{L}}(c)
\\
\mathsf{int}_{\mathbf{L}}( \bigwedge_{\mathbf{L}}C ) 
& = &
\left( \bigcup_{c{\in}C} \, \mathsf{int}_{\mathbf{L}}(c) \right)^{\bullet}
\end{array}$
&
$\begin{array}{rcl}
\mathsf{ext}_{\mathbf{L}}( \bigvee_{\mathbf{L}}C ) 
& = & 
\left( \bigcup_{c{\in}C} \, \mathsf{ext}_{\mathbf{L}}(c) \right)^{\bullet}
\\
\mathsf{int}_{\mathbf{L}}( \bigvee_{\mathbf{L}}C ) 
& = & 
\bigcap_{c{\in}C} \, \mathsf{int}_{\mathbf{L}}(c) 
\end{array}$
\end{tabular}
\end{center}
for any collection of concepts $C \subseteq \mathsf{elem}(\mathbf{L})$.
This implies that meet and join are definable in terms of the extent, intent, closure and interior operators
(plus the union and intersection of subsets of instances and types). 
This allows a more abstract definition of the notion of a concept lattice 
--- a concept lattice 
$\mathbf{L}
= \langle \mathsf{elem}(\mathbf{L}), \mathsf{inst}(\mathbf{L}), \mathsf{typ}(\mathbf{L}),
\mathsf{extent}_{\mathbf{L}}, \mathsf{intent}_{\mathbf{L}} \rangle$
consists of 
a preorder $\mathsf{elem}(\mathbf{L})$,
an instance set $\mathsf{inst}(\mathbf{L})$,
a type set $\mathsf{typ}(\mathbf{L})$,
an extent reflection
$\mathsf{extent}_{\mathbf{L}}$,
and an intent coreflection
$\mathsf{intent}_{\mathbf{L}}$
\[ 
{\wp}\,\mathsf{inst}(\mathbf{L}) 
\stackrel{\mathsf{extent}_{\mathbf{L}}}{\rightleftharpoons} 
\mathsf{elem}(\mathbf{L})
\stackrel{\mathsf{intent}_{\mathbf{L}}}{\rightleftharpoons} 
{\wp}\,\mathsf{typ}(\mathbf{L})^{\mathrm{op}}.
\]

\subsubsection{Old Definition.}

A concept lattice $\mathbf{L} = \clg{L}$,
consists of 
a complete lattice $\mathsf{lat}(\mathbf{L}) = \cl{L}$,
an instance set $\mathsf{inst}(\mathbf{L})$,
a type set $\mathsf{typ}(\mathbf{L})$,
a set of lattice elements (formal concepts) $\mathsf{elem}(\mathbf{L})$,
a lattice order ${\leq}_{\mathbf{L}} \subseteq \mathsf{elem}(\mathbf{L}) {\times} \mathsf{elem}(\mathbf{L})$,
meet and join operations $\bigwedge_{\mathbf{L}}, \bigvee_{\mathbf{L}} : {\wp}\,\mathsf{elem}(\mathbf{L}) \rightarrow \mathsf{elem}(\mathbf{L})$, 
an instance embedding function $\iota_{\mathbf{L}} : \mathsf{inst}(\mathbf{L}) \rightarrow L$ and
a type embedding function $\tau_{\mathbf{L}} : \mathsf{typ}(\mathbf{L}) \rightarrow L$, 
such that $\iota_{\mathbf{L}}[\mathsf{inst}(\mathbf{L})]$ is join-dense (up to equivalence) in $\mathsf{lat}(\mathbf{L})$ 
and $\tau_{\mathbf{L}}[\mathsf{typ}(\mathbf{L})]$ is meet-dense (up to equivalence) in $\mathsf{lat}(\mathbf{L})$.

We can define an instance embedding relation 
$\iota_{\mathbf{L}} \subseteq \mathsf{inst}(\mathbf{L}) {\times} \mathsf{elem}(\mathbf{L})$ as follows: 
for every instance $x \in \mathsf{inst}(\mathbf{L})$ and every element $c \in \mathsf{elem}(\mathbf{L})$ 
the relationship $x\iota_{\mathbf{L}}c$ holds when $\iota_{\mathbf{L}}(x) \leq_{\mathbf{L}} c$.
This relation is closed on the right with respect to lattice order:
$\iota_{\mathbf{L}} \circ {\leq}_{\mathbf{L}} = \iota_{\mathbf{L}}$. 
The given function can be expressed in terms of the defined relation as the meet 
$\iota_{\mathbf{L}}(x) = \bigwedge_{\mathbf{L}} x\iota_{\mathbf{L}}$.
We can also define a type embedding relation
$\tau_{\mathbf{L}} \subseteq \mathsf{elem}(\mathbf{L}) {\times} \mathsf{typ}(\mathbf{L})$ as follows:
for every element $c \in \mathsf{elem}(\mathbf{L})$ and every type $y \in \mathsf{typ}(\mathbf{L})$ 
the relationship $c\tau_{\mathbf{L}}y$ holds when $c \leq_{\mathbf{L}} \tau_{\mathbf{L}}(x)$.
This relation is closed on the left respect to lattice order:
${\leq}_{\mathbf{L}} \circ \tau_{\mathbf{L}} = \tau_{\mathbf{L}}$.
The given function can be expressed in terms of the defined relation as the join 
$\tau_{\mathbf{L}}(y) = \bigvee_{\mathbf{L}} \tau_{\mathbf{L}}y$.

\subsubsection{Examples.}

Systemic examples of concept lattices abound. 
Given any set $X$, 
the instance power concept lattice 
$\check{\wp}\,{X} 
= \langle 
\langle {\wp}{X}, \subseteq, \bigcap, \bigcup \rangle, 
X, {\wp}{X}, \{{-}\}_X, \mathrm{id}_{{\wp}{X}} \rangle$ 
associated with $X$,
has elements of $X$ as instances, subsets of $X$ as concepts and types,
with order being subset inclusion,
meet being intersection and join being union,
with membership serving as the instance embedding relation and identity serving as the type embedding function.
Join-density is proven by the set-theoretic identity
$A = \bigcup\, \{ \{x\} \mid x \in A \}$
for any instance subset $A \subseteq X$. 
Dually,
given any set $Y$,
the type power concept lattice 
$\hat{\wp}\,{Y} 
= \langle 
\langle {\wp}{Y}, \supseteq, \bigcup, \bigcap \rangle, 
{\wp}{Y}, Y, \mathrm{id}_{{\wp}{Y}}, \{{-}\}_Y \rangle$ 
associated with $Y$, 
has subsets of $Y$ as instances and concepts, 
elements of $Y$ as types, 
with order being reverse subset inclusion,
meet being union and join being intersection,
with identity serving as the instance embedding function and membership transpose serving as the type embedding relation.
Meet-density is proven by the set-theoretic identity
$B = \bigcap\, \{ B \}$
for any type subset $B \subseteq Y$. 

Given any concept lattice $\mathbf{L}$,
the dual (or transpose) concept lattice (or involution)
${\mathbf{L}}^{\!\propto} = \langle {\mathsf{lat}(\mathbf{L})}^{\!\propto}, \mathsf{typ}(\mathbf{L}), \mathsf{inst}(\mathbf{L}), \tau_{\mathbf{L}}, \iota_{\mathbf{L}} \rangle$
with
${\mathsf{lat}(\mathbf{L})}^{\!\propto}
=
{\langle \mathsf{elem}(\mathbf{L}), \geq_{\mathbf{L}}, \bigvee_{\mathbf{L}}, \bigwedge_{\mathbf{L}} \rangle}$
has 
elements of $\mathbf{L}$ as elements, types of $\mathbf{L}$ as instances, instances of $\mathbf{L}$ as types, 
with order being the transpose $\mathsf{lat}(\mathbf{L})$-order, meets being $\mathsf{lat}(\mathbf{L})$-joins and joins being $\mathsf{lat}(\mathbf{L})$-meets.
The transpose operator is idempotent:
${\mathbf{L}}^{\propto\propto} = \mathbf{L}$.
The transpose of instance power is type power, and vice versa:
${(\check{\wp}\,{X})}^{\!\propto} = \hat{\wp}\,{X}$
and
${(\hat{\wp}\,{Y})}^{\!\propto} = \check{\wp}\,{Y}$.

Types $y, y^{\prime} \in \mathsf{typ}(\mathbf{L})$ are coextensive in $\mathbf{L}$ 
when $\tau_{\mathbf{L}}(y) = \tau_{\mathbf{L}}(y^{\prime})$.
Instances $x, x^{\prime} \in \mathsf{inst}(\mathbf{L})$ are indistinguishable in $\mathbf{L}$ 
when $\iota_{\mathbf{L}}(x) = \iota_{\mathbf{L}}(x^{\prime})$.
A concept lattice $\mathbf{L}$ is extensional when there are no distinct coextensive types;
that is, when the type embedding function $\tau_{\mathbf{L}}$ is injective.
A concept lattice $\mathbf{L}$ is separated when there are no distinct indistinguishable instances;
that is, when the intent function $\iota_{\mathbf{L}}(x)$ is injective.
Instance and power concept lattices are both extensional and separated.

\subsubsection{Classifications ($\mathcal{K}_{00}$).}
\label{subsubsec:classification:functor:objects}

Any concept lattice $\mathbf{L} = \clg{L}$ has an associated classification $\mathsf{clsn}(\mathbf{L}) = \clsn{L}$,
which has $\mbit{L}$-instances as its instance set,
$\mbit{L}$-types as its type set,
and the relational composition
$\models_{\mathbf{L}} 
= \iota_{\mathbf{L}} \circ \leq_{\mathbf{L}} \circ \tau_{\mathbf{L}}
= \iota_{\mathbf{L}} \circ \tau_{\mathbf{L}}$
as its classification relation.
In more detail,
for any instance $x \in \mathsf{inst}(\mathbf{L})$ and any type $y \in \mathsf{typ}(\mathbf{L})$
$x \models_{\mathbf{L}} y$ iff $\iota_{\mathbf{L}}(x) \leq_{\mathbf{L}} \tau_{\mathbf{L}}(y)$.

Any concept lattice defines a classification by composition,
either relational composition or composition of Galois connections.
Given any concept lattice $\mathbf{L}$ there is a classification 
$\mathsf{clsn}(\mathbf{L}) = \clsn{L}$.
An instance $x \in \mathsf{inst}(\mathbf{L})$ has a type $y \in \mathsf{typ}(\mathbf{L})$ 
when there is a concept $c \in \mathsf{elem}(\mathbf{L})$ that connects the two;
that is,
when $x$ is an instance of $c$ and $c$ is of type $y$;
or symbolically,
$x \models_{\mathbf{L}} y$ 
when 
$\exists_{c \in \mathsf{elem}(\mathbf{L})}\,(x\,\iota_{\mathbf{L}}\,c) \& (c\,\tau_{\mathbf{L}}\,y)$
\underline{iff}
$\iota_{\mathbf{L}}(x) \leq_{\mathbf{L}} \tau_{\mathbf{L}}(y)$
\underline{iff}
$\iota_{\mathbf{L}}(x) \,\tau_{\mathbf{L}}\, y$
\underline{iff}
$x \,\iota_{\mathbf{L}}\, \tau_{\mathbf{L}}(y)$;
or abstractly,
$\models_{\mathbf{L}} = \iota_{\mathbf{L}} \,\circ\, \tau_{\mathbf{L}}$.
The discussion above develop a concept lattice as a pair of compsable Galois connections,
an extent reflection $\mathsf{extent}_{\mathbf{L}}$ and an intent coreflection $\mathsf{intent}_{\mathbf{L}}$.
What is the composition of these two Galois connections?
The left adjoint 
$\mathsf{iota}_{\mathbf{L}} \cdot \mathsf{int}_{\mathbf{L}}
: {\wp}\,\mathsf{inst}(\mathbf{L}) \rightarrow \mathsf{elem}(\mathbf{L}) \rightarrow {\wp}\,\mathsf{typ}(\mathbf{L})$
is the composition of the instance concept generator (iota) function and the intent function.
For any subset $X \subseteq \mathsf{inst} (\mathbf{L})$,
$\mathsf{int}_{\mathbf{L}}(\mathsf{iota}_{\mathbf{L}}(X)) 
= \{ y \in \mathsf{typ}(\mathbf{L}) \mid \mathsf{iota}_{\mathbf{L}}(X) \leq_{\mathbf{L}} \tau_{\mathbf{L}}(y) \}
= \{ y \in \mathsf{typ}(\mathbf{L}) \mid \bigwedge\, \{ c \in \mathsf{elem}(\mathbf{L}) \mid \forall_{x \in X}\, (x \,\iota_{\mathbf{L}}\, c) \} \leq_{\mathbf{L}} \tau_{\mathbf{L}}(y) \}
= \{ y \in \mathsf{typ}(\mathbf{L}) \mid \forall_{x \in X}\, (x \,\iota_{\mathbf{L}}\, \tau_{\mathbf{L}}(y)) \}
= \{ y \in \mathsf{typ}(\mathbf{L}) \mid \forall_{x \in X}\, (\iota_{\mathbf{L}}(x) \leq_{\mathbf{L}} \tau_{\mathbf{L}}(y)) \}
= \{ y \in \mathsf{typ}(\mathbf{L}) \mid \forall_{x \in X}\, (x \models_{\mathbf{L}} y) \}
= X^{\mathsf{clsn}(\mathbf{L})}$
the derivation of $X$ in $\mathsf{clsn}(\mathbf{L})$ the classification of $\mathbf{L}$;
that is,
the left adjoint of $\mathsf{iota}_{\mathbf{L}} \cdot \mathsf{int}_{\mathbf{L}}$ is the left derivation function of $\mathsf{clsn}(\mathbf{L})$.
The right adjoint
$\mathsf{tau}_{\mathbf{L}} \cdot \mathsf{ext}_{\mathbf{L}}
: {\wp}\,\mathsf{typ}(\mathbf{L}) \rightarrow \mathsf{elem}(\mathbf{L}) \rightarrow {\wp}\,\mathsf{inst}(\mathbf{L})$
is the composition of the type concept generator (tau) function and the extent function.
For any subset $Y \subseteq \mathsf{typ} (\mathbf{L})$,
$\mathsf{ext}_{\mathbf{L}}(\mathsf{tau}_{\mathbf{L}}(Y)) 
= \{ x \in \mathsf{inst}(\mathbf{L}) \mid \iota_{\mathbf{L}}(x) \leq_{\mathbf{L}} \mathsf{tau}_{\mathbf{L}}(Y) \}
= \{ x \in \mathsf{inst}(\mathbf{L}) \mid \iota_{\mathbf{L}}(x) \leq_{\mathbf{L}} \bigvee\, \{ c \in \mathsf{elem}(\mathbf{L}) \mid \forall_{y \in Y}\, (c \,\tau_{\mathbf{L}}\, y) \}  \}
= \{ x \in \mathsf{inst}(\mathbf{L}) \mid \forall_{y \in Y}\, (\iota_{\mathbf{L}}(x) \,\tau_{\mathbf{L}}\, y) \}
= \{ x \in \mathsf{inst}(\mathbf{L}) \mid \forall_{y \in Y}\, (\iota_{\mathbf{L}}(x) \leq_{\mathbf{L}} \tau_{\mathbf{L}}(y)) \}
= \{ x \in \mathsf{inst}(\mathbf{L}) \mid \forall_{y \in Y}\, (x \models_{\mathbf{L}} y) \}
= Y^{\mathsf{clsn}(\mathbf{L})}$
the derivation of $Y$ in $\mathsf{clsn}(\mathbf{L})$ the classification of $\mathbf{L}$;
that is,
the right adjoint of $\mathsf{tau}_{\mathbf{L}} \cdot \mathsf{ext}_{\mathbf{L}}$ is the right derivation function of $\mathsf{clsn}(\mathbf{L})$.
Hence,
the derivation Galois connection is the composition
\[\mathsf{deriv}_{\mathsf{clsn}(\mathbf{L})}
= \mathsf{extent}_{\mathbf{L}} \circ \mathsf{intent}_{\mathbf{L}}.\]

\subsection{Concept Morphisms}\label{subsec:concept:lattice:category:morphisms}

\subsubsection{Extension.}

A morphism between the extensional aspect of conceptual structures 
consists of 
a Galois connection between the hierarchies of concepts 
and 
a function between the sets of instances,
which together respect the instance-of relations.
More precisely, 
the extensional aspect of a concept lattice morphism 
$\mathbf{h} : \mathbf{L}_1 \rightleftharpoons \mathbf{L}_2$ 
consists of 
a Galois connection 
$\mathsf{adj}(\mathbf{h}) = \langle \mathsf{left}(\mathbf{h}), \mathsf{right}(\mathbf{h}) \rangle
: \mathsf{ord}(\mathbf{L}_2) \rightleftharpoons \mathsf{ord}(\mathbf{L}_1)$
(in the reverse direction) between the generalization-specialization hierarchies (concept orders) called the concept connection,
and a function
$\mathsf{inst}(\mathbf{h}) : \mathsf{inst}(\mathbf{L}_2) \rightarrow \mathsf{inst}(\mathbf{L}_1)$ 
(in the reverse direction) between instance sets 
called the instance function,
which respect the instance-of relations in the sense that,
the instance function image of a target instance $x_2$ belongs to a source concept $c_1$ 
\underline{iff}
the target instance belongs to the right adjoint image of the source concept;
in symbols,
$\mathsf{inst}(\mathbf{h})(x_2) \,\iota_{\mathbf{L}_1}\, c_1$
\underline{iff}
$x_2 \,\iota_{\mathbf{L}_2}\, \mathsf{right}(\mathbf{h})(c_1)$
for any target instance $x_2 \in \mathsf{inst}(\mathbf{L}_2)$
and any source concept $c_1 \in  \mathsf{elem}(\mathbf{L}_1)$.
Call this the extensional condition.
This closely resembles the fundamental conditions for Galois connections and infomorphisms.

The sequence of equivalences,
$x_2 \in \mathsf{ext}_{\mathbf{L}_2}(\mathsf{right}(\mathbf{h})(c_1))$
\underline{iff}
$\iota_{\mathbf{L}_2}(x_2) \leq_{\mathbf{L}_2} \mathsf{right}(\mathbf{h})(c_1)$
\underline{iff}
$x_2 \,\iota_{\mathbf{L}_2}\, \mathsf{right}(\mathbf{h})(c_1)$
\underline{iff}
$\mathsf{inst}(\mathbf{h})(x_2) \,\iota_{\mathbf{L}_1}\, c_1$
\underline{iff}
$\iota_{\mathbf{L}_1}(\mathsf{inst}(\mathbf{h})(x_2)) \leq_{\mathbf{L}_1} c_1$
\underline{iff}
$\mathsf{inst}(\mathbf{h})(x_2) \in \mathsf{ext}_{\mathbf{L}_1}(c_1)$
\underline{iff}
$x_2 \in {\mathsf{inst}(\mathbf{h})}^{-1}(\mathsf{ext}_{\mathbf{L}_1}(c_1))$
for any target instance $x_2 \in  \mathsf{inst}(\mathbf{L}_2)$
and any source concept $c_1 \in \mathsf{elem}(\mathbf{L}_1)$,
demonstrates the extent function identity
$\mathsf{right}(\mathbf{h}) \cdot \mathsf{ext}_{\mathbf{L}_2}
= \mathsf{ext}_{\mathbf{L}_1} \cdot {\mathsf{inst}(\mathbf{h})}^{-1}$;
in turn,
this implies the extensional condition.
The sequence of equivalences,
$\mathsf{left}(\mathbf{h})(\iota_{\mathbf{L}_2}(x_2)) \leq_{\mathbf{L}_1} c_1$
\underline{iff}
$\iota_{\mathbf{L}_2}(x_2) \leq_{\mathbf{L}_2} \mathsf{right}(\mathbf{h})(c_1)$ 
\underline{iff}
$x_2 \,\iota_{\mathbf{L}_2}\, \mathsf{right}(\mathbf{h})(c_1)$
\underline{iff}
$\mathsf{inst}(\mathbf{h})(x_2) \,\iota_{\mathbf{L}_2}\, c_1$
\underline{iff}
$\iota_{\mathbf{L}_1}(\mathsf{inst}(\mathbf{h})(x_2)) \leq_{\mathbf{L}_1} c_1$
for any target instance $x_2 \in \mathsf{inst}(\mathbf{L}_2)$
and any source concept $c_1 \in \mathsf{elem}(\mathbf{L}_1)$,
demonstates the instance embedding function identity
$\iota_{\mathbf{L}_2} \cdot \mathsf{left}(\mathbf{h})
= \mathsf{inst}(\mathbf{h}) \cdot \iota_{\mathbf{L}_1}$;
in turn,
this implies the extensional condition.
The sequence of identities,
$\mathsf{left}(\mathbf{h})(\mathsf{iota}_{\mathbf{L}_1}(X_2))
= \mathsf{left}(\mathbf{h})\left( \bigvee\, \{ \iota_{\mathbf{L}_2}(x_2)) \mid x_2 \in X_2 \} \right)
= \bigvee\, \{ \mathsf{left}(\mathbf{h})(\iota_{\mathbf{L}_2}(x_2)) \mid x_2 \in X_2 \}
= \bigvee\, \{ \iota_{\mathbf{L}_1}(\mathsf{inst}(\mathbf{h})(x_2)) \mid x_2 \in X_2 \}
= \mathsf{iota}_{\mathbf{L}_1}(\{ \mathsf{inst}(\mathbf{h})(x_2) \mid x_2 \in X_2 \})
= \mathsf{iota}_{\mathbf{L}_1}( \exists\,\mathsf{inst}(\mathbf{h})(X_2))$
demonstates the iota function identity
$\mathsf{iota}_{\mathbf{L}_2} \cdot \mathsf{left}(\mathbf{h})
= \exists\,\mathsf{inst}(\mathbf{h}) \cdot \mathsf{iota}_{\mathbf{L}_1}$;
in turn,
this implies the extensional condition.
The iota and extent function identities are equivalent to the Galois connection identity
$\mathsf{extent}_{\mathbf{L}_2} \cdot \mathsf{adj}(\mathbf{h})
= \mathsf{dir}(\mathsf{inst}(\mathbf{h})) \cdot \mathsf{extent}_{\mathbf{L}_1}$;
in turn,
this implies the extensional condition.

\begin{figure}
\begin{center}
\begin{tabular}{c@{\hspace{25pt}}c@{\hspace{30pt}}c}
\setlength{\unitlength}{0.85pt}
\begin{picture}(80,70)(0,-10)
\put(14,45){\makebox(60,30){$\mathsf{Clg}_{\iota}^{\mathrm{op}}$}}
\put(-30,-15){\makebox(60,30){$\mathsf{Set}$}}
\put(50,-15){\makebox(60,30){$\mathsf{Adj}$}}
\put(-20,20){\makebox(60,30){\footnotesize{$\mathsf{inst}$}}}
\put(40,20){\makebox(60,30){\footnotesize{$\mathsf{adj}$}}}
\put(10,-22){\makebox(60,30){\footnotesize{$\mathsf{dir}$}}}
\put(10,10){\makebox(60,30){\footnotesize{$\mathsf{extent}$}}}
\put(10,0){\makebox(60,30){\Large{$\Rightarrow$}}}
\put(32,48){\vector(-2,-3){24}}
\put(48,48){\vector(2,-3){24}}
\put(20,0){\vector(1,0){40}}
\end{picture}
& 
\begin{picture}(0,0)(0,-30)
\put(0,0){\makebox(0,0){\large{$=$}}}
\end{picture}
&
\begin{picture}(120,70)(0,-30)
\put(33,25){\makebox(60,30){$\mathsf{Clg}_{\iota}^{\mathrm{op}}$}}
\put(-30,-15){\makebox(60,30){$\mathsf{Set}$}}
\put(30,-15){\makebox(60,30){$\mathsf{Refl}$}}
\put(90,-15){\makebox(60,30){$\mathsf{Adj}$}}
\put(0,-45){\makebox(60,30){$\mathsf{Adj}$}}
\put(60,-45){\makebox(60,30){$\mathsf{Adj}$}}
\put(-6,10){\makebox(60,30){\footnotesize{$\mathsf{inst}$}}}
\put(21,6){\makebox(60,30){\footnotesize{$\mathsf{incl}$}}}
\put(66,10){\makebox(60,30){\footnotesize{$\mathsf{adj}$}}}
\put(-20,-35){\makebox(60,30){\footnotesize{$\mathsf{dir}$}}}
\put(80,-35){\makebox(60,30){\footnotesize{$\mathrm{id}$}}}
\put(13,-23){\makebox(60,30){\footnotesize{$\partial_0^{\mathrm{h}}$}}}
\put(47,-23){\makebox(60,30){\footnotesize{$\partial_1^{\mathrm{h}}$}}}
\put(30,-30){\makebox(60,30){\footnotesize{$\mathsf{refl}$}}}
\put(31,-38){\makebox(60,30){\Large{$\Rightarrow$}}}
\put(48,32){\vector(-3,-2){36}}
\put(72,32){\vector(3,-2){36}}
\put(60,30){\vector(0,-1){20}}
\put(7,-7){\vector(1,-1){16}}
\put(53,-7){\vector(-1,-1){16}}
\put(67,-7){\vector(1,-1){16}}
\put(113,-7){\vector(-1,-1){16}}
\put(45,-30){\vector(1,0){30}}
\put(30,-50){\makebox(60,30){\footnotesize{$\mathrm{id}$}}}
\put(20,-15){\begin{picture}(20,10)(0,0)
\put(10,10){\line(-1,-1){10}}
\put(10,10){\line(1,-1){10}}
\end{picture}}
\put(80,-15){\begin{picture}(20,10)(0,0)
\put(10,10){\line(-1,-1){10}}
\put(10,10){\line(1,-1){10}}
\end{picture}}
\end{picture}
\\ \\ \\
\setlength{\unitlength}{0.85pt}
\begin{picture}(80,70)(0,-10)
\put(14,45){\makebox(60,30){$\mathsf{Clg}_{\tau}^{\mathrm{op}}$}}
\put(-30,-15){\makebox(60,30){$\mathsf{Adj}$}}
\put(53,-15){\makebox(60,30){$\mathsf{Set}^{\mathrm{op}}$}}
\put(-20,20){\makebox(60,30){\footnotesize{$\mathsf{adj}$}}}
\put(43,20){\makebox(60,30){\footnotesize{$\mathsf{typ}^{\mathrm{op}}$}}}
\put(10,-22){\makebox(60,30){\footnotesize{$\mathsf{inv}$}}}
\put(10,10){\makebox(60,30){\footnotesize{$\mathsf{intent}$}}}
\put(10,0){\makebox(60,30){\Large{$\Rightarrow$}}}
\put(32,48){\vector(-2,-3){24}}
\put(48,48){\vector(2,-3){24}}
\put(60,0){\vector(-1,0){40}}
\end{picture}
& 
\begin{picture}(0,0)(0,-30)
\put(0,0){\makebox(0,0){\large{$=$}}}
\end{picture}
&
\begin{picture}(120,70)(0,-30)
\put(33,25){\makebox(60,30){$\mathsf{Clg}_{\tau}^{\mathrm{op}}$}}
\put(-30,-15){\makebox(60,30){$\mathsf{Adj}$}}
\put(30,-15){\makebox(60,30){$\mathsf{coRefl}$}}
\put(93,-15){\makebox(60,30){$\mathsf{Set}^{\mathrm{op}}$}}
\put(0,-45){\makebox(60,30){$\mathsf{Adj}$}}
\put(60,-45){\makebox(60,30){$\mathsf{Adj}$}}
\put(-6,10){\makebox(60,30){\footnotesize{$\mathsf{adj}$}}}
\put(21,6){\makebox(60,30){\footnotesize{$\mathsf{incl}$}}}
\put(69,10){\makebox(60,30){\footnotesize{$\mathsf{typ}^{\mathrm{op}}$}}}
\put(-20,-35){\makebox(60,30){\footnotesize{$\mathrm{id}$}}}
\put(80,-35){\makebox(60,30){\footnotesize{$\mathsf{inv}$}}}
\put(13,-23){\makebox(60,30){\footnotesize{$\partial_0^{\mathrm{h}}$}}}
\put(47,-23){\makebox(60,30){\footnotesize{$\partial_1^{\mathrm{h}}$}}}
\put(30,-30){\makebox(60,30){\footnotesize{$\mathsf{corefl}$}}}
\put(31,-38){\makebox(60,30){\Large{$\Rightarrow$}}}
\put(48,32){\vector(-3,-2){36}}
\put(72,32){\vector(3,-2){36}}
\put(60,30){\vector(0,-1){20}}
\put(7,-7){\vector(1,-1){16}}
\put(53,-7){\vector(-1,-1){16}}
\put(67,-7){\vector(1,-1){16}}
\put(113,-7){\vector(-1,-1){16}}
\put(45,-30){\vector(1,0){30}}
\put(30,-50){\makebox(60,30){\footnotesize{$\mathrm{id}$}}}
\put(20,-15){\begin{picture}(20,10)(0,0)
\put(10,10){\line(-1,-1){10}}
\put(10,10){\line(1,-1){10}}
\end{picture}}
\put(80,-15){\begin{picture}(20,10)(0,0)
\put(10,10){\line(-1,-1){10}}
\put(10,10){\line(1,-1){10}}
\end{picture}}
\end{picture}
\\ \\ \\
\multicolumn{3}{c}{\setlength{\unitlength}{1.1pt}
\begin{picture}(120,60)(0,0)
\put(30,45){\makebox(60,30){$\mathsf{Clg}^{\mathrm{op}}$}}
\put(72,45){\makebox(60,30){$= \mathsf{Clg}_{\iota}^{\mathrm{op}} \circ \mathsf{Clg}_{\tau}^{\mathrm{op}}$}}
\put(0,15){\makebox(60,30){$\mathsf{Clg}_{\iota}^{\mathrm{op}}$}}
\put(60,15){\makebox(60,30){$\mathsf{Clg}_{\tau}^{\mathrm{op}}$}}
\put(-30,-15){\makebox(60,30){$\mathsf{Set}$}}
\put(30,-15){\makebox(60,30){$\mathsf{Adj}$}}
\put(90,-15){\makebox(60,30){$\mathsf{Set}^{\mathrm{op}}$}}
\put(9,35){\makebox(60,30){\footnotesize{$\pi_0^{\mathrm{h}}$}}}
\put(51,35){\makebox(60,30){\footnotesize{$\pi_1^{\mathrm{h}}$}}}
\put(-22,5){\makebox(60,30){\footnotesize{$\mathsf{inst}$}}}
\put(17,7){\makebox(60,30){\footnotesize{$\mathsf{adj}$}}}
\put(43,7){\makebox(60,30){\footnotesize{$\mathsf{adj}$}}}
\put(84,5){\makebox(60,30){\footnotesize{$\mathsf{typ}^{\mathrm{op}}$}}}
\put(0,-21){\makebox(60,30){\footnotesize{$\mathsf{dir}$}}}
\put(60,-21){\makebox(60,30){\footnotesize{$\mathsf{inv}$}}}
\put(0,-1){\makebox(60,30){\footnotesize{$\mathsf{extent}$}}}
\put(0,-9){\makebox(60,30){\Large{$\Rightarrow$}}}
\put(60,-1){\makebox(60,30){\footnotesize{$\mathsf{intent}$}}}
\put(60,-9){\makebox(60,30){\Large{$\Rightarrow$}}}
\put(23,23){\vector(-1,-1){16}}
\put(37,23){\vector(1,-1){16}}
\put(83,23){\vector(-1,-1){16}}
\put(97,23){\vector(1,-1){16}}
\put(53,53){\vector(-1,-1){16}}
\put(67,53){\vector(1,-1){16}}
\put(15,0){\vector(1,0){30}}
\put(105,0){\vector(-1,0){30}}
\put(50,15){\begin{picture}(20,10)(0,0)
\put(10,10){\line(-1,-1){10}}
\put(10,10){\line(1,-1){10}}
\end{picture}}
\end{picture}}
\end{tabular}
\end{center}
\caption{Extension, Intension and Combined Conceptual Structure}
\label{conceptual-structure}
\end{figure}
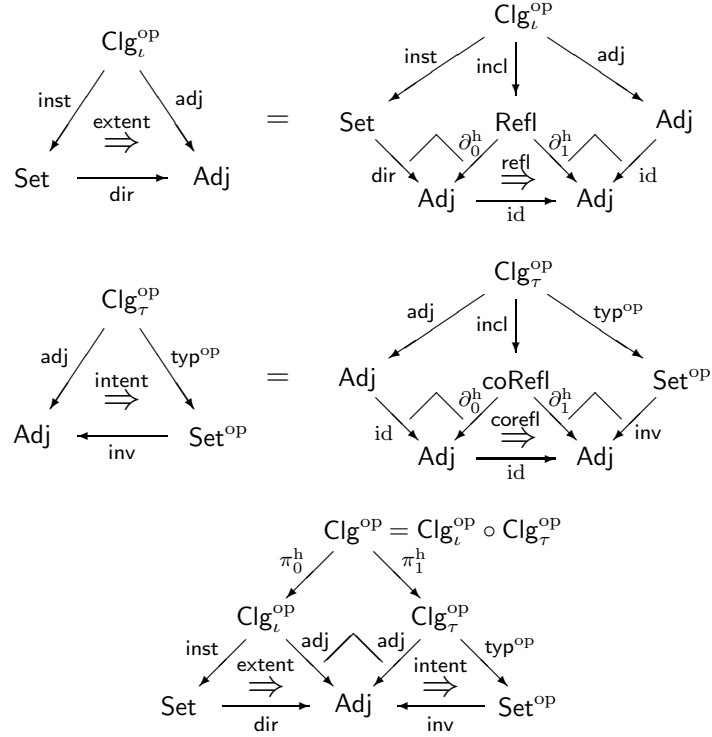

\begin{fact}
The extensional aspect of morphisms of conceptual structure is an opposite morphism of conceptual structure reflections $\mathbf{h} : \mathbf{L}_1 \rightleftharpoons \mathbf{L}_2$ 
\begin{center}
\begin{tabular}{c@{\hspace{110pt}}c}
\\ \\
\begin{picture}(80,50)(0,0)
\put(-25,37.5){\makebox(50,25){${\wp}\,\mathsf{inst}(\mathbf{L}_1)$}}
\put(-25,-12.5){\makebox(50,25){${\wp}\,\mathsf{typ}(\mathbf{L}_2)$}}
\put(60,37.5){\makebox(50,25){$\mathsf{ord}(\mathbf{L}_1)$}}
\put(60,-12.5){\makebox(50,25){$\mathsf{ord}(\mathbf{L}_2)$}}
\put(-50,12.5){\makebox(50,25){\footnotesize{$\mathsf{dir}(\mathsf{inst}(\mathbf{h}))$}}}
\put(73,12.5){\makebox(50,25){\footnotesize{$\mathsf{adj}(\mathbf{h})$}}}
\put(22,45.5){\makebox(50,25){\footnotesize{$\mathsf{extent}_{\mathbf{L}_1}$}}}
\put(22,-21.5){\makebox(50,25){\footnotesize{$\mathsf{extent}_{\mathbf{L}_2}$}}}
\put(0,17){\begin{picture}(0,16)(0,0)
\thinlines
\put(-1,0){\line(0,1){16}}
\put(-3,16){\oval(4,6)[br]}
\put(1,16){\line(0,-1){16}}
\put(3,0){\oval(4,6)[tl]}
\end{picture}}
\put(80,17){\begin{picture}(0,16)(0,0)
\thinlines
\put(-1,0){\line(0,1){16}}
\put(-3,16){\oval(4,6)[br]}
\put(1,16){\line(0,-1){16}}
\put(3,0){\oval(4,6)[tl]}
\end{picture}}
\put(34.5,50){\begin{picture}(16,0)(0,0)
\thinlines
\put(0,1){\line(1,0){16}}
\put(16,3){\oval(6,4)[bl]}
\put(16,-1){\line(-1,0){16}}
\put(0,-3){\oval(6,4)[tr]}
\end{picture}}
\put(34.5,0){\begin{picture}(16,0)(0,0)
\thinlines
\put(0,1){\line(1,0){16}}
\put(16,3){\oval(6,4)[bl]}
\put(16,-1){\line(-1,0){16}}
\put(0,-3){\oval(6,4)[tr]}
\end{picture}}
\end{picture}
&
\begin{picture}(100,50)(0,0)
\put(-25,37.5)
{\makebox(50,25){$\langle {\wp}\,\mathsf{inst}(\mathbf{L}_1), \subseteq \rangle$}}
\put(-25,-12.5)
{\makebox(50,25){$\langle {\wp}\,\mathsf{inst}(\mathbf{L}_2), \subseteq \rangle$}} 
\put(75,37.5)
{\makebox(50,25){$\langle \mathsf{elem}(\mathbf{L}_1), \leq_{\mathbf{L}_1} \rangle$}}
\put(75,-12.5)
{\makebox(50,25){$\langle \mathsf{elem}(\mathbf{L}_2), \leq_{\mathbf{L}_2} \rangle$}}
\put(-47,12.5){\makebox(50,25){\footnotesize{$\exists\mathsf{inst}(\mathbf{h})$}}}
\put(1,12.5){\makebox(50,25){\footnotesize{${\mathsf{inst}(\mathbf{h})}^{-1}$}}}
\put(56,12.5){\makebox(50,25){\footnotesize{$\mathsf{left}(\mathbf{h})$}}}
\put(99,11){\makebox(50,25){\footnotesize{$\mathsf{right}(\mathbf{h})$}}}
\put(27,48.5){\makebox(50,25){\footnotesize{$\mathsf{iota}_{\mathbf{L}_1}$}}}
\put(29,27.5){\makebox(50,25){\footnotesize{$\mathsf{ext}_{\mathbf{L}_1}$}}}
\put(27,-1.5){\makebox(50,25){\footnotesize{$\mathsf{iota}_{\mathbf{L}_2}$}}}
\put(29,-22.5){\makebox(50,25){\footnotesize{$\mathsf{ext}_{\mathbf{L}_2}$}}}
\put(37,54){\vector(1,0){26}}
\put(63,46){\vector(-1,0){26}}
\put(37,4){\vector(1,0){26}}
\put(63,-4){\vector(-1,0){26}}
\put(-5,12.5){\vector(0,1){25}}
\put(5,37.5){\vector(0,-1){25}}
\put(95,12.5){\vector(0,1){25}}
\put(105,37.5){\vector(0,-1){25}}
\end{picture}
\\ \\ \\
$(\underline{\mathrm{iconic}})$ & $(\underline{\mathrm{detailed}})$
\\
\end{tabular}
\end{center}
\end{fact}
The extensional aspect is abstracted as 
a category $\mathsf{Clg}_\iota$ of extensional conceptual structures,
with a contravariant instance component functor
$\mathsf{inst} : \mathsf{Clg}_\iota^{\mathrm{op}} \rightarrow \mathsf{Set}$,
a contravariant adjoint component functor
$\mathsf{adj} : \mathsf{Clg}_\iota^{\mathrm{op}} \rightarrow \mathsf{Adj}$,
and a natural transformation
$\mathsf{extent} 
: \mathsf{inst} \circ \mathsf{dir} \Rightarrow \mathsf{adj}
: \mathsf{Clg}_\iota^{\mathrm{op}} \rightarrow \mathsf{Adj}$.
The category $\mathsf{Clg}_\iota^{\mathrm{op}}$ is a subcategory of 
$\mathsf{Refl}$ the category of reflections (Figure~\ref{conceptual-structure}),
with inclusion functor
$\mathsf{incl} : \mathsf{Clg}_\iota^{\mathrm{op}} \rightarrow \mathsf{Refl}$.
The components are related as
$\mathsf{inst} \circ \mathsf{dir} = \mathsf{incl} \circ \partial_0^{\mathrm{h}}$,
$\mathsf{adj} = \mathsf{incl} \circ \partial_1^{\mathrm{h}}$
and $\mathsf{extent} = \mathsf{incl} \circ \mathsf{refl}$.

\subsubsection{Intension.}

A morphism between the intensional aspect of conceptual structures 
consists of 
a Galois connection between the hierarchies of concepts 
and 
a function between the sets of types,
which together respect the of-type relations.
More precisely, 
the intensional aspect of a concept lattice morphism 
$\mathbf{h} : \mathbf{L}_1 \rightleftharpoons \mathbf{L}_2$ 
consists of 
a Galois connection 
$\mathsf{adj}(\mathbf{h}) = \langle \mathsf{left}(\mathbf{h}), \mathsf{right}(\mathbf{h}) \rangle
: \mathsf{ord}(\mathbf{L}_2) \rightleftharpoons \mathsf{ord}(\mathbf{L}_1)$
(in the reverse direction) between the generalization-specialization hierarchies (concept orders) called the concept connection,
and a function
$\mathsf{typ}(\mathbf{h}) : \mathsf{typ}(\mathbf{L}_1) \rightarrow \mathsf{typ}(\mathbf{L}_2)$ 
(in the forward direction) between type sets 
called the type function,
which respect the of-type relations in the sense that,
the left adjoint image of a target concept $c_2$ has a source type $y_1$ 
\underline{iff}
the target concept has the type function image of the source type;
in symbols,
$\mathsf{left}(\mathbf{h})(c_2) \,\tau_{\mathbf{L}_1}\, y_1$
\underline{iff}
$c_2 \,\tau_{\mathbf{L}_2}\, \mathsf{typ}(\mathbf{h})(y_1)$
for any target concept $c_2 \in \mathsf{elem}(\mathbf{L}_2)$
and any source type $y_1 \in  \mathsf{typ}(\mathbf{L}_1)$.
Call this the intensional condition.
This closely resembles the fundamental conditions for Galois connections and infomorphisms.

The sequence of equivalences,
$y_1 \in \mathsf{int}_{\mathbf{L}_1}(\mathsf{left}(\mathbf{h})(c_2))$
\underline{iff}
$\mathsf{left}(\mathbf{h})(c_2) \leq_{\mathbf{L}_1} \tau_{\mathbf{L}_1}(y_1)$
\underline{iff}
$\mathsf{left}(\mathbf{h})(c_2) \,\tau_{\mathbf{L}_1}\, y_1$
\underline{iff}
$c_2 \,\tau_{\mathbf{L}_2}\, \mathsf{typ}(\mathbf{h})(y_1)$
\underline{iff}
$c_2 \leq_{\mathbf{L}_2} \tau_{\mathbf{L}_2}(\mathsf{typ}(\mathbf{h})(y_1))$
\underline{iff}
$\mathsf{typ}(\mathbf{h})(y_1) \in \mathsf{int}_{\mathbf{L}_2}(c_2)$
\underline{iff}
$y_1 \in {\mathsf{typ}(\mathbf{h})}^{-1}(\mathsf{int}_{\mathbf{L}_2}(c_2))$
for any target concept $c_2 \in \mathsf{elem}(\mathbf{L}_2)$
and any source type $y_1 \in  \mathsf{typ}(\mathbf{L}_1)$,
demonstrates the intent function identity
$\mathsf{left}(\mathbf{h}) \cdot \mathsf{int}_{\mathbf{L}_1}
= \mathsf{int}_{\mathbf{L}_2} \cdot {\mathsf{typ}(\mathbf{h})}^{-1}$;
in turn,
this implies the intensional condition.
The sequence of equivalences,
$c_2 \leq_{\mathbf{L}_2} \mathsf{right}(\mathbf{h})(\tau_{\mathbf{L}_1}(y_1))$
\underline{iff}
$\mathsf{left}(\mathbf{h})(c_2) \leq_{\mathbf{L}_1} \tau_{\mathbf{L}_1}(y_1)$ 
\underline{iff}
$\mathsf{left}(\mathbf{h})(c_2) \,\tau_{\mathbf{L}_1}\, y_1$
\underline{iff}
$c_2 \,\tau_{\mathbf{L}_2}\, \mathsf{typ}(\mathbf{h})(y_1)$
\underline{iff}
$c_2 \leq_{\mathbf{L}_2} \tau_{\mathbf{L}_2}(\mathsf{typ}(\mathbf{h})(y_1))$
for any target concept $c_2 \in \mathsf{elem}(\mathbf{L}_2)$
and any source type $y_1 \in  \mathsf{typ}(\mathbf{L}_1)$,
demonstates the type embedding function identity
$\tau_{\mathbf{L}_1} \cdot \mathsf{right}(\mathbf{h})
= \mathsf{typ}(\mathbf{h}) \cdot \tau_{\mathbf{L}_2}$;
in turn,
this implies the intensional condition.
The sequence of identities,
$\mathsf{right}(\mathbf{h})(\mathsf{tau}_{\mathbf{L}_1}(Y_1))
= \mathsf{right}(\mathbf{h})\left( \bigwedge\, \{ \tau_{\mathbf{L}_1}(y_1)) \mid y_1 \in Y_1 \} \right)
= \bigwedge\, \{ \mathsf{right}(\mathbf{h})(\tau_{\mathbf{L}_1}(y_1)) \mid y_1 \in Y_1 \}
= \bigwedge\, \{ \tau_{\mathbf{L}_2}(\mathsf{typ}(\mathbf{h})(y_1)) \mid y_1 \in Y_1 \}
= \mathsf{tau}_{\mathbf{L}_2}(\{ \mathsf{typ}(\mathbf{h})(y_1) \mid y_1 \in Y_1 \})
= \mathsf{tau}_{\mathbf{L}_2}( \exists\,\mathsf{typ}(\mathbf{h})(Y_1))$
demonstates the tau function identity
$\mathsf{tau}_{\mathbf{L}_1} \cdot \mathsf{right}(\mathbf{h})
= \exists\,\mathsf{typ}(\mathbf{h}) \cdot \mathsf{tau}_{\mathbf{L}_2}$;
in turn,
this implies the intensional condition.
The intent and tau function identities are equivalent to the Galois connection identity
$\mathsf{adj}(\mathbf{h}) \cdot \mathsf{intent}_{\mathbf{L}_1}
= \mathsf{intent}_{\mathbf{L}_2} \cdot \mathsf{inv}(\mathsf{typ}(\mathbf{h}))$;
in turn,
this implies the intensional condition.

\begin{fact}
The intensional aspect of morphisms of conceptual structure is an opposite morphism of conceptual structure coreflections $\mathbf{h} : \mathbf{L}_1 \rightleftharpoons \mathbf{L}_2$ 
\begin{center}
\begin{tabular}{c@{\hspace{110pt}}c}
\\ \\
\begin{picture}(80,50)(0,0)
\put(-25,37.5){\makebox(50,25){$\mathsf{ord}(\mathbf{L}_1)$}}
\put(-25,-12.5){\makebox(50,25){$\mathsf{ord}(\mathbf{L}_2)$}}
\put(60,37.5){\makebox(50,25){${{\wp}\,\mathsf{typ}(\mathbf{L}_1)}^{\mathrm{op}}$}}
\put(60,-12.5){\makebox(50,25){${{\wp}\,\mathsf{typ}(\mathbf{L}_2)}^{\mathrm{op}}$}}
\put(-40,12.5){\makebox(50,25){\footnotesize{$\mathsf{adj}(\mathbf{h})$}}}
\put(81,12.5){\makebox(50,25){\footnotesize{$\mathsf{inv}(\mathsf{typ}(\mathbf{h}))$}}}
\put(15,45.5){\makebox(50,25){\footnotesize{$\mathsf{intent}_{\mathbf{L}_1}$}}}
\put(15,-21.5){\makebox(50,25){\footnotesize{$\mathsf{intent}_{\mathbf{L}_2}$}}}

\put(0,17){\begin{picture}(0,16)(0,0)
\thinlines
\put(-1,0){\line(0,1){16}}
\put(-3,16){\oval(4,6)[br]}
\put(1,16){\line(0,-1){16}}
\put(3,0){\oval(4,6)[tl]}
\end{picture}}
\put(80,17){\begin{picture}(0,16)(0,0)
\thinlines
\put(-1,0){\line(0,1){16}}
\put(-3,16){\oval(4,6)[br]}
\put(1,16){\line(0,-1){16}}
\put(3,0){\oval(4,6)[tl]}
\end{picture}}
\put(30,50){\begin{picture}(16,0)(0,0)
\thinlines
\put(0,1){\line(1,0){16}}
\put(16,3){\oval(6,4)[bl]}
\put(16,-1){\line(-1,0){16}}
\put(0,-3){\oval(6,4)[tr]}
\end{picture}}
\put(30,0){\begin{picture}(16,0)(0,0)
\thinlines
\put(0,1){\line(1,0){16}}
\put(16,3){\oval(6,4)[bl]}
\put(16,-1){\line(-1,0){16}}
\put(0,-3){\oval(6,4)[tr]}
\end{picture}}
\end{picture}
&
\begin{picture}(100,50)(0,0)
\put(-25,37.5)
{\makebox(50,25){$\langle \mathsf{elem}(\mathbf{L}_1), \leq_{\mathbf{L}_1} \rangle$}}
\put(-25,-12.5)
{\makebox(50,25){$\langle \mathsf{elem}(\mathbf{L}_2), \leq_{\mathbf{L}_2} \rangle$}} 
\put(75,37.5){\makebox(50,25){$\langle {\wp}\,\mathsf{typ}(\mathbf{L}_1), \supseteq \rangle$}}
\put(75,-12.5){\makebox(50,25){$\langle {\wp}\,\mathsf{typ}(\mathbf{L}_2), \supseteq \rangle$}}
\put(-44,12.5){\makebox(50,25){\footnotesize{$\mathsf{left}(\mathbf{h})$}}}
\put(-3,12.5){\makebox(50,25){\footnotesize{$\mathsf{right}(\mathbf{h})$}}}
\put(52,12.5){\makebox(50,25){\footnotesize{${\mathsf{typ}(\mathbf{h})}^{-1}$}}}
\put(99,11){\makebox(50,25){\footnotesize{$\exists\mathsf{typ}(\mathbf{h})$}}}
\put(27,48.5){\makebox(50,25){\footnotesize{$\mathsf{int}_{\mathbf{L}_1}$}}}
\put(29,27.5){\makebox(50,25){\footnotesize{$\mathsf{tau}_{\mathbf{L}_1}$}}}
\put(27,-1.5){\makebox(50,25){\footnotesize{$\mathsf{int}_{\mathbf{L}_2}$}}}
\put(29,-22.5){\makebox(50,25){\footnotesize{$\mathsf{tau}_{\mathbf{L}_2}$}}}
\put(37,54){\vector(1,0){26}}
\put(63,46){\vector(-1,0){26}}
\put(37,4){\vector(1,0){26}}
\put(63,-4){\vector(-1,0){26}}
\put(-5,12.5){\vector(0,1){25}}
\put(5,37.5){\vector(0,-1){25}}
\put(95,12.5){\vector(0,1){25}}
\put(105,37.5){\vector(0,-1){25}}
\end{picture}
\\ \\ \\
$(\underline{\mathrm{iconic}})$ & $(\underline{\mathrm{detailed}})$
\\
\end{tabular}
\end{center}
\end{fact}
The intensional aspect is abstracted as 
a category $\mathsf{Clg}_\tau$ of intensional conceptual structures,
with a contravariant adjoint component functor
$\mathsf{adj} : \mathsf{Clg}_\tau^{\mathrm{op}} \rightarrow \mathsf{Adj}$,
a covariant type component functor
$\mathsf{typ} : \mathsf{Clg}_\tau \rightarrow \mathsf{Set}$,
and a natural transformation
$\mathsf{intent} 
: \mathsf{adj} \Rightarrow {\mathsf{typ}}^{\mathrm{op}} \circ \mathsf{inv}
: \mathsf{Clg}_\tau^{\mathrm{op}} \rightarrow \mathsf{Adj}$.
The category $\mathsf{Clg}_\tau^{\mathrm{op}}$ is a subcategory of 
$\mathsf{coRefl}$ the category of coreflections (Figure~\ref{conceptual-structure}),
with inclusion functor
$\mathsf{incl} : \mathsf{Clg}_\tau^{\mathrm{op}} \rightarrow \mathsf{coRefl}$.
The components are related as
$\mathsf{adj} = \mathsf{incl} \circ \partial_0^{\mathrm{h}}$,
${\mathsf{typ}}^{\mathrm{op}} \circ \mathsf{inv} = \mathsf{incl} \circ \partial_1^{\mathrm{h}}$
and $\mathsf{intent} = \mathsf{incl} \circ \mathsf{corefl}$.

\begin{figure}
\begin{center}
\begin{picture}(100,50)(0,0)
\put(-25,-12.5){\makebox(50,25){${\wp}\,\mathsf{inst}(\mathbf{L})$}}
\put(25,37.5){\makebox(50,25){$\mathsf{ord}(\mathbf{L})$}}
\put(80,-12.5){\makebox(50,25){${{\wp}\,\mathsf{typ}(\mathbf{L})}^{\mathrm{op}}$}}
\put(-2,25){\makebox(25,12.5){\footnotesize{$\mathsf{iota}_{\mathbf{L}}$}}}
\put(29,15){\makebox(25,12.5){\footnotesize{$\mathsf{ext}_{\mathbf{L}}$}}}
\put(78,25){\makebox(25,12.5){\footnotesize{$\mathsf{int}_{\mathbf{L}}$}}}
\put(48,15){\makebox(25,12.5){\footnotesize{$\mathsf{tau}_{\mathbf{L}}$}}}
\put(37.5,3.75){\makebox(25,12.5){\footnotesize{$\mathsf{left}(\mathsf{deriv}(\mathbf{L}))$}}}
\put(37.5,-16.25){\makebox(25,12.5){\footnotesize{$\mathsf{right}(\mathsf{deriv}(\mathbf{L}))$}}}
\put(-33,20){\makebox(25,12.5){\footnotesize{$\mathsf{clo}_{\mathbf{L}}$}}}
\put(112,20){\makebox(25,12.5){\footnotesize{$\mathsf{clo}_{\mathbf{L}}$}}}
\put(-25,10){\oval(20,20)[bl]}
\put(-22.5,10){\oval(25,20)[t]}
\put(-10,7.5){\vector(0,-1){0}}
\put(125,10){\oval(20,20)[br]}
\put(122.5,10){\oval(25,20)[t]}
\put(110,7.5){\vector(0,-1){0}}
\put(25,4){\vector(1,0){50}}
\put(75,-4){\vector(-1,0){50}}
\put(8.5,15){\vector(1,1){25}}
\put(41.5,35){\vector(-1,-1){25}}
\put(68.5,40){\vector(1,-1){25}}
\put(81.5,10){\vector(-1,1){25}}
\end{picture}
\end{center}
\caption{Core elements for concept lattices}
\label{core-concept-elements}
\end{figure}
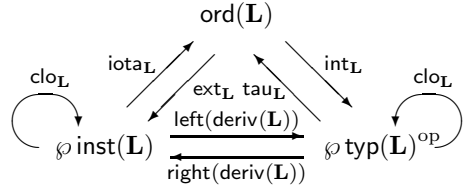

\subsubsection{Old Definition.}

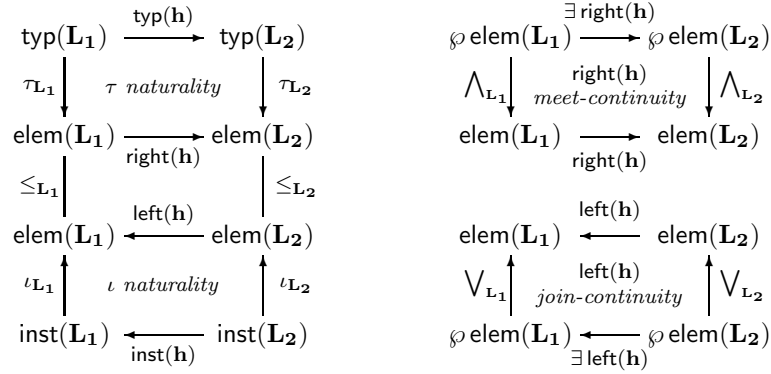
\begin{figure}
\begin{center}
\setlength{\unitlength}{0.75pt}
\begin{picture}(100,180)(0,0)
\put(-75,0){\begin{picture}(100,150)(25,0)
\put(-25,137.5){\makebox(50,25){$\mathsf{typ}(\mathbf{L_1})$}}
\put(-25,87.5){\makebox(50,25){$\mathsf{elem}(\mathbf{L_1})$}}
\put(-25,37.5){\makebox(50,25){$\mathsf{elem}(\mathbf{L_1})$}}
\put(-25,-12.5){\makebox(50,25){$\mathsf{inst}(\mathbf{L_1})$}}
\put(75,137.5){\makebox(50,25){$\mathsf{typ}(\mathbf{L_2})$}}
\put(75,87.5){\makebox(50,25){$\mathsf{elem}(\mathbf{L_2})$}}
\put(75,37.5){\makebox(50,25){$\mathsf{elem}(\mathbf{L_2})$}}
\put(75,-12.5){\makebox(50,25){$\mathsf{inst}(\mathbf{L_2})$}}
\put(25,148){\makebox(50,25){\footnotesize{$\mathsf{typ}(\mathbf{h})$}}}
\put(25,77){\makebox(50,25){\footnotesize{$\mathsf{right}(\mathbf{h})$}}}
\put(25,48){\makebox(50,25){\footnotesize{$\mathsf{left}(\mathbf{h})$}}}
\put(25,-23){\makebox(50,25){\footnotesize{$\mathsf{inst}(\mathbf{h})$}}}
\put(-37,112.5){\makebox(50,25){\footnotesize{$\tau_{\mathbf{L_1}}$}}}
\put(-37,62.5){\makebox(50,25){\footnotesize{$\leq_{\mathbf{L_1}}$}}}
\put(-37,12.5){\makebox(50,25){\footnotesize{$\iota_{\mathbf{L_1}}$}}}
\put(92,112.5){\makebox(50,25){\footnotesize{$\tau_{\mathbf{L_2}}$}}}
\put(92,62.5){\makebox(50,25){\footnotesize{$\leq_{\mathbf{L_2}}$}}}
\put(92,12.5){\makebox(50,25){\footnotesize{$\iota_{\mathbf{L_2}}$}}}
\put(0,140){\vector(0,-1){30}}
\put(0,60){\line(0,1){30}}
\put(0,10){\vector(0,1){30}}
\put(30,150){\vector(1,0){40}}
\put(30,100){\vector(1,0){40}}
\put(70,50){\vector(-1,0){40}}
\put(70,0){\vector(-1,0){40}}
\put(100,140){\vector(0,-1){30}}
\put(100,60){\line(0,1){30}}
\put(100,10){\vector(0,1){30}}
\put(25,112.5){\makebox(50,25){\footnotesize{$\tau$ \emph{naturality}}}}
\put(25,12.5){\makebox(50,25){\footnotesize{$\iota$ \emph{naturality}}}}
\end{picture}}
\put(125,12.5){\begin{picture}(100,150)(0,0)
\put(0,87.5){\begin{picture}(100,50)(0,0)
\put(25,18.75){\makebox(50,25){\footnotesize{$\mathsf{right}(\mathbf{h})$}}}
\put(25,6.25){\makebox(50,25){\footnotesize{\emph{meet-continuity}}}}
\put(-25,37.5){\makebox(50,25){${\wp}\,\mathsf{elem}(\mathbf{L_1})$}}
\put(-25,-12.5){\makebox(50,25){$\mathsf{elem}(\mathbf{L_1})$}}
\put(75,37.5){\makebox(50,25){${\wp}\,\mathsf{elem}(\mathbf{L_2})$}}
\put(75,-12.5){\makebox(50,25){$\mathsf{elem}(\mathbf{L_2})$}}
\put(25,50){\makebox(50,25){\footnotesize{${\exists}\,\mathsf{right}(\mathbf{h})$}}}
\put(25,-25){\makebox(50,25){\footnotesize{$\mathsf{right}(\mathbf{h})$}}}
\put(-37,12.5){\makebox(50,25){\scriptsize{$\bigwedge_{\mathbf{L_1}}$}}}
\put(92,12.5){\makebox(50,25){\scriptsize{$\bigwedge_{\mathbf{L_2}}$}}}
\put(0,40){\vector(0,-1){30}}
\put(100,40){\vector(0,-1){30}}
\put(35,50){\vector(1,0){30}}
\put(35,0){\vector(1,0){30}}
\end{picture}}
\put(0,-12.5){\begin{picture}(100,50)(0,0)
\put(25,18.75){\makebox(50,25){\footnotesize{$\mathsf{left}(\mathbf{h})$}}}
\put(25,6.25){\makebox(50,25){\footnotesize{\emph{join-continuity}}}}
\put(-25,37.5){\makebox(50,25){$\mathsf{elem}(\mathbf{L_1})$}}
\put(-25,-12.5){\makebox(50,25){${\wp}\,\mathsf{elem}(\mathbf{L_1})$}}
\put(75,37.5){\makebox(50,25){$\mathsf{elem}(\mathbf{L_2})$}}
\put(75,-12.5){\makebox(50,25){${\wp}\,\mathsf{elem}(\mathbf{L_2})$}}
\put(25,50){\makebox(50,25){\footnotesize{$\mathsf{left}(\mathbf{h})$}}}
\put(25,-25){\makebox(50,25){\footnotesize{${\exists}\,\mathsf{left}(\mathbf{h})$}}}
\put(-37,12.5){\makebox(50,25){\scriptsize{$\bigvee_{\mathbf{L_1}}$}}}
\put(92,12.5){\makebox(50,25){\scriptsize{$\bigvee_{\mathbf{L_2}}$}}}
\put(0,10){\vector(0,1){30}}
\put(100,10){\vector(0,1){30}}
\put(65,50){\vector(-1,0){30}}
\put(65,0){\vector(-1,0){30}}
\end{picture}}
\end{picture}}
\end{picture}
\end{center}
\caption{Concept (Lattice) Morphism}
\label{concept-morphism}
\end{figure}

A concept (lattice) morphism
$\mathbf{h} 
= \langle \mathsf{left}(\mathbf{h}), \mathsf{right}(\mathbf{h}), \mathsf{inst}(\mathbf{h}), \mathsf{typ}(\mathbf{h}) \rangle : \mathbf{L}_1 \rightleftharpoons \mathbf{L}_2$
(Figure~\ref{concept-morphism})
from source concept lattice $\mathbf{L}_1$ to target concept lattice $\mathbf{L}_2$ \cite{kent:02},
consists of 
an instance function 
$\mathsf{inst}(\mathbf{h}) : \mathsf{inst}(\mathbf{L}_1) \leftarrow \mathsf{inst}(\mathbf{L}_2)$,
a type function
$\mathsf{typ}(\mathbf{h}) : \mathsf{typ}(\mathbf{L}_1) \rightarrow \mathsf{typ}(\mathbf{L}_2)$,
and a Galois connection (adjoint pair)
$\langle \mathsf{left}(\mathbf{h}), \mathsf{right}(\mathbf{h}) \rangle : 
\langle \mathsf{elem}(\mathbf{L_2}), \leq_{\mathbf{L}_2} \rangle \rightleftharpoons \langle \mathsf{elem}(\mathbf{L_1}), \leq_{\mathbf{L}_1} \rangle$,
where 
the left adjoint 
$\mathsf{left}(\mathbf{h}) : \mathsf{elem}(\mathbf{L_1}) \leftarrow \mathsf{elem}(\mathbf{L_2})$
is a monotonic function in the backward direction that preserves instances
$\iota_{\mathbf{L}_2} \cdot \mathsf{left}(\mathbf{h}) = \mathsf{inst}(\mathbf{h}) \cdot \iota_{\mathbf{L}_1}$, and
the right adjoint 
$\mathsf{right}(\mathbf{h}) : \mathsf{elem}(\mathbf{L_1}) \rightarrow \mathsf{elem}(\mathbf{L_2})$
is a monotonic function in the forward direction that preserves types 
$\tau_{\mathbf{L}_1} \cdot \mathsf{right}(\mathbf{h}) = \mathsf{typ}(\mathbf{h}) \cdot \tau_{\mathbf{L}_2}$.
Since $\mathsf{left}(\mathbf{h})$ is a left adjoint,
it is join-continuous
${\bigvee}_{\mathbf{L}_2}\!\! \cdot \mathsf{left}(\mathbf{h}) = 
{\exists}\,\mathsf{left}(\mathbf{h}) \cdot {\bigvee}_{\mathbf{L}_1}$.
Since $\mathsf{right}(\mathbf{h})$ is a right adjoint,
it is meet-continuous
${\bigwedge}_{\mathbf{L}_1}\!\! \cdot \mathsf{right}(\mathbf{h}) 
= {\exists}\,\mathsf{right}(\mathbf{h}) \cdot {\bigwedge}_{\mathbf{L}_2}$.

Given any two concept morphisms 
$\mathbf{f} : \mathbf{L}_1 \rightleftharpoons \mathbf{L}_2$
and 
$\mathbf{g} : \mathbf{L}_2 \rightleftharpoons \mathbf{L}_3$,
there is a composite infomorphism
$\mathbf{f} \circ \mathbf{g} : \mathbf{L}_1 \rightleftharpoons \mathbf{L}_3$
defined by composing the left, right, instance and type functions:
$\mathsf{left}(\mathbf{f} \circ \mathbf{g}) = \mathsf{left}(\mathbf{g}) \cdot \mathsf{left}(\mathbf{f})$,
$\mathsf{right}(\mathbf{f} \circ \mathbf{g}) = \mathsf{right}(\mathbf{f}) \cdot \mathsf{right}(\mathbf{g})$,
$\mathsf{inst}(\mathbf{f} \circ \mathbf{g}) = \mathsf{inst}(\mathbf{g}) \cdot \mathsf{inst}(\mathbf{f})$
and
$\mathsf{typ}(\mathbf{f} \circ \mathbf{g}) = \mathsf{typ}(\mathbf{f}) \cdot \mathsf{typ}(\mathbf{g})$.
Given any concept lattice $\mathbf{L}$,
there is an identity concept morphism
$\mathrm{id}_{\mathbf{L}} : \mathbf{L} \rightleftharpoons \mathbf{L}$
(with respect to composition)
defined in terms of the identity functions on instances, types and concepts:
$\mathsf{left}(\mathrm{id}_{\mathbf{L}}) 
= \mathsf{right}(\mathrm{id}_{\mathbf{L}}) 
= \mathrm{id}_{\mathsf{lat}(\mathbf{L})}$,
$\mathsf{inst}(\mathrm{id}_{\mathbf{L}}) = \mathrm{id}_{\mathsf{inst}(\mathbf{L})}$
and
$\mathsf{typ}(\mathrm{id}_{\mathbf{L}}) = \mathrm{id}_{\mathsf{typ}(\mathbf{L})}$.
Using these notions of composition and identity,
concept lattices and concept morphisms form the category $\mathsf{Clg}$. 
This is the category  of complete lattices with two-sided generators.

\begin{figure}
\begin{center}
\begin{tabular}{c}
\\
\begin{picture}(50,90)(0,0)
\put(7.5,78.75){\makebox(30,22.5){$\mathsf{Set}$}}
\put(-15,33.75){\makebox(30,22.5){$\mathsf{Adj}^{\mathrm{op}}$}}
\put(35,33.75){\makebox(30,22.5){$\mathsf{Clg}$}}
\put(12.5,-11.25){\makebox(30,22.5){$\mathsf{Set}^{\mathrm{op}}$}}
\put(15,48.0){\makebox(20,12.5){\small{${\mathsf{adj}}^{\mbox{\tiny{op}}}$}}}
\put(-12,63){\makebox(20,12.5){\small{$\mathsf{right}$}}}
\put(39,63){\makebox(20,12.5){\small{$\mathsf{right}$}}}
\put(-12,15){\makebox(20,12.5){\small{${\mathsf{left}}^{\mbox{\tiny{op}}}$}}}
\put(39,15){\makebox(20,12.5){\small{${\mathsf{left}}^{\mbox{\tiny{op}}}$}}}
\put(6.25,57.5){\vector(1,2){11.25}}
\put(38.75,57.5){\vector(-1,2){11.25}}
\put(35,45){\vector(-1,0){25}}
\put(6.25,32.5){\vector(1,-2){11.25}}
\put(38.75,32.5){\vector(-1,-2){11.25}}
\end{picture} 
\\ \\
\small{$\begin{array}[t]{r@{\hspace{5pt}}c@{\hspace{5pt}}l}
{\mathsf{adj}}^{\mbox{\tiny{op}}} \circ \mathsf{right} & = & \mathsf{right} \\
{\mathsf{adj}}^{\mbox{\tiny{op}}} \circ {\mathsf{left}}^{\mbox{\tiny{op}}} & = & {\mathsf{left}}^{\mbox{\tiny{op}}}
\end{array}$}
\end{tabular}
\end{center}
\caption{Definition of Adjoint Pair Component}
\label{adjoint-pair-component}
\end{figure}
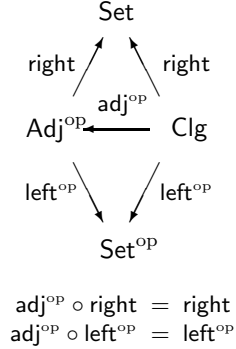

\begin{figure}
\begin{center}

\begin{tabular}{c@{\hspace{10pt}}c@{\hspace{2pt}}c}

\begin{tabular}[b]{c}
\\
\setlength{\unitlength}{0.8pt}
\begin{picture}(60,140)(0,0)
\put(15,117.5){\vector(2,1){30}}
\put(45,85){\vector(-3,2){30}}
\put(60,85){\vector(0,1){40}}
\put(0,80){\vector(0,1){20}}
\put(23,110){\makebox(30,15){\scriptsize{$\bigwedge$}}}
\put(25,98){\makebox(30,15){\Large{$\Rightarrow$}}}
\put(11,123){\makebox(30,15){\footnotesize{$\exists$}}}
\put(60,97.5){\makebox(30,15){\footnotesize{$\mathsf{right}$}}}
\put(9,81){\makebox(30,15){\footnotesize{$\mathsf{right}$}}}
\put(-27,82.5){\makebox(30,15){\footnotesize{$\mathsf{left}$}}}
\put(21,75){\vector(1,0){24}}
\put(19,72){\makebox(30,15){\footnotesize{$\propto$}}}
\put(18,62.5){\makebox(30,15){\footnotesize{$\cong$}}}
\put(23,53){\makebox(30,15){\footnotesize{${\propto}^{\mathrm{op}}$}}}
\put(45,65){\vector(-1,0){24}}
\put(60,42.5){\makebox(30,15){\footnotesize{$\mathsf{right}$}}}
\put(22,44){\makebox(30,15){\footnotesize{$\mathsf{left}$}}}
\put(-27,27.5){\makebox(30,15){\footnotesize{$\mathsf{left}$}}}
\put(19,1){\makebox(30,15){\footnotesize{$\exists$}}}
\put(7,30){\makebox(30,15){\scriptsize{$\bigvee$}}}
\put(5,18){\makebox(30,15){\Large{$\Leftarrow$}}}
\put(60,60){\vector(0,-1){20}}
\put(0,55){\vector(0,-1){40}}
\put(15,55){\vector(3,-2){30}}
\put(45,22.5){\vector(-2,-1){30}}
\put(30,125){\makebox(60,30){$\mathsf{Set}$}}
\put(-30,95){\makebox(60,30){$\mathsf{Set}$}}
\put(-25,55){\makebox(60,30){$\mathsf{Clg}^{\mathrm{op}}$}}
\put(30,55){\makebox(60,30){$\mathsf{Clg}$}}
\put(30,15){\makebox(60,30){$\mathsf{Set}$}}
\put(-30,-15){\makebox(60,30){$\mathsf{Set}$}}
\end{picture}
\\ \\
\small{$\begin{array}[t]{r@{\hspace{5pt}}c@{\hspace{5pt}}l}
\mathsf{right} \circ \exists & \stackrel{\wedge}{\Rightarrow} & \mathsf{right} \\
\mathsf{left} \circ \exists  & \stackrel{\vee}{\Rightarrow} & \mathsf{left} \\
{\propto}^{\mbox{\tiny{op}}} \circ \bigvee & = & \bigwedge \\
{\propto} \circ \bigwedge & = & \bigvee
\end{array}$}
\\ \\
\underline{meet and join}
\end{tabular}

&

\begin{tabular}[b]{c}
\\
\setlength{\unitlength}{0.8pt}
\begin{picture}(60,140)(0,0)
\put(15,117.5){\vector(2,1){30}}
\put(45,85){\vector(-3,2){30}}
\put(60,85){\vector(0,1){40}}
\put(0,80){\vector(0,1){20}}
\put(24,108){\makebox(30,15){\scriptsize{$\mathsf{tau}$}}}
\put(25,98){\makebox(30,15){\Large{$\Rightarrow$}}}
\put(11,123){\makebox(30,15){\footnotesize{$\exists$}}}
\put(60,97.5){\makebox(30,15){\footnotesize{$\mathsf{right}$}}}
\put(12,82){\makebox(30,15){\footnotesize{$\mathsf{typ}$}}}
\put(-27,82.5){\makebox(30,15){\footnotesize{$\mathsf{inst}$}}}
\put(21,75){\vector(1,0){24}}
\put(19,72){\makebox(30,15){\footnotesize{$\propto$}}}
\put(18,62.5){\makebox(30,15){\footnotesize{$\cong$}}}
\put(23,53){\makebox(30,15){\footnotesize{${\propto}^{\mathrm{op}}$}}}
\put(42,65){\vector(-1,0){24}}
\put(57,42.5){\makebox(30,15){\footnotesize{$\mathsf{typ}$}}}
\put(18,44){\makebox(30,15){\footnotesize{$\mathsf{inst}$}}}
\put(-27,27.5){\makebox(30,15){\footnotesize{$\mathsf{left}$}}}
\put(19,1){\makebox(30,15){\footnotesize{$\exists$}}}
\put(6,28){\makebox(30,15){\scriptsize{$\mathsf{iota}$}}}
\put(5,18){\makebox(30,15){\Large{$\Leftarrow$}}}
\put(60,60){\vector(0,-1){20}}
\put(0,55){\vector(0,-1){40}}
\put(15,55){\vector(3,-2){30}}
\put(45,22.5){\vector(-2,-1){30}}
\put(30,125){\makebox(60,30){$\mathsf{Set}$}}
\put(-30,95){\makebox(60,30){$\mathsf{Set}$}}
\put(-25,55){\makebox(60,30){$\mathsf{Clg}^{\mathrm{op}}$}}
\put(30,55){\makebox(60,30){$\mathsf{Clg}$}}
\put(30,15){\makebox(60,30){$\mathsf{Set}$}}
\put(-30,-15){\makebox(60,30){$\mathsf{Set}$}}
\end{picture}
\\ \\
\small{$\begin{array}[t]{r@{\hspace{5pt}}c@{\hspace{5pt}}l}
\mathsf{typ} \circ \exists & \stackrel{\mathsf{tau}}{\Rightarrow} & \mathsf{right} \\
\mathsf{inst} \circ \exists  & \stackrel{\mathsf{iota}}{\Rightarrow} & \mathsf{left} \\
\mathsf{tau}  & \doteq & (\tau \circ \exists) \bullet \bigwedge \\
\mathsf{iota} & \doteq & (\iota \circ \exists) \bullet \bigvee
\end{array}$}
\\ \\
\underline{iota amd tau}
\end{tabular}

& 

\begin{tabular}[b]{c}
\\
\setlength{\unitlength}{0.8pt}
\begin{picture}(60,140)(0,0)
\put(15,117.5){\vector(2,1){30}}
\put(45,85){\vector(-3,2){30}}
\put(60,85){\vector(0,1){40}}
\put(0,80){\vector(0,1){20}}
\put(25,108){\makebox(30,15){\scriptsize{$\mathsf{int}$}}}
\put(24,98){\makebox(30,15){\Large{$\Leftarrow$}}}
\put(13,126){\makebox(30,15){\scriptsize{${(-)}^{\mathrm{-1}}$}}}
\put(58,97.5){\makebox(30,15){\footnotesize{$\mathsf{left}$}}}
\put(12,82){\makebox(30,15){\footnotesize{$\mathsf{typ}^{\mathrm{op}}$}}}
\put(-32,82.5){\makebox(30,15){\footnotesize{$\mathsf{inst}^{\mathrm{op}}$}}}
\put(18,75){\vector(1,0){24}}
\put(20,72){\makebox(30,15){\footnotesize{${\propto}^{\mathrm{op}}$}}}
\put(15,62.5){\makebox(30,15){\footnotesize{$\cong$}}}
\put(16,53){\makebox(30,15){\footnotesize{$\propto$}}}
\put(42,65){\vector(-1,0){24}}
\put(62,42.5){\makebox(30,15){\footnotesize{$\mathsf{typ}^{\mathrm{op}}$}}}
\put(18,44){\makebox(30,15){\footnotesize{$\mathsf{inst}^{\mathrm{op}}$}}}
\put(-28,27.5){\makebox(30,15){\footnotesize{$\mathsf{right}$}}}
\put(27,1){\makebox(30,15){\scriptsize{${(-)}^{\mathrm{-1}}$}}}
\put(6,28){\makebox(30,15){\scriptsize{$\mathsf{ext}$}}}
\put(8,18){\makebox(30,15){\Large{$\Rightarrow$}}}
\put(60,60){\vector(0,-1){20}}
\put(0,55){\vector(0,-1){40}}
\put(15,55){\vector(3,-2){30}}
\put(45,22.5){\vector(-2,-1){30}}
\put(30,125){\makebox(60,30){$\mathsf{Set}$}}
\put(-25,95){\makebox(60,30){$\mathsf{Set}^{\mathrm{op}}$}}
\put(-30,55){\makebox(60,30){$\mathsf{Clg}$}}
\put(35,55){\makebox(60,30){$\mathsf{Clg}^{\mathrm{op}}$}}
\put(35,15){\makebox(60,30){$\mathsf{Set}^{\mathrm{op}}$}}
\put(-30,-15){\makebox(60,30){$\mathsf{Set}$}}
\end{picture}
\\ \\
\small{$\begin{array}[t]{r@{\hspace{5pt}}c@{\hspace{5pt}}l}
\mathsf{typ}^{\mathrm{op}} \circ {(-)}^{\mathrm{-1}} & \stackrel{\mathsf{int}}{\Leftarrow} & \mathsf{left} \\
\mathsf{inst}^{\mathrm{op}} \circ {(-)}^{\mathrm{-1}} & \stackrel{\mathsf{ext}}{\Leftarrow} & \mathsf{right} \\
{\propto} \circ \mathsf{ext} & = & \mathsf{int} \\
{\propto}^{\mbox{\tiny{op}}} \circ \mathsf{int} & = & \mathsf{ext}
\end{array}$}
\\ \\
\underline{extent and intent}
\end{tabular}

\\ && \\

&

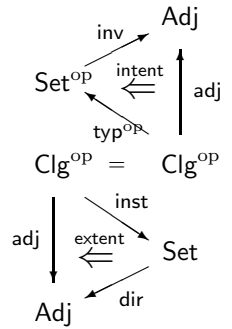
\begin{tabular}[b]{c}
\\
\setlength{\unitlength}{0.8pt}
\begin{picture}(60,140)(0,0)
\put(15,117.5){\vector(2,1){30}}
\put(45,85){\vector(-3,2){30}}
\put(60,85){\vector(0,1){40}}
\put(26,108){\makebox(30,15){\scriptsize{$\mathsf{intent}$}}}
\put(25,98){\makebox(30,15){\Large{$\Leftarrow$}}}
\put(12,126){\makebox(30,15){\scriptsize{\footnotesize{$\mathsf{inv}$}}}}
\put(58,97.5){\makebox(30,15){\footnotesize{$\mathsf{adj}$}}}
\put(16,78){\makebox(30,15){\footnotesize{$\mathsf{typ}^{\mathrm{op}}$}}}
\put(15,62.5){\makebox(30,15){\footnotesize{$=$}}}
\put(22,46){\makebox(30,15){\footnotesize{$\mathsf{inst}$}}}
\put(-28,27.5){\makebox(30,15){\footnotesize{$\mathsf{adj}$}}}
\put(22,0){\makebox(30,15){\scriptsize{\footnotesize{$\mathsf{dir}$}}}}
\put(7,28){\makebox(30,15){\scriptsize{$\mathsf{extent}$}}}
\put(5,18){\makebox(30,15){\Large{$\Leftarrow$}}}
\put(0,55){\vector(0,-1){40}}
\put(15,55){\vector(3,-2){30}}
\put(45,22.5){\vector(-2,-1){30}}
\put(30,125){\makebox(60,30){$\mathsf{Adj}$}}
\put(-25,95){\makebox(60,30){$\mathsf{Set}^{\mathrm{op}}$}}
\put(-25,55){\makebox(60,30){$\mathsf{Clg}^{\mathrm{op}}$}}
\put(35,55){\makebox(60,30){$\mathsf{Clg}^{\mathrm{op}}$}}
\put(30,15){\makebox(60,30){$\mathsf{Set}$}}
\put(-30,-15){\makebox(60,30){$\mathsf{Adj}$}}
\end{picture}
\\ \\
\small{$\begin{array}[t]{r@{\hspace{5pt}}c@{\hspace{5pt}}l}
\mathsf{typ}^{\mathrm{op}} \circ \mathsf{inv} & \stackrel{\mathsf{intent}}{\Leftarrow} & \mathsf{adj} \\
\mathsf{inst} \circ \mathsf{dir} & \stackrel{\mathsf{extent}}{\Rightarrow} & \mathsf{adj} \\
\mathsf{intent} \circ \mathsf{left}                & = & \mathsf{int} \\
\mathsf{intent} \circ \mathsf{right}^{\mathrm{op}} & = & \mathsf{tau}^{\mathrm{op}} \\
\mathsf{extent} \circ \mathsf{left}                & = & \mathsf{iota} \\
\mathsf{extent} \circ \mathsf{right}^{\mathrm{op}} & = & \mathsf{ext}^{\mathrm{op}} \\
\propto \circ\; {\mathsf{extent}}^{\mathrm{op}} \circ \propto & = & \mathsf{intent} \\
\propto \circ\; {\mathsf{intent}}^{\mathrm{op}} \circ \propto & = & \mathsf{extent}
\end{array}$}
\\ \\
\underline{extent and intent}
\end{tabular}

&

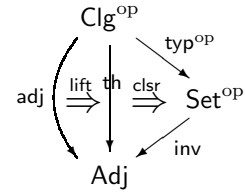
\begin{tabular}[b]{c}
\\
\setlength{\unitlength}{0.5pt}
\begin{picture}(120,120)(0,-60)
\put(10,105){\makebox(60,30){$\mathsf{Clg}^{\mathrm{op}}$}}
\put(90,45){\makebox(60,30){$\mathsf{Set}^{\mathrm{op}}$}}
\put(10,-15){\makebox(60,30){$\mathsf{Adj}$}}
\put(-35,52.5){\makebox(30,15){\footnotesize{$\mathsf{adj}$}}}
\put(27,65){\makebox(30,15){\footnotesize{$\mathsf{th}$}}}
\put(86,93){\makebox(30,15){\footnotesize{$\mathsf{typ}^{\mathrm{op}}$}}}
\put(84,15){\makebox(30,15){\scriptsize{\footnotesize{$\mathsf{inv}$}}}}
\put(3,62.5){\makebox(30,15){\scriptsize{$\mathsf{lift}$}}}
\put(5,47.5){\makebox(30,15){\Large{$\Rightarrow$}}}
\put(53,62.5){\makebox(30,15){\scriptsize{$\mathsf{clsr}$}}}
\put(55,47.5){\makebox(30,15){\Large{$\Rightarrow$}}}
\qbezier(15,105)(-20,60)(15,15)
\put(15,15){\vector(3,-4){0}}
\put(40,100){\vector(0,-1){80}}
\put(60,105){\vector(4,-3){40}}
\put(100,45){\vector(-4,-3){40}}
\end{picture}
\\ \\
\small{$\begin{array}[t]{r@{\hspace{5pt}}c@{\hspace{5pt}}l}
\mathsf{adj} & \stackrel{\mathsf{intent}}{\Rightarrow} & \mathsf{typ}^{\mathrm{op}} \circ \mathsf{inv} \\
\mathsf{adj} & \stackrel{\mathsf{lift}}{\Rightarrow} & \mathsf{th} \\
\mathsf{th} & \stackrel{\mathsf{clsr}}{\Rightarrow} & \mathsf{typ}^{\mathrm{op}} \circ \mathsf{inv} \\
\mathsf{intent} & = & \mathsf{lift} \circ \mathsf{clsr} \\
\\ \\ \\ \\
\end{array}$}
\\ \\
\underline{factorization of intent}
\end{tabular}

\end{tabular}

\end{center}
\caption{The Component Architecture for Concept Lattices}
\label{concept-lattice-component-architecture}
\end{figure}

Concept lattices and concept morphisms resolve into components
(top middle Figure~\ref{concept-lattice-component-architecture}):
there is an instance functor
$\mathsf{inst} : \mathsf{Clg}^{\mathrm{op}} \rightarrow \mathsf{Set}$,
a type functor
$\mathsf{typ} : \mathsf{Clg} \rightarrow \mathsf{Set}$,
a left functor
$\mathsf{left} : \mathsf{Clg}^{\mathrm{op}} \rightarrow \mathsf{Set}$ 
and
a right functor
$\mathsf{right} : \mathsf{Clg} \rightarrow \mathsf{Set}$.
Instance preservation can be viewed as a naturality condition.
Hence,
there is an instance embedding natural transformation
$\iota : \mathsf{inst} \Rightarrow \mathsf{left} : \mathsf{Clg}^{\mathrm{op}} \rightarrow \mathsf{Set}$
(top middle Figure~\ref{concept-lattice-component-architecture}).
Type preservation can be viewed as a naturality condition.
Hence,
there is a type embedding natural transformation
$\tau : \mathsf{typ} \Rightarrow \mathsf{right} : \mathsf{Clg} \rightarrow \mathsf{Set}$
(top middle Figure~\ref{concept-lattice-component-architecture}).
The continuity condition of the right functions can be viewed as a naturality condition for meet.
Hence,
there is an meet natural transformation
$\bigwedge : \mathsf{right} \circ \exists \Rightarrow \mathsf{right} : \mathsf{Clg}^{\mathrm{op}} \rightarrow \mathsf{Set}$
(bottom left Figure~\ref{concept-lattice-component-architecture}).
The continuity condition of the left functions can be viewed as a naturality condition for join.
Hence,
there is an join natural transformation
$\bigvee : \mathsf{left} \circ \exists \Rightarrow \mathsf{left} : \mathsf{Clg} \rightarrow \mathsf{Set}$.

\subsubsection{Examples.}

Power examples.
Involution example.

\subsubsection{Extent and Intent.}

For any concept lattice $\mathbf{L}$ and any element $c \in \mathsf{elem}(\mathbf{L})$, 
the extent (or instance set) of $c$ is the set 
$\mathsf{ext}_{\mathbf{L}}(c) 
= \{ x \in \mathsf{inst}(\mathbf{L}) \mid \iota_{\mathbf{L}}(x) \leq_{\mathbf{L}} c \}$,
and the intent (or type set) of $c$ is the set 
$\mathsf{int}_{\mathbf{L}}(c) 
= \{ y \in \mathsf{typ}(\mathbf{L}) \mid c \leq_{\mathbf{L}} \tau_{\mathbf{L}}(y) \}$.
These define the extent and intent functions
$\mathsf{ext}_{\mathbf{L}} : \mathsf{elem}(\mathbf{L}) \rightarrow {\wp}\,\mathsf{inst}(\mathbf{L})$
and 
$\mathsf{int}_{\mathbf{L}} : \mathsf{elem}(\mathbf{L}) \rightarrow
 {\wp}\, \mathsf{typ} (\mathbf{L})$.
For any concept morphism
$\mathbf{h} : \mathbf{L}_1 \rightleftharpoons \mathbf{L}_2$,
the following naturality conditions hold:
$\mathsf{ext}_{\mathbf{L}_1} \cdot_{\mathsf{Set}} {\mathsf{inst}(\mathbf{h})}^{{-}1} 
= \mathsf{right}(\mathbf{h}) \cdot_{\mathsf{Set}} \mathsf{ext}_{\mathbf{L}_2}$
and
$\mathsf{int}_{\mathbf{L}_2} \cdot_{\mathsf{Set}} {\mathsf{typ}(\mathbf{h})}^{{-}1} 
= \mathsf{left}(\mathbf{h}) \cdot_{\mathsf{Set}} \mathsf{int}_{\mathbf{L}_1}$.
Hence,
there are two natural transformations
$\mathsf{ext} : \mathsf{right} \Rightarrow \mathsf{inst}^{\mathrm{op}} \circ {(-)}^{-1} : \mathsf{Clg} \rightarrow \mathsf{Set}$
and
$\mathsf{int} : \mathsf{left} \Rightarrow \mathsf{typ}^{\mathrm{op}} \circ {(-)}^{-1} : \mathsf{Clg}^{\mathrm{op}} \rightarrow \mathsf{Set}$
(bottom middle Figure~\ref{concept-lattice-component-architecture}).

\subsubsection{Subset Embeddings.}

Let $\mathbf{L}$ be any concept lattice.
For any instance subset $X \in {\wp}\,\mathsf{inst}(\mathbf{L})$, 
the embedding of $X$ is the smallest concept whose extent contains $X$;
that is,
the embedding of $X$ is the concept
$\mathsf{iota}_{\mathbf{L}}(X) 
= \bigwedge_{\mathbf{L}} \{ c \in \mathsf{elem}(\mathbf{L}) \mid X \subseteq \mathsf{ext}_{\mathbf{L}}(c) \}
= \bigwedge_{\mathbf{L}} \{ c \in \mathsf{elem}(\mathbf{L}) \mid (\forall_{x \in X})\; x \in \mathsf{ext}_{\mathbf{L}}(c) \}
= \bigwedge_{\mathbf{L}} \{ c \in \mathsf{elem}(\mathbf{L}) \mid (\forall_{x \in X})\; \iota_{\mathbf{L}}(x) \leq_{\mathbf{L}} c \}
= \bigvee_{\mathbf{L}} \{ \iota_{\mathbf{L}}(x) \mid x \in X \}
= \bigvee_{\mathbf{L}}\iota_{\mathbf{L}}[X]$.
This defines the instance subset embedding function
$\mathsf{iota}_{\mathbf{L}} = {\exists}\iota_{\mathbf{L}} \cdot \bigvee_{\mathbf{L}} : {\wp}\,\mathsf{inst}(\mathbf{L}) \rightarrow {\wp}\,\mathsf{elem}(\mathbf{L}) \rightarrow \mathsf{elem}(\mathbf{L})$
as the composition of existential direct image of instance embedding with join.
For any concept morphism
$\mathbf{h} : \mathbf{L}_1 \rightleftharpoons \mathbf{L}_2$,
the following naturality condition holds:
$\exists\mathsf{inst}(\mathbf{h}) \cdot_{\mathsf{Set}} \mathsf{iota}_{\mathbf{L}_1} 
= \mathsf{iota}_{\mathbf{L}_2} \cdot_{\mathsf{Set}} \mathsf{left}(\mathbf{h})$.
Hence,
instance subset embedding is a natural transformation
$\mathsf{iota} = (\iota \circ \exists) \bullet \bigvee : \mathsf{inst} \circ \exists \Rightarrow \mathsf{left} \circ \exists \Rightarrow \mathsf{left} : \mathsf{Clg}^{\mathrm{op}} \rightarrow \mathsf{Set}$.
The join-dense assumption for concept lattices is equivalent to the condition
$c 
= \bigvee_{x{\in}\mathsf{ext}(c)} \iota_{\mathbf{L}}(x)
= \bigvee_{\mathbf{L}} \iota_{\mathbf{L}}[\mathsf{ext}_{\mathbf{L}}(c)]
= \mathsf{iota}_{\mathbf{L}}(\mathsf{ext}_{\mathbf{L}}(c))
= ( \mathsf{ext}_{\mathbf{L}} \cdot \mathsf{iota}_{\mathbf{L}} )(c)$
for every element $c \in \mathsf{elem}(\mathbf{L})$;
that is,
$\mathsf{ext}_{\mathbf{L}} \cdot \mathsf{iota}_{\mathbf{L}} = \mathrm{id}_{\mathsf{elem}(\mathbf{L})}$.
On the other hand,
$\mathsf{ext}_{\mathbf{L}}(\mathsf{iota}_{\mathbf{L}}(X))
= \{ x \in \mathsf{inst}(\mathbf{L}) \mid \iota_{\mathbf{L}}(x) \leq_{\mathbf{L}} \bigvee_{\mathbf{L}}\iota_{\mathbf{L}}[X] \}
\supseteq X$
for every instance subset $X \in {\wp}\,\mathsf{inst}(\mathbf{L})$.
Hence,
there is a reflection
$\mathsf{extent}_{\mathbf{L}}
= \langle \mathsf{iota}_{\mathbf{L}}, \mathsf{ext}_{\mathbf{L}} \rangle : {\wp}\,\mathsf{inst}(\mathbf{L}) \rightleftharpoons \mathsf{elem}(\mathbf{L})$
called extent.
For any concept morphism
$\mathbf{h} : \mathbf{L}_1 \rightleftharpoons \mathbf{L}_2$,
the following naturality condition holds 
(see the lower parts of Figure~\ref{ext-int-embeddings}):
$\mathsf{dir}(\mathsf{inst}(\mathbf{h})) \circ_{\mathsf{Adj}} \mathsf{extent}_{\mathbf{L}_1} 
= \mathsf{extent}_{\mathbf{L}_2} \circ_{\mathsf{Adj}} \mathsf{adj}(\mathbf{h})$.
Hence,
extent is a natural transformation
$\mathsf{extent} : \mathsf{inst} \circ \mathsf{dir} \Rightarrow \mathsf{adj} : \mathsf{Clg}^{\mathrm{op}} \rightarrow \mathsf{Adj}$,
defined by
$\mathsf{extent} \circ \mathsf{right}^{\mathrm{op}} = \mathsf{ext}^{\mathrm{op}}$
and $\mathsf{extent} \circ \mathsf{left} = \mathsf{iota}$.

For any type subset $Y \in {\wp}\,\mathsf{typ}(\mathbf{L})$, 
the embedding of $Y$ is the largest concept whose intent contains $Y$;
that is,
the embedding of $Y$ is the concept
$\mathsf{tau}_{\mathbf{L}}(Y) 
= \bigvee_{\mathbf{L}} \{ c \in \mathsf{elem}(\mathbf{L}) \mid Y \subseteq \mathsf{int}_{\mathbf{L}}(c) \}
= \bigvee_{\mathbf{L}} \{ c \in \mathsf{elem}(\mathbf{L}) \mid (\forall_{y \in Y})\; y \in \mathsf{int}_{\mathbf{L}}(c) \}
= \bigvee_{\mathbf{L}} \{ c \in \mathsf{elem}(\mathbf{L}) \mid (\forall_{y \in Y})\; c \leq_{\mathbf{L}} \tau_{\mathbf{L}}(y) \}
= \bigwedge_{\mathbf{L}} \{ \tau_{\mathbf{L}}(y) \mid y \in Y \}
= \bigwedge_{\mathbf{L}}\tau_{\mathbf{L}}[Y]$.
This defines the type subset embedding function
$\mathsf{tau}_{\mathbf{L}} = {\exists}\tau_{\mathbf{L}} \cdot \bigwedge_{\mathbf{L}} : {\wp}\,\mathsf{typ}(\mathbf{L}) \rightarrow {\wp}\,\mathsf{elem}(\mathbf{L}) \rightarrow \mathsf{elem}(\mathbf{L})$
as the composition of existential direct image of type embedding with meet.
For any concept morphism
$\mathbf{h} : \mathbf{L}_1 \rightleftharpoons \mathbf{L}_2$,
the following naturality condition holds:
$\exists\mathsf{typ}(\mathbf{h}) \cdot_{\mathsf{Set}} \mathsf{tau}_{\mathbf{L}_2} 
= \mathsf{tau}_{\mathbf{L}_1} \cdot_{\mathsf{Set}} \mathsf{right}(\mathbf{h})$.
Hence,
type subset embedding is a natural transformation
$\mathsf{tau} = (\tau \circ \exists) \bullet \bigwedge : \mathsf{typ} \circ \exists \Rightarrow \mathsf{right} \circ \exists \Rightarrow \mathsf{right} : \mathsf{Clg} \rightarrow \mathsf{Set}$.
The meet-dense assumption for concept lattices is equivalent to the condition
$c 
= \bigwedge_{y{\in}\mathsf{int}(c)} \tau_{\mathbf{L}}(y)
= \bigwedge_{\mathbf{L}} \tau_{\mathbf{L}}[\mathsf{int}_{\mathbf{L}}(c)]
= \mathsf{tau}_{\mathbf{L}}(\mathsf{int}_{\mathbf{L}}(c))
= ( \mathsf{int}_{\mathbf{L}} \cdot \mathsf{tau}_{\mathbf{L}} )(c)$
for every element $c \in \mathsf{elem}(\mathbf{L})$;
that is,
$\mathsf{int}_{\mathbf{L}} \cdot \mathsf{tau}_{\mathbf{L}} = \mathrm{id}_{\mathsf{elem}(\mathbf{L})}$.
On the other hand,
$\mathsf{int}_{\mathbf{L}}(\mathsf{tau}_{\mathbf{L}}(Y))
= \{ y \in \mathsf{typ}(\mathbf{L}) \mid \tau_{\mathbf{L}}(y) \geq_{\mathbf{L}} \bigwedge_{\mathbf{L}}\tau_{\mathbf{L}}[Y] \}
\supseteq Y$
for every type subset $Y \in {\wp}\,\mathsf{typ}(\mathbf{L})$.
Hence,
there is a coreflection
$\mathsf{intent}_{\mathbf{L}}
= \langle \mathsf{int}_{\mathbf{L}}, \mathsf{tau}_{\mathbf{L}} \rangle : \mathsf{elem}(\mathbf{L}) \rightleftharpoons {\wp}\,\mathsf{typ}(\mathbf{L})^{\mathrm{op}}$
called intent.
For any concept morphism
$\mathbf{h} : \mathbf{L}_1 \rightleftharpoons \mathbf{L}_2$,
the following naturality condition holds
(see the upper parts of Figure~\ref{ext-int-embeddings}):
$\mathsf{intent}_{\mathbf{L}_2} \circ_{\mathsf{Adj}} \mathsf{inv}(\mathsf{typ}(\mathbf{h})) 
= \mathsf{adj}(\mathbf{h}) \circ_{\mathsf{Adj}} \mathsf{intent}_{\mathbf{L}_1}$.
Hence,
intent is a natural transformation
$\mathsf{intent} : \mathsf{adj} \Rightarrow {\mathsf{inst}}^{\mathrm{op}} \circ \mathsf{inv} : \mathsf{Clg}^{\mathrm{op}} \rightarrow \mathsf{Adj}$,
defined by
$\mathsf{intent} \circ \mathsf{right}^{\mathrm{op}} = \mathsf{tau}^{\mathrm{op}}$
and $\mathsf{intent} \circ \mathsf{left} = \mathsf{int}$.

\subsubsection{Intent Factorization.}

We can define an type subset embedding relation 
$\mathsf{tau}_{\mathbf{L}} \subseteq \mathsf{elem}(\mathbf{L}) 
{\times} {\wp}\,\mathsf{typ}(\mathbf{L})$ as follows: 
for every every element $c \in \mathsf{elem}(\mathbf{L})$ 
and every type subset $Y \subseteq \mathsf{typ}(\mathbf{L})$
the relationship $c\,\mathsf{tau}_{\mathbf{L}}\,Y$ holds 
when 
$c \leq_{\mathbf{L}} \mathsf{tau}_{\mathbf{L}}(Y)$;
or equivalently when
$c \leq_{\mathbf{L}} \tau_{\mathbf{L}}(y)$ for all types $y \in Y$;
or equivalently when
$c \tau_{\mathbf{L}} y$ for all types $y \in Y$.
This relation is closed on the left with respect to lattice order:
${\leq}_{\mathbf{L}} \circ \mathsf{tau}_{\mathbf{L}} = \mathsf{tau}_{\mathbf{L}}$
and on the right with respect to reverse type subset order
$\mathsf{tau}_{\mathbf{L}} \circ \supseteq_{{\wp}\,\mathsf{typ}(\mathbf{L})} = \mathsf{tau}_{\mathbf{L}}$. 
The given function can be expressed in terms of the defined relation as the join 
$\mathsf{tau}_{\mathbf{L}}(Y) = \bigvee_{\mathbf{L}} \mathsf{tau}_{\mathbf{L}}Y$.

The (Galois connection) interior of $\mathsf{intent}_{\mathbf{L}}$ is the function
$\mathsf{clo}_{\mathbf{L}}
= {(-)}^{\bullet} \doteq \mathsf{tau}_{\mathbf{L}} \cdot \mathsf{int}_{\mathbf{L}}
: {\wp}\,\mathsf{typ}(\mathbf{L}) \rightarrow {\wp}\,\mathsf{typ}(\mathbf{L})$,
which maps a type subset 
$Y \subset {\wp}\,\mathsf{typ}(\mathbf{L})$ 
to the type subset
$\mathsf{clo}_{\mathbf{L}}(Y) 
= {Y}^{\bullet}
= \{ y \in \mathsf{typ}(\mathbf{L}) 
\mid \forall_{c \in \mathsf{elem}(\mathbf{L})}, 
c \,\mathsf{tau}_{\mathbf{L}}\, Y \;\mbox{implies}\; c \tau_{\mathbf{L}} y \}
= \{ y \in \mathsf{typ}(\mathbf{L}) 
\mid \forall_{x \in \mathsf{inst}(\mathbf{L})}, 
x \models_{\mathbf{L}} Y \;\mbox{implies}\; x \models_{\mathbf{L}} y \}$.
Note: we use the closure notation for compatibility with logic and model theory;
that is,
the closure of theories (subsets of types) is the usual closure used in logic and model theory, but is the interior for Galois connections.

Since $\mathsf{intent}_{\mathbf{L}}$ is a coreflection
(see Section~\ref{par:morphisms:coreflections}),
it factors in terms of the kernel 
$\mathsf{th}(\mathbf{L}) 
\doteq \mathsf{ker}(\mathsf{tau}_{\mathbf{L}})
= \mathsf{ker}(\mathsf{tau}_{\mathbf{L}} \cdot \mathsf{int}_{\mathbf{L}})
= \mathsf{ker}(\mathsf{clo}_{\mathbf{L}})$
of its right adjoint:
$\mathsf{intent}_{\mathbf{L}}
= \mathsf{lift}_{\mathbf{L}} \circ \mathsf{clsr}_{\mathbf{L}}
: \mathsf{elem}(\mathbf{L}) \rightleftharpoons \mathsf{th}(\mathbf{L})
\rightleftharpoons {\wp}\,\mathsf{typ}(\mathbf{L})^{\mathrm{op}}$
(Figure~\ref{intensional-conceptual-structure}),
where the left and right adjoint form the lift Galois connection
$\mathsf{lift}_{\mathbf{L}}
= \langle \mathsf{int}_{\mathbf{L}}, \mathsf{tau}_{\mathbf{L}} \rangle 
: \mathsf{elem}(\mathbf{L}) \rightleftharpoons \mathsf{th}(\mathbf{L})$
and
closure and target identity form the closure Galois connection
$\mathsf{clsr}_{\mathbf{L}} 
= \langle \mathsf{clo}_{\mathbf{L}}, \mathrm{id}_{{\wp}\,\mathsf{typ}(\mathbf{L})} \rangle
: \mathsf{th}(\mathbf{L}) \rightleftharpoons {\wp}\,\mathsf{typ}(\mathbf{L})^{\mathrm{op}}$.
The kernel
$\mathsf{th}(\mathbf{L}) = \langle {\wp}\,\mathsf{typ}(\mathbf{L}), \vdash_{\mathbf{L}} \rangle$
is a complete preorder.
The elements of $\mathsf{th}(\mathbf{L})$, called theories, are subsets of types.
The order of $\mathsf{th}(\mathbf{L})$, called entailment order between theories, is define by 
$Y_1 \vdash_{\mathbf{L}} Y_2$ 
when  
$\mathsf{clo}_{\mathbf{L}}(Y_1) \supseteq \mathsf{clo}_{\mathbf{L}}(Y_2)$.

\begin{figure}
\begin{center}
\begin{tabular}{c@{\hspace{100pt}}c}
\setlength{\unitlength}{0.7pt}
\begin{picture}(120,120)(0,-20)
\put(-60,90){\makebox(120,60) {${\wp}\mathsf{typ}(\mathbf{L}_1)^{\mathrm{op}}$}}
\put(60,90){\makebox(120,60)  {${\wp}\mathsf{typ}(\mathbf{L}_2)^{\mathrm{op}}$}}
\put(-60,30){\makebox(120,60) {$\mathsf{elem}(\mathbf{L}_1)$}}
\put(60,30){\makebox(120,60)  {$\mathsf{elem}(\mathbf{L}_2)$}}
\put(-60,-30){\makebox(120,60){${\wp}\mathsf{inst}(\mathbf{L}_1)$}}
\put(60,-30){\makebox(120,60) {${\wp}\mathsf{inst}(\mathbf{L}_2)$}}
\put(-60,60){\makebox(60,60){\footnotesize{$\mathsf{intent}_{\mathbf{L}_1}$}}}
\put(125,60){\makebox(60,60){\footnotesize{$\mathsf{intent}_{\mathbf{L}_2}$}}}
\put(-60,0){\makebox(60,60){\footnotesize{$\mathsf{extent}_{\mathbf{L}_1}$}}}
\put(125,0){\makebox(60,60){\footnotesize{$\mathsf{extent}_{\mathbf{L}_2}$}}}
\put(10,105){\makebox(120,60){\footnotesize{$\mathsf{inv}(\mathsf{typ}(\mathbf{h}))$}}}
\put(0,45){\makebox(120,60){\footnotesize{$\mathsf{adj}(\mathbf{h})$}}}
\put(4,-15){\makebox(120,60){\footnotesize{$\mathsf{dir}(\mathsf{inst}(\mathbf{h}))$}}}
\put(0,78){\begin{picture}(0,24)(0,0)
\thinlines
\put(-1.5,0){\line(0,1){24}}
\put(-4.5,24){\oval(6,9)[br]}
\put(1.5,24){\line(0,-1){24}}
\put(4.5,0){\oval(6,9)[tl]}
\end{picture}}
\put(0,18){\begin{picture}(0,24)(0,0)
\thinlines
\put(-1.5,0){\line(0,1){24}}
\put(-4.5,24){\oval(6,9)[br]}
\put(1.5,24){\line(0,-1){24}}
\put(4.5,0){\oval(6,9)[tl]}
\end{picture}}
\put(120,78){\begin{picture}(0,24)(0,0)
\thinlines
\put(-1.5,0){\line(0,1){24}}
\put(-4.5,24){\oval(6,9)[br]}
\put(1.5,24){\line(0,-1){24}}
\put(4.5,0){\oval(6,9)[tl]}
\end{picture}}
\put(120,18){\begin{picture}(0,24)(0,0)
\thinlines
\put(-1.5,0){\line(0,1){24}}
\put(-4.5,24){\oval(6,9)[br]}
\put(1.5,24){\line(0,-1){24}}
\put(4.5,0){\oval(6,9)[tl]}
\end{picture}}
\put(47,120){\begin{picture}(24,0)(0,0)
\thinlines
\put(0,1.5){\line(1,0){24}}
\put(24,4.5){\oval(9,6)[bl]}
\put(24,-1.5){\line(-1,0){24}}
\put(0,-4.5){\oval(9,6)[tr]}
\end{picture}}
\put(47,60){\begin{picture}(24,0)(0,0)
\thinlines
\put(0,1.5){\line(1,0){24}}
\put(24,4.5){\oval(9,6)[bl]}
\put(24,-1.5){\line(-1,0){24}}
\put(0,-4.5){\oval(9,6)[tr]}
\end{picture}}
\put(47,0){\begin{picture}(24,0)(0,0)
\thinlines
\put(0,1.5){\line(1,0){24}}
\put(24,4.5){\oval(9,6)[bl]}
\put(24,-1.5){\line(-1,0){24}}
\put(0,-4.5){\oval(9,6)[tr]}
\end{picture}}
\end{picture}
&
\setlength{\unitlength}{0.7pt}
\begin{picture}(160,160)(0,0)
\put(-10,100){\vector(0,1){40}}
\put(10,140){\vector(0,-1){40}}
\put(150,100){\vector(0,1){40}}
\put(170,140){\vector(0,-1){40}}
\put(-10,20){\vector(0,1){40}}
\put(10,60){\vector(0,-1){40}}
\put(150,20){\vector(0,1){40}}
\put(170,60){\vector(0,-1){40}}
\put(50,170){\vector(1,0){60}}
\put(110,150){\vector(-1,0){60}}
\put(50,90){\vector(1,0){60}}
\put(110,70){\vector(-1,0){60}}
\put(50,10){\vector(1,0){60}}
\put(110,-10){\vector(-1,0){60}}
\put(-50,100){\makebox(40,40){\footnotesize{$\mathsf{int}_{\mathbf{L}_1}$}}}
\put(10,100){\makebox(40,40){\footnotesize{$\mathsf{tau}_{\mathbf{L}_1}$}}}
\put(110,100){\makebox(40,40){\footnotesize{$\mathsf{int}_{\mathbf{L}_2}$}}}
\put(170,100){\makebox(40,40){\footnotesize{$\mathsf{tau}_{\mathbf{L}_2}$}}}
\put(-50,20){\makebox(40,40){\footnotesize{$\mathsf{iota}_{\mathbf{L}_1}$}}}
\put(10,20){\makebox(40,40){\footnotesize{$\mathsf{ext}_{\mathbf{L}_1}$}}}
\put(110,20){\makebox(40,40){\footnotesize{$\mathsf{iota}_{\mathbf{L}_2}$}}}
\put(170,20){\makebox(40,40){\footnotesize{$\mathsf{ext}_{\mathbf{L}_2}$}}}
\put(48,160){\makebox(60,40){\footnotesize{${\exists}\mathsf{typ}(\mathbf{h})$}}}
\put(58,120){\makebox(60,40){\footnotesize{${\mathsf{typ}(\mathbf{h})}^{-1}$}}}
\put(50,80){\makebox(60,40){\footnotesize{$\mathsf{right}(\mathbf{h})$}}}
\put(50,40){\makebox(60,40){\footnotesize{$\mathsf{left}(\mathbf{h})$}}}
\put(58,0){\makebox(60,40){\footnotesize{${\mathsf{inst}(\mathbf{h})}^{-1}$}}}
\put(50,-41){\makebox(60,40){\footnotesize{${\exists}\mathsf{inst}(\mathbf{h})$}}}
\put(-80,120){\makebox(160,80) {$\langle {\wp}\mathsf{typ}(\mathbf{L}_1), \supseteq \rangle$}}
\put(80,120){\makebox(160,80)  {$\langle {\wp}\mathsf{typ}(\mathbf{L}_2), \supseteq \rangle$}}
\put(-80,40){\makebox(160,80) {$\langle \mathsf{elem}(\mathbf{L}_1), \leq_{\mathbf{L}_1} \rangle$}}
\put(80,40){\makebox(160,80)  {$\langle \mathsf{elem}(\mathbf{L}_2), \leq_{\mathbf{L}_2} \rangle$}}
\put(-80,-40){\makebox(160,80){$\langle {\wp}\mathsf{inst}(\mathbf{L}_1), \subseteq \rangle$}}
\put(80,-40){\makebox(160,80) {$\langle {\wp}\mathsf{inst}(\mathbf{L}_2), \subseteq \rangle$}}
\end{picture}
\\ \\ \\
(iconic) & (detailed)
\end{tabular}
\end{center}
\caption{The Galois connection version of a concept morphism}
\label{ext-int-embeddings}
\end{figure}

\subsubsection{Unit and Counit.}

Any classification can be compared with the power of its instance.
Given any classification $\mathbf{A}$,
the eta infomorphism 
$\eta_{\mathbf{A}} 
= \langle \mathrm{id}_{\mathsf{inst}(\mathbf{A})}, \mathsf{ext}_{\mathbf{A}} \rangle 
: \mathbf{A} \rightleftharpoons \check{\wp}\,(\mathsf{inst}(\mathbf{A}))$ 
from $\mathbf{A}$ to the instance power classification of the instance set of $\mathbf{A}$,
is the identity function on instances and the extent function on types.
For any infomorphism
$\mathbf{f} 
= \langle \mathsf{inst}(\mathbf{f}), \mathsf{typ}(\mathbf{f}) \rangle : \mathbf{A}_1 \rightleftharpoons \mathbf{A}_2$,
the following naturality conditions holds:
$\eta_{\mathbf{A}_1} \circ_{\mathsf{Clsn}} \check{\wp}\,{\mathsf{inst}(\mathbf{f})}
= \mathbf{f} \circ_{\mathsf{Clsn}} \eta_{\mathbf{A}_2}$.
Hence,
there is an eta natural transformation
$\eta : \mathrm{id}_{\mathsf{Clsn}} \Rightarrow \mathsf{inst}^{\mathrm{op}} \circ \check{\wp} : \mathsf{Clsn} \rightarrow \mathsf{Clsn}$
(bottom middle Figure~\ref{concept-lattice-component-architecture}),
with
$\eta \circ \mathsf{inst}^{\mathrm{op}} = 1_{\mathsf{inst}^{\mathrm{op}}}$
and
$\eta \circ \mathsf{typ} = \mathsf{ext}$.

Dually,
any concept lattice can be compared with the power of its type.
Given any concept lattice $\mathbf{L}$,
the epsilon infomorphism 
$\varepsilon_{\mathbf{L}} 
= \langle \mathsf{int}_{\mathbf{L}}, {\exists}\tau_{\mathbf{L}} {\cdot} \bigwedge_{\mathbf{L}}, \mathsf{int}_{\mathsf{clsn}(\mathbf{L})}, \mathrm{id}_{\mathsf{typ}(\mathbf{L})} \rangle 
: \hat{\wp}\,(\mathsf{typ}(\mathbf{L})) \rightleftharpoons \mathbf{L}$ 
to $\mathbf{L}$ from the type power concept lattice of the type set of $\mathbf{L}$,
has 
the intent function on elements (concepts) as its left function,
the composition of the direct image of type embedding with the meet operation as its right function,
the intent function of its underlying classification as its instance function
and 
the identity function as its type function.
The left function of $\varepsilon_{\mathbf{L}}$ preserves instances,
since the intent function of the classification is the restriction to instances of the intent function:
$\mathsf{int}_{\mathsf{clsn}(\mathbf{L})} = \iota_{\mathbf{L}} \cdot \mathsf{int}_{\mathbf{L}}$.
The right function of $\varepsilon_{\mathbf{L}}$ preserves types,
since 
$\{{-}\}_{\mathbf{L}} \cdot \exists \tau_{\mathbf{L}} \cdot \bigwedge_{\mathbf{L}} = \tau_{\mathbf{L}}$.
The left and right functions of $\varepsilon_{\mathbf{L}}$ form a Galois connection,
since
$\mathsf{int}_{\mathbf{L}}(c) \supseteq Y$ iff $c \leq_{\mathbf{L}} \bigwedge_{\mathbf{L}}\tau_{\mathbf{L}}[Y]$
for any concept $c \in \mathsf{elem}(\mathbf{L})$ and any type subset $Y \in {\wp}\,\mathsf{typ}(\mathbf{L})$.
For any concept morphism
$\mathbf{h} 
= \langle \mathsf{left}(\mathbf{h}), \mathsf{right}(\mathbf{h}), \mathsf{inst}(\mathbf{h}), \mathsf{typ}(\mathbf{h}) \rangle : \mathbf{L}_1 \rightleftharpoons \mathbf{L}_2$,
the following naturality conditions holds:
$\varepsilon_{\mathbf{L}_1} \circ_{\mathsf{Clg}} \mathbf{h}
= \hat{\wp}\,{\mathsf{typ}(\mathbf{h})} \circ_{\mathsf{Clg}} \varepsilon_{\mathbf{L}_2}$.
Hence,
there is an epsilon natural transformation
$\varepsilon : \mathsf{typ} \circ \hat{\wp} \Rightarrow \mathrm{id}_{\mathsf{Clg}} : \mathsf{Clg} \rightarrow \mathsf{Clg}$
(bottom middle Figure~\ref{concept-lattice-component-architecture}),
with
$\varepsilon \circ \mathsf{left}^{\mathrm{op}} = \mathsf{int}^{\mathrm{op}}$,
$\varepsilon \circ \mathsf{right} = (\tau \circ \exists) \bullet \bigwedge$,
$\varepsilon \circ \mathsf{inst}^{\mathrm{op}} = \mathsf{clsn} \circ \mathsf{int}^{\mathrm{op}}$
and
$\varepsilon \circ \mathsf{typ} = 1_{\mathsf{typ}}$.

\begin{figure}
\begin{center}
\begin{tabular}{c}
\begin{tabular}{c@{\hspace{110pt}}c}
\setlength{\unitlength}{0.7pt}
\begin{picture}(120,200)(0,-20)
\put(-60,150){\makebox(120,60) {${\wp}\mathsf{typ}(\mathbf{L}_1)^{\mathrm{op}}$}}
\put(60,150){\makebox(120,60)  {${\wp}\mathsf{typ}(\mathbf{L}_2)^{\mathrm{op}}$}}
\put(-60,90){\makebox(120,60) {$\mathsf{th}(\mathbf{L}_1)$}}
\put(60,90){\makebox(120,60)  {$\mathsf{th}(\mathbf{L}_2)$}}
\put(-60,30){\makebox(120,60) {$\mathsf{elem}(\mathbf{L}_1)$}}
\put(60,30){\makebox(120,60)  {$\mathsf{elem}(\mathbf{L}_2)$}}
\put(-60,-30){\makebox(120,60){${\wp}\mathsf{inst}(\mathbf{L}_1)$}}
\put(60,-30){\makebox(120,60) {${\wp}\mathsf{inst}(\mathbf{L}_2)$}}
\put(-10,120){\makebox(60,60){\footnotesize{$\mathsf{clsr}_{\mathbf{L}_1}$}}}
\put(75,120){\makebox(60,60){\footnotesize{$\mathsf{clsr}_{\mathbf{L}_2}$}}}
\put(-10,60){\makebox(60,60){\footnotesize{$\mathsf{lift}_{\mathbf{L}_1}$}}}
\put(75,60){\makebox(60,60){\footnotesize{$\mathsf{lift}_{\mathbf{L}_2}$}}}
\put(-80,70){\makebox(60,60){\footnotesize{$\mathsf{intent}_{\mathbf{L}_1}$}}}
\put(142,115){\makebox(60,60){\footnotesize{$\mathsf{intent}_{\mathbf{L}_2}$}}}
\put(-57,0){\makebox(60,60){\footnotesize{$\mathsf{extent}_{\mathbf{L}_1}$}}}
\put(120,0){\makebox(60,60){\footnotesize{$\mathsf{extent}_{\mathbf{L}_2}$}}}
\put(0,163){\makebox(120,60){\footnotesize{$\mathsf{inv}(\mathsf{typ}(\mathbf{h}))$}}}
\put(0,101){\makebox(120,60){\footnotesize{$\mathsf{th}(\mathbf{h})$}}}
\put(0,41){\makebox(120,60){\footnotesize{$\mathsf{adj}(\mathbf{h})$}}}
\put(0,-18){\makebox(120,60){\footnotesize{$\mathsf{dir}(\mathsf{inst}(\mathbf{h}))$}}}
\thicklines
\put(-15,165){\vector(1,2){0}}
\qbezier(-15,75)(-40,120)(-15,165)
\put(135,165){\vector(-1,2){0}}
\qbezier(135,75)(160,120)(135,165)
\put(0,135){\vector(0,1){30}}
\put(120,135){\vector(0,1){30}}
\put(0,75){\vector(0,1){30}}
\put(120,75){\vector(0,1){30}}
\put(0,15){\vector(0,1){30}}
\put(120,15){\vector(0,1){30}}
\put(80,180){\vector(-1,0){40}}
\put(85,120){\vector(-1,0){50}}
\put(85,60){\vector(-1,0){50}}
\put(80,0){\vector(-1,0){40}}
\end{picture}

&

\setlength{\unitlength}{0.65pt}
\begin{picture}(160,160)(0,0)
\put(-62,227){\vector(1,2){0}}
\qbezier(-62,93)(-92,160)(-62,227)
\qbezier(-44,100)(-71,160)(-44,220)
\put(-44,100){\vector(1,-2){0}}
\put(222,227){\vector(-1,2){0}}
\qbezier(222,93)(252,160)(222,227)
\qbezier(204,100)(231,160)(204,220)
\put(204,100){\vector(-1,-2){0}}
\put(-10,180){\vector(0,1){40}}
\put(10,220){\vector(0,-1){40}}
\put(150,180){\vector(0,1){40}}
\put(170,220){\vector(0,-1){40}}
\put(-10,100){\vector(0,1){40}}
\put(10,140){\vector(0,-1){40}}
\put(150,100){\vector(0,1){40}}
\put(170,140){\vector(0,-1){40}}
\put(-10,20){\vector(0,1){40}}
\put(10,60){\vector(0,-1){40}}
\put(150,20){\vector(0,1){40}}
\put(170,60){\vector(0,-1){40}}
\put(55,250){\vector(1,0){50}}
\put(105,230){\vector(-1,0){50}}
\put(55,170){\vector(1,0){50}}
\put(105,150){\vector(-1,0){50}}
\put(55,90){\vector(1,0){50}}
\put(105,70){\vector(-1,0){50}}
\put(55,10){\vector(1,0){50}}
\put(105,-10){\vector(-1,0){50}}
\put(-45,180){\makebox(40,40){\footnotesize{$\mathsf{clo}_{\mathbf{L}_1}$}}}
\put(2,180){\makebox(40,40){\footnotesize{$\mathrm{id}$}}}
\put(123,180){\makebox(20,40){\footnotesize{$\mathsf{clo}_{\mathbf{L}_2}$}}}
\put(161,180){\makebox(40,40){\footnotesize{$\mathrm{id}$}}}
\put(-45,95){\makebox(40,40){\footnotesize{$\mathsf{int}_{\mathbf{L}_1}$}}}
\put(10,100){\makebox(40,40){\footnotesize{$\mathsf{tau}_{\mathbf{L}_1}$}}}
\put(8,90){\makebox(40,40){\scriptsize{$(\mathrm{iso})$}}}
\put(115,95){\makebox(40,40){\footnotesize{$\mathsf{int}_{\mathbf{L}_2}$}}}
\put(170,100){\makebox(40,40){\footnotesize{$\mathsf{tau}_{\mathbf{L}_2}$}}}
\put(168,90){\makebox(40,40){\scriptsize{$(\mathrm{iso})$}}}
\put(-48,20){\makebox(40,40){\footnotesize{$\mathsf{iota}_{\mathbf{L}_1}$}}}
\put(10,20){\makebox(40,40){\footnotesize{$\mathsf{ext}_{\mathbf{L}_1}$}}}
\put(112,20){\makebox(40,40){\footnotesize{$\mathsf{iota}_{\mathbf{L}_2}$}}}
\put(170,20){\makebox(40,40){\footnotesize{$\mathsf{ext}_{\mathbf{L}_2}$}}}
\put(-109,114){\makebox(40,40){\footnotesize{$\mathsf{int}_{\mathbf{L}_1}$}}}
\put(-55,114){\makebox(40,40){\footnotesize{$\mathsf{tau}_{\mathbf{L}_1}$}}}
\put(181,164){\makebox(40,40){\footnotesize{$\mathsf{int}_{\mathbf{L}_2}$}}}
\put(235,164){\makebox(40,40){\footnotesize{$\mathsf{tau}_{\mathbf{L}_2}$}}}
\put(48,240){\makebox(60,40){\footnotesize{${\exists}\mathsf{typ}(\mathbf{h})$}}}
\put(58,200){\makebox(60,40){\footnotesize{${\mathsf{typ}(\mathbf{h})}^{-1}$}}}
\put(48,160){\makebox(60,40){\footnotesize{${\exists}\mathsf{typ}(\mathbf{h})$}}}
\put(60,120){\makebox(60,40){\footnotesize{$\mathsf{clo}_{\mathbf{L}_2} {\cdot}\, {\mathsf{typ}(\mathbf{h})}^{-1}$}}}
\put(50,80){\makebox(60,40){\footnotesize{$\mathsf{right}(\mathbf{h})$}}}
\put(50,40){\makebox(60,40){\footnotesize{$\mathsf{left}(\mathbf{h})$}}}
\put(58,0){\makebox(60,40){\footnotesize{${\mathsf{inst}(\mathbf{h})}^{-1}$}}}
\put(50,-41){\makebox(60,40){\footnotesize{${\exists}\mathsf{inst}(\mathbf{h})$}}}
\put(-80,200){\makebox(160,80) {$\langle {\wp}\mathsf{typ}(\mathbf{L}_1), \supseteq \rangle$}}
\put(80,200){\makebox(160,80)  {$\langle {\wp}\mathsf{typ}(\mathbf{L}_2), \supseteq \rangle$}}
\put(-80,120){\makebox(160,80) {$\langle {\wp}\mathsf{typ}(\mathbf{L}_1), \vdash_{\mathbf{L}_1} \rangle$}}
\put(80,120){\makebox(160,80)  {$\langle {\wp}\mathsf{typ}(\mathbf{L}_2), \vdash_{\mathbf{L}_2} \rangle$}}
\put(-80,40){\makebox(160,80) {$\langle \mathsf{elem}(\mathbf{L}_1), \leq_{\mathbf{L}_1} \rangle$}}
\put(80,40){\makebox(160,80)  {$\langle \mathsf{elem}(\mathbf{L}_2), \leq_{\mathbf{L}_2} \rangle$}}
\put(-80,-40){\makebox(160,80){$\langle {\wp}\mathsf{inst}(\mathbf{L}_1), \subseteq \rangle$}}
\put(80,-40){\makebox(160,80) {$\langle {\wp}\mathsf{inst}(\mathbf{L}_2), \subseteq \rangle$}}
\end{picture}
\\ \\ \\
\underline{iconic} & \underline{detailed} 
\end{tabular}

\\ \\

\begin{tabular}{c@{\hspace{20pt}}c}
\footnotesize{$\begin{array}[t]{rcl}
\mathbf{A}_2                     & \mapsto & \mathsf{elem}(\mathbf{L}_1)                   \\
\mathbf{B}_2                     & \mapsto & {\wp}\mathsf{typ}(\mathbf{L}_1)^{\mathrm{op}} \\
\mathsf{ker}(\hat{\mathbf{g}}_2) & \mapsto & \mathsf{th}(\mathbf{L}_1)                     \\
\mathbf{g}_2 & \mapsto           & \mathsf{intent}_{\mathbf{L}_1}                          \\
\check{\mathbf{g}}_2             & \mapsto & \mathsf{int}_{\mathbf{L}_1}                   \\
\hat{\mathbf{g}}_2               & \mapsto & \mathsf{tau}_{\mathbf{L}_1}                   \\
\mathsf{lift}(\mathbf{g}_2)      & \mapsto & \mathsf{lift}_{\mathbf{L}_1}                  \\
\mathsf{int}(\mathbf{g}_2)       & \mapsto & \mathsf{clsr}_{\mathbf{L}_1}
\end{array}$}
&
\footnotesize{$\begin{array}[t]{rcl}
\mathbf{a}                       & \mapsto & \mathsf{adj}(\mathbf{h})                      \\
\check{\mathbf{a}}               & \mapsto & \mathsf{left}(\mathbf{h})                     \\
\hat{\mathbf{a}}                 & \mapsto & \mathsf{right}(\mathbf{h})                    \\
\mathbf{b}                       & \mapsto & \mathsf{inv}(\mathsf{typ}(\mathbf{h}))        \\
\check{\mathbf{b}}               & \mapsto & {\mathsf{typ}(\mathbf{h})}^{-1}               \\
\hat{\mathbf{b}}                 & \mapsto & {\exists}\mathsf{typ}(\mathbf{h})             \\
\mathbf{d}                       & \mapsto & \mathsf{th}(\mathbf{h}) \\
\check{\mathbf{d}}               & \mapsto & \mathsf{clo}_{\mathbf{L}_2} {\cdot}\, {\mathsf{typ}(\mathbf{h})}^{-1} \\
\hat{\mathbf{d}}                 & \mapsto & {\exists}\mathsf{typ}(\mathbf{h})
\end{array}$}
\end{tabular}
\\ \\
\underline{bindings to Figure~\ref{quartet-factorization} elements}

\end{tabular}
\end{center}
\caption{Intensional Conceptual Structure}
\label{intensional-conceptual-structure}
\end{figure}

\subsubsection{Infomorphisms}\label{subsubsec:classification:functor:morphisms}

Any concept lattice morphism $\mathbf{h} = \clgmor{h}{L}{K}$
from concept lattice $\mathbf{L}$ to concept lattice $\mathbf{K}$
has an associated infomorphism 
$\mathsf{clsn}(\mathbf{h})
= \langle \mathsf{inst}(\mathbf{h}), \mathsf{typ}(\mathbf{h}) \rangle : \mathsf{clsn}(\mathbf{L}) \rightleftharpoons \mathsf{clsn}(\mathbf{K})$.
The fundamental property of infomorphisms is an easy translation of the adjointness condition for 
$\mathsf{adj}(\mathbf{h}) : \mathsf{lat}(\mathbf{K}) \rightleftharpoons \mathsf{lat}(\mathbf{L})$
and the commutativity of the instance/type functions with the adjoint pair of monotonic functions (left/right functions).

\subsubsection{The Classification Functor}\label{subsubsec:classification:functor}

The classification functor 
$\mathsf{clsn} : \mathsf{Clg} \rightarrow \mathsf{Clsn}$,
from the category of concept lattices and concept morphisms
to the category of classifications and infomorphisms,
is defined below via its object and morphism class functions.

\subsubsection{Equivalence}\label{subsubsec:equivalence} 

In a concept preorder,
if two elements $c_1$ and $c_2$ are equivalent $c_1 \equiv c_2$,
then they have the same extent and intent.
Conversely,
if two elements $c_1$ and $c_2$ have the same extent and intent,
then 
$c_1 \equiv \bigvee_{\mathbf{L}} [\mathsf{ext}(c_1)] = \bigvee_{\mathbf{L}} [\mathsf{ext}(c_2)] \equiv c_2$ 
and
$c_1 \equiv \bigwedge_{\mathbf{L}} [\mathsf{int}(c_1)] = \bigwedge_{\mathbf{L}} [\mathsf{int}(c_2)] \equiv c_2$.

The functor composition $\mathsf{clsn} \circ \mathsf{clg}$ is naturally isomorphic to the identity functor $\mathrm{id}_{\mathsf{Clg}}$.
To see this, 
let $\mathbf{L} = \clg{L}$ be a concept lattice with associated classification $\mathsf{clsn}(\mathbf{L}) = \clsn{L}$.
We assert the isomorphism $\mathbf{L} \cong \mathsf{clg}(\mathsf{clsn}(\mathbf{L}))$\footnote{This is part of the fundamental theorem of formal concept analysis \cite{ganter:wille:99}.}.
The instance derivation operator in $\mathsf{clsn}(\mathbf{L})$ is given by
$X^{\mathsf{clsn}(\mathbf{L})} 
= \mathsf{int}_{\mathbf{L}}\left(\bigvee_{\mathbf{L}}\iota_{\mathbf{L}}[X]\right) 
= \{ y \in \mathsf{typ}(\mathbf{L}) 
  \mid \bigvee_{\mathbf{L}}\iota_{\mathbf{L}}[X] \leq_{\mathbf{L}} \tau_{\mathbf{L}}(y) \}$
for any subset of instances $X \subseteq \mathsf{inst}(\mathbf{L})$.
Dually,
the type derivation operator in $\mathsf{clsn}(\mathbf{L})$ is given by
$Y^{\mathsf{clsn}(\mathbf{L})} 
= \mathsf{ext}_{\mathbf{L}}\left(\bigwedge_{\mathbf{L}}\tau_{\mathbf{L}}[Y]\right) 
= \{ x \in \mathsf{inst}(\mathbf{L}) 
  \mid \iota_{\mathbf{L}}(x) \leq_{\mathbf{L}} \bigwedge_{\mathbf{L}}\tau_{\mathbf{L}}[Y] \}$
for any subset of types $Y \subseteq \mathsf{typ}(\mathbf{L})$.
An element of $(X, Y) \in \mathsf{clg}(\mathsf{clsn}(\mathbf{L}))$,
a formal concept of $\mathsf{clsn}(\mathbf{L})$,
is a pair $(X, Y)$, 
where $X \subseteq \mathsf{inst}(\mathbf{L})$,
where $Y \subseteq \mathsf{typ}(\mathbf{L})$,
$X 
= Y^{\mathsf{clsn}(\mathbf{L})}
= \mathsf{ext}_{\mathbf{L}}\left(\bigwedge_{\mathbf{L}}\tau_{\mathbf{L}}[Y]\right)$ 
and 
$Y 
= X^{\mathsf{clsn}(\mathbf{L})}
= \mathsf{int}_{\mathbf{L}}\left(\bigvee_{\mathbf{L}}\iota_{\mathbf{L}}[X]\right)$. 
But,
$\bigvee_{\mathbf{L}} \iota_{\mathbf{L}}[\mathsf{ext}_{\mathbf{L}}(c)]
= c 
= \bigwedge_{\mathbf{L}} \tau_{\mathbf{L}}[\mathsf{int}_{\mathbf{L}}(c)]$
for every element $c \in L$.
Hence,
$\bigvee_{\mathbf{L}} \iota_{\mathbf{L}}[X] = \bigwedge_{\mathbf{L}} \tau_{\mathbf{L}}[Y]$
for the pair $(X, Y)$.
Hence,
the map 
$\mathsf{clg}(\mathsf{clsn}(\mathbf{L})) \rightarrow \mathbf{L}$ 
defined by 
$(X, Y) \mapsto \bigvee_{\mathbf{L}} \iota_{\mathbf{L}}[X] = \bigwedge_{\mathbf{L}} \tau_{\mathbf{L}}[Y]$ 
is a well-defined monotonic function.
Its inverse is the monotonic function
$\mathbf{L} \rightarrow \mathsf{clg}(\mathsf{clsn}(\mathbf{L}))$
defined by
$c \mapsto \left(\mathsf{ext}_{\mathbf{L}}(c), \mathsf{int}_{\mathbf{L}}(c)\right)$ 
for every lattice element $c \in \mathbf{L}$.
Let $\mathbf{h} = \clgmor{h}{L}{K}$ be a concept lattice morphism from concept lattice $\mathbf{L}$ to concept lattice $\mathbf{K}$,
with associated infomorphism 
$\mathsf{clsn}(\mathbf{h}) = \langle \mathsf{inst}(\mathbf{h}), \mathsf{typ}(\mathbf{h}) \rangle : \mathsf{clsn}(\mathbf{L}) \rightleftharpoons \mathsf{clsn}(\mathbf{K})$.
Then, up to isomorphism, $\mathsf{clg}(\mathsf{clsn}(\mathbf{h})) = \mathbf{h}$.
This defines the natural isomorphism: $\mathsf{clsn} \circ \mathsf{clg} \cong \mathrm{id}_{\mathsf{Clg}}$.

The functor composition $\mathsf{clg} \circ \mathsf{clsn}$ is equal to the identity functor $\mathrm{id}_{\mathsf{Clsn}}$.
To see this,
consider whether 
$\mathbf{A} = \mathsf{clsn}(\mathsf{clg}(\mathbf{A}))$ for any classification $\mathbf{A}$.
Obviously,
the type and instance sets are the same.
What about the classification relations?
For any instance $x \in \mathsf{inst}(\mathbf{A})$ and any type $y \in \mathsf{typ}(\mathbf{A})$,
any classification $x \models_{\mathbf{L}} y$ can be expressed in terms of the instance/type mappings and concept order as
$\iota_{\mathbf{L}}(x) \leq_{\mathbf{L}} \tau_{\mathbf{L}}(y)$,
since
$x \models_{\mathbf{A}} y$ iff $y \in \{x\}^{\mathbf{A}}$ 
iff 
$\{x\}^{\mathbf{A}} \supseteq \{y\}^{\bullet}$ 
iff
$\iota_{\mathbf{A}}(x) \leq_{\mathbf{A}} \tau_{\mathbf{A}}(y)$.
More completely,
the classification relation decomposes as the relational composition:
$\models_{\mathbf{A}} 
= \iota_{\mathbf{A}} \circ \leq_{\mathbf{A}} \circ \tau_{\mathbf{A}}
= \iota_{\mathbf{A}} \circ \tau_{\mathbf{A}}$.
Hence, $\mathbf{A} = \mathsf{clsn}(\mathsf{clg}(\mathbf{A}))$.
What about infomorphisms?
The infomorphisms $\mathbf{f}$ and $\mathsf{clsn}(\mathsf{clg}(\mathbf{f}))$ are equal.

For any classification $\mathsf{clsn}(\mathbf{L}) = \clsn{L}$,
two other identities may also be useful:
the intent function for $\mathbf{A}$ is 
the composition of the instance embedding function with the intent function for $\mathsf{clg}(\mathbf{A})$,
$\mathsf{int}_{\mathbf{A}} = \iota_{\mathbf{A}} \cdot \mathsf{int}_{\mathbf{A}}$; and
the extent function for $\mathbf{A}$ is 
the composition of the type embedding function with the extent function for $\mathsf{clg}(\mathbf{A})$,
$\mathsf{ext}_{\mathbf{A}} = \tau_{\mathbf{A}} \cdot \mathsf{ext}_{\mathbf{A}}$.
Both can be verified using the fact that 
$x \models_{\mathbf{A}} y$ is equivalent to $\iota_{\mathbf{A}}(x) \leq_{\mathbf{A}} \tau_{\mathbf{A}}(y)$
for any instance $x \in \mathsf{inst}(\mathbf{A})$ and any type $y \in \mathsf{typ}(\mathbf{A})$.

\end{document}